\documentclass[preprint,10pt]{aastex}

\begin{document}

\title{Ultraviolet-Selected Field and Pre-Main-Sequence Stars Towards Taurus and Upper Scorpius
\let\thefootnote\relax\footnotetext{Accepted for publication in \textit{The Astronomical Journal}. \copyright\ 2010. The American Astronomical Society. All rights reserved.}}

\author{K. Findeisen, L. Hillenbrand}
\affil{California Institute of Technology, MC 249-17, Pasadena, CA 91125}
\email{krzys@astro.caltech.edu, lah@astro.caltech.edu}

\begin{abstract}
We have carried out a Galaxy Evolution Explorer (GALEX) Cycle~1 guest investigator program covering 56~square degrees near the Taurus T association and 12~square degrees along the northern edge of the Upper Scorpius OB association. We combined photometry in the GALEX FUV and NUV bands with data from the Two Micron All Sky Survey to identify candidate young ($\lesssim 100$~Myr old) stars as those with an ultraviolet excess relative to older main sequence stars. Follow-up spectroscopy of a partial sample of these candidates suggest 5 new members of Taurus, with 8-20 expected from additional observations, and 5 new members of Upper Scorpius, with 3-6 expected from additional observations. These candidate new members appear to represent a distributed, non-clustered population in either region, although our sample statistics are as of yet too poor to constrain the nature or extent of this population. Rather, our study demonstrates the ability of GALEX observations to identify young stellar populations distributed over a wide area of the sky. We also highlight the necessity of a better understanding of the Galactic ultraviolet source population to support similar investigations. In particular, we report a large population of stars with an ultraviolet excess but no optical indicators of stellar activity or accretion, and briefly argue against several interpretations of these sources.
\end{abstract}


\keywords{stars: pre-main sequence -- stars: activity -- ultraviolet: stars -- Galaxy: stellar content}

\section{Introduction}

Stellar populations associated with most nearby star-forming regions tend to be a few Myr old, while molecular clouds are asserted to have lifetimes of 10~Myr or more. If clouds are long-lived, a subject of current debate \citep[and references therein]{sf_review}, and they form stars for a significant fraction of their lifetimes, known young stars should represent only a small portion of a given association. The absence of an observed population of 3-10~Myr old stars associated with star-forming regions is known as the post-T-Tauri star problem \citep{ptts_first, ptts_intro}. Confirmation of a lack of older association members would place strong constraints on the evolution of molecular clouds, while discovery of a $\sim 10$~Myr old population would allow new probes of clouds' star formation histories, overall star formation efficiencies, and circumstellar disk evolution \citep{xray_review}. However, many surveys are biased toward few Myr old objects, or are restricted to small areas around molecular clouds where the population should be dominated by very young stars that have not had time to migrate away.

All stars cooler than type F and younger than $\sim 100$~Myr are prominent soft X-ray sources \citep{xray_review}, and the advent of sensitive X-ray observatories has made large area X-ray surveys the natural tool to probe young stellar populations of all ages \citep[among many others]{rosat, coup, xest}. Because stars show little evolution in their X-ray properties until they are 100~Myr old, X-ray data alone cannot distinguish a 3~Myr old pre-main sequence star from a 100~Myr old main sequence star. As a result, X-ray surveys of an association can be heavily contaminated by unrelated field stars \citep{ptts_xray} and any such survey requires spectroscopic confirmation.

X-ray emission is a tracer of coronal activity, and other stellar activity indicators can in principle be used to identify young stars as well. In particular, the chromospheres and transition regions of active stars produce ultraviolet emission from a variety of lines \citep{chromo_review}. 
Chromospheric activity, as traced by the Ca~II lines at 3934 and 3968\AA, has a similar dependence on age as X-ray emission, remaining constant up to $\sim 100$~Myr and decaying together with rotation speed in older stars \citep{chro_levels}. This makes it a viable youth indicator except among early F or hotter stars, where chromospheric activity shows no correlation with either rotation \citep{bluerot} or age over the range 50~Myr-3~Gyr \citep{blueyouth}, and among M stars, where activity remains constant for almost a Gyr \citep{redyouth}. Ultraviolet surveys of chromosphere emission, like X-ray surveys and most other young star search techniques, require spectroscopic follow-up to distinguish the youngest stars from 100 Myr old active field stars.

The Galaxy Evolution Explorer (GALEX), launched in 2003, is an ideal instrument for wide-field ultraviolet surveys. GALEX images the sky in a far-ultraviolet (FUV, 1344-1786\AA) and a near-ultraviolet (NUV, 1771-2831\AA) band over a circular field of view more than a degree across. The observatory's microchannel plate detectors can reach 20th magnitude point sources in 100~seconds, while the $\sim 5$\arcsec~PSF provides acceptable resolution for the sparse fields of young stellar associations. GALEX can observe for at most 28~minutes of each 99~minute orbit. 

As we show in section~\ref{galexmax}, a 300~s GALEX exposure can detect the chromospheres of active main sequence stars at $3\sigma$ down to type K2 in the nearest star-forming regions. GALEX thus requires 316~s~deg$^{-2}$ to carry out a survey. For comparison, the EPIC-pn camera on XMM-Newton, the most sensitive soft X-ray imager currently in orbit, can detect the same K2 star ($f_X \sim 3 \times 10^{-13}$ erg cm s$^{-2}$, assuming $L_X \sim 10^{-3} L_\textrm{bol}$) in 260~s \citep[section~3.3.8]{XMM}. However, EPIC-pn has a much smaller field of view, requiring 1290~s~deg$^{-2}$ to survey a given region. Even correcting for GALEX's poor duty cycle, and the resulting shortage of observing time, GALEX effectively requires 1120~s~deg$^{-2}$, while XMM (given 40 usable hours each 48-hour orbit) requires 1550~s~deg$^{-2}$. For solar-type stars in nearby star forming regions, then, GALEX has a slightly higher survey efficiency than X-ray observatories.

The Taurus T association and the Upper Scorpius OB association represent the two nearest regions of recent star formation, both about 140~pc away and 35~pc across. Taurus is the prototype low-mass star forming region, with known members mostly G, K, or M stars less than 2~Myr old \citep{ptts_intro, hsfr_tau}. Taurus still contains $3-4 \times 10^4 M_\sun$ of molecular gas with a highly filamentary structure; the T association is dominated by subgroups of stars associated with the densest regions of this gas, although there are a handful of members scattered over a larger area. \citet{slesnick1} have identified a new population of pre-main-sequence stars east of the known Taurus association ($\alpha \gtrsim 80\degr$) and suggest the full spatial distribution has not yet been explored. Upper Scorpius, in contrast, is a classic high-mass star forming region. Stellar ages are consistent with a single burst of star formation 5~Myr ago \citep{slesnick2}, possibly triggered by the nearby Upper Centaurus-Lupus association \citep{hsfr_sco}, and any gas has already been blown out. Unlike Taurus, the Upper Scorpius association has no apparent structure. 

We construct a toy model to estimate the number of post-T-Tauri stars hotter than type K2 we might find in a survey of the sky around these two regions. We assume that stars form at a constant rate $dN/dt$ over a disk on the sky of angular radius $r_0$. A generation of stars of age $t$ will have spread into a uniform disk of radius $\sqrt{r_0^2 + (\sigma t)^2}$, where $\sigma$ is the velocity dispersion of the association. If the region has been constantly forming stars since some time $t_0$ before the present, the density of stars seen at some distance $r$ from the center of the association is 
\begin{equation}
\Sigma(r) = \left\{ \begin{array}{lcll}
\int_{\sqrt{r^2-r_0^2}/\sigma}^{t_0} \frac{dN}{dt} \frac{dt}{\pi(r_0^2 + (\sigma t)^2)} & = & 
	\frac{dN}{dt} \frac{1}{\pi r_0 \sigma} \left(\arctan{\left(\frac{t_0\sigma}{r_0}\right)} - \arctan{\sqrt{(r/r_0)^2-1}}\right) & (r > r_0)	\\
\int_{0}^{t_0} \frac{dN}{dt} \frac{dt}{\pi(r_0^2 + (\sigma t)^2)} & = & 
	\frac{dN}{dt} \frac{1}{\pi r_0 \sigma} \arctan{\left(\frac{t_0\sigma}{r_0}\right)} & (r \leq r_0)
\end{array} \right.
\end{equation}
This model can also be applied to a region that stopped forming stars at some time $t_f$ before the present, as long as hypothetical stars younger than $t_f$ have not had time to migrate to $r$.

We approximate the Upper Scorpius region as a disk with $r(\textrm{5~Myr}) = 6\degr$ centered on $(\alpha = 243\degr, \delta = -23\degr)$, although this excludes some outlying members; \citet{memberlist} find 34 K2 or hotter members within this circle, giving a mean star formation rate of 7~Myr$^{-1}$. Most of our survey fields are 8\degr\ from the center of the circle. The model predicts that a significant population will appear in these fields only if the region has been forming stars for at least 11~Myr. After 15~Myr, the density there would be 0.1~deg$^{-2}$; after 30~Myr, it would be 0.3~deg$^{-2}$. We can likewise model Taurus as a disk with $r(\textrm{2~Myr}) = 5\degr$ centered on $(67\degr, 26\degr)$, containing 6 members of type K2 or earlier, though this not only cuts off outliers but ignores the significant internal structure of the association. Many of our fields are also $\sim 8\degr$ from the disk center; a significant population should appear in them after 8~Myr. After 15~Myr, the density of K2 or earlier stars should be 0.07~deg$^{-2}$; by 30~Myr, the density approaches its asymptotic value of 0.15~deg$^{-2}$. In the case of both Taurus and Upper Scorpius, a survey covering several dozen square degrees should detect post-T-Tauri stars if they are present.

\section{Our GALEX Survey}

\subsection{Sensitivity}

Modest exposures with GALEX are sensitive to stars as late as G- or K-type at 140~pc, depending on the degree of ultraviolet excess and the degree of extinction. To quantify this limit, we found GALEX $3\sigma$ magnitude limits (Table~\ref{gxdetect}) using the GALEX Exposure Time Calculator\footnote{\url{http://sherpa.caltech.edu/gips/tools/expcalc.html}} for exposures of 300~s, 1000~s, and 1500~s, which cover the range of exposure times used by our observations presented in section~\ref{thedata}. 
For comparison we also give the sensitivity for the GALEX All-Sky-Survey (AIS) exposure time of 100~s.

In section~\ref{intro2mass}, we use the Two-Micron All-Sky Survey \citep[2MASS,][]{2MASS_cat} to characterize our GALEX sources, introducing a second set of sensitivity limits. 
The Explanatory Supplement to the 2MASS All Sky Data Release \citep{2MASS} gives completeness limits both for all detected point sources and for a subset with a photometric quality of A. Both limits are given in Table~\ref{2mdetect}. 
Since we require only C quality photometry from our 2MASS sources, we expect the completeness limit of our sample to be somewhere between the two values in Table~\ref{2mdetect}. 
By comparing the location of our fields to figures 7, 9, 22, and 24 from \citet[section VI.7.a]{2MASS}, we find these limits are unaffected by our fields' proximity to the Galactic plane.\label{maglimits}

We compare the limiting magnitudes of the GALEX and 2MASS data to the main sequence magnitudes found in appendix~\ref{mags_ms}. Assuming a distance to both Taurus and Upper Scorpius of 140~pc, a 300~s GALEX exposure can detect unextinguished photospheres down to spectral type F8 in the FUV and K5 in the NUV. However, both Taurus and Upper Scorpius have significant extinction. Assuming $A_\textrm{FUV}/A_V = 2.68$ and $A_\textrm{NUV}/A_V = 2.63$ for $R_V = 3.1$ (Gil de Paz 2007, private communication), we can detect stellar photospheres as long as they have an extinction less than that given in the first two columns of Table~\ref{maxav}. The 2MASS data are sensitive down to spectral type M, even with extinction, so they do not affect our overall sensitivity.

As we show in section~\ref{minex}, our procedure cannot reliably identify excesses below $\sim 5.3$~mag in the FUV and $\sim 2.6$~mag in the NUV. Therefore, the sensitivity of GALEX does not begin to affect our results until stars with this minimum excess cease to be detectable. The extinctions at which this occurs are given in the last two columns of Table~\ref{maxav}. $A_V$ in our Taurus fields is typically less than 3 \citep{tau_av}, with the exception of the field TauAur\_MOS23, where the L1529 dark cloud has extinctions ranging from 3-6. $A_V$ in our Upper Scorpius fields typically less than 2 \citep{hsfr_sco}. Therefore, our observations of stars of type K and later are limited by GALEX sensitivity. Under 2 magnitudes of extinction, GALEX can detect only those K4 stars with an NUV excess of at least 3.7 magnitudes, those M0 stars with an excess of at least 6.9 magnitudes, and those M2 stars with an excess of at least 8.7 magnitudes.\label{galexmax}

To estimate the excesses we expect from young stars, we take from \citet{valenti} narrow-band continuum fluxes of T~Tauri stars measured by IUE at 1760\AA\ and 1958\AA\ and assume the stars' FUV and NUV magnitudes correspond to the 1760\AA\ and 1958\AA\ fluxes, respectively. The magnitudes are underestimates because broad-band photometry will pick up additional flux from emission lines.
We estimate that those stars that \citet{valenti} designate as accreting classical T~Tauri stars (CTTS) typically have FUV excesses of 10-18~magnitudes and NUV excesses of 2-7~magnitudes. These stars should be detectable in either the FUV or the NUV down to spectral type K0 under $\sim 2$ magnitudes of extinction. 
Stars marked as non-accreting naked T~Tauri stars (NTTS) can have excesses of up to 10~magnitudes in the FUV and up to 2~magnitudes in the NUV, although some show no continuum excess at all. A 2~magnitude NUV excess is too small for us to easily distinguish, but we can detect sources with an FUV excess of 5-10 magnitudes down to spectral type F8-G8.

\subsection{Observations}\label{thedata}

\begin{figure*}
\includegraphics[angle=90,width=0.9\textwidth]{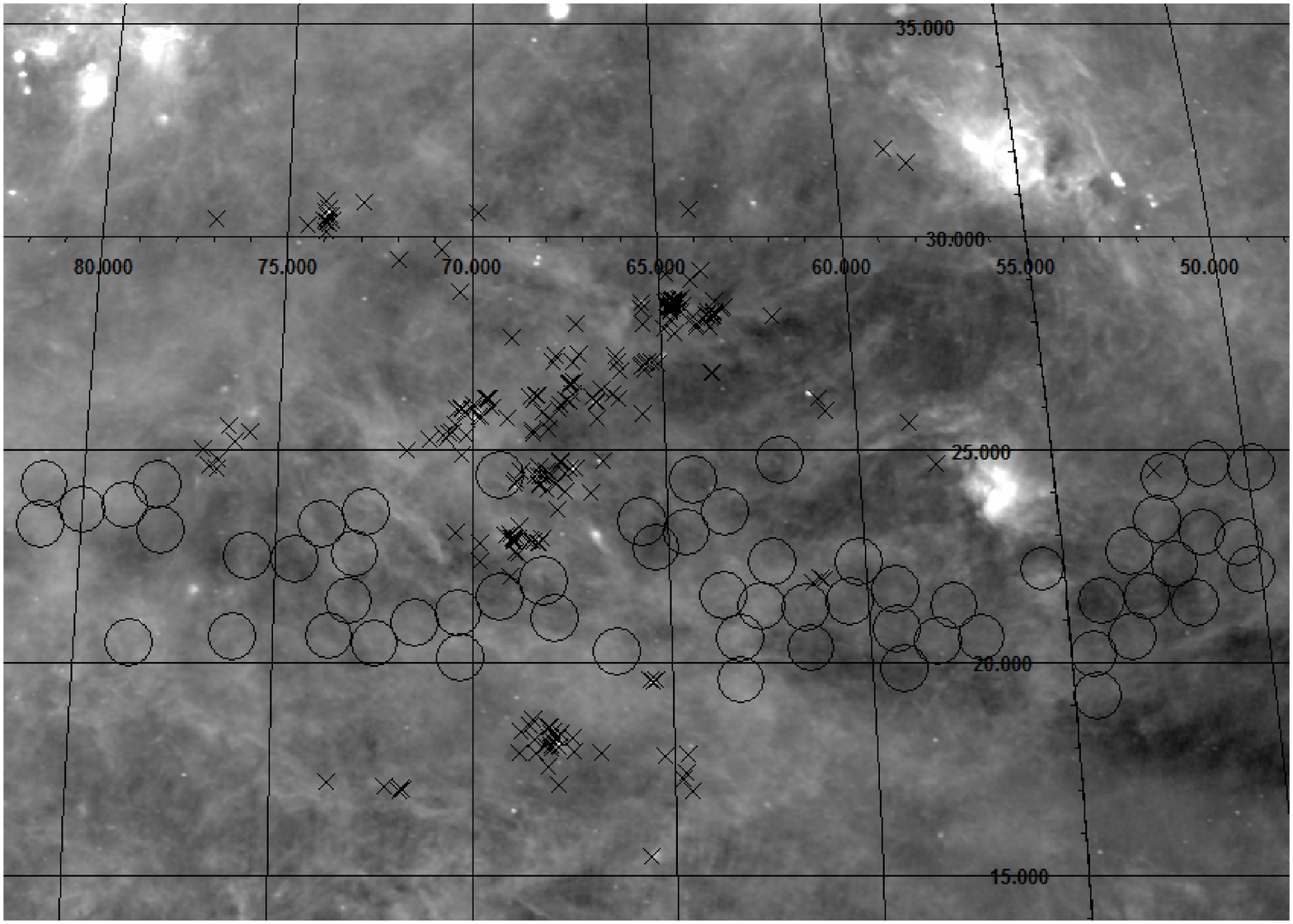}
\caption{IRAS 60~$\mu$m map of Taurus, with our GALEX fields overplotted as 0.55\degr\ radius circles. Known Taurus members from the compilation of \citet{memberlist} and from \citet{TAPcheck} are marked as X's. The areas with dense concentrations of sources could not be observed because they might overload the GALEX detectors.}\label{context_tau}
\end{figure*}

\begin{figure*}
\includegraphics[width=\textwidth]{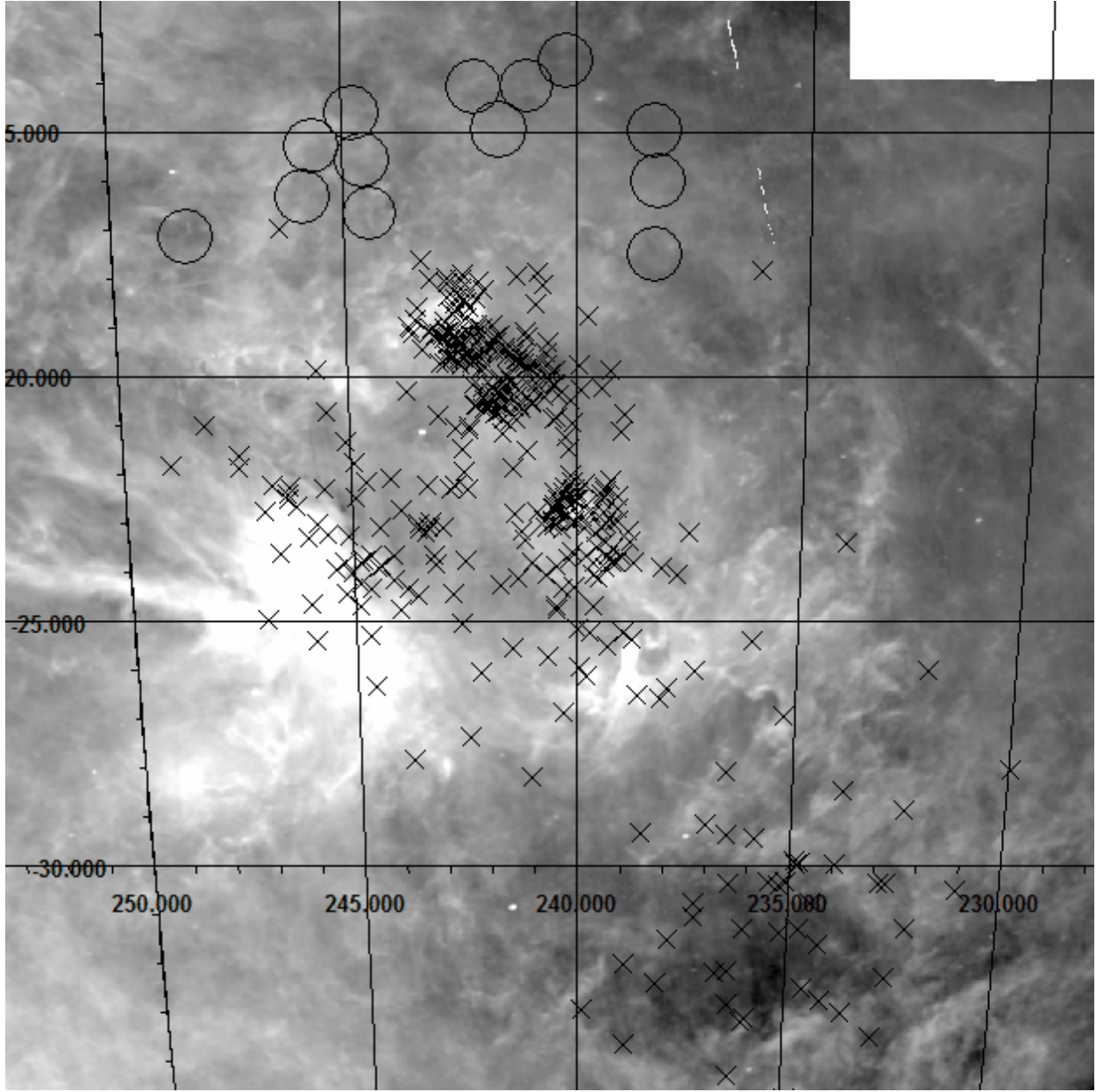}
\caption{IRAS 60~$\mu$m map of Upper Scorpius, with our GALEX fields overplotted as 0.55\degr\ radius circles. Known Upper Sco members from the compilation of \citet{memberlist} are marked as X's. The areas with dense concentrations of sources could not be observed because they might overload the GALEX detectors.}\label{context_sco}
\end{figure*}

As part of our Cycle~1 proposal, GALEX imaged 61~fields, covering 56~square degrees, in southern Taurus in the FUV and NUV bands and 13~fields, covering 12.2~square degrees, in northern Upper Scorpius in only the NUV band. The fields were chosen to coincide with the survey area of \citet{slesnick1} and \citet{slesnick2}. The observed fields are shown in figures~\ref{context_tau} and~\ref{context_sco}. 
Because of concerns that bright stars or dense groups of stars might damage the detectors, GALEX avoids fields containing either; unfortunately, these are exactly the fields containing most of the known young stars in these regions. In particular, the observed Taurus fields are located south of the molecular cloud L1527, but north of L1551. We catch the eastern edge of L1529 but avoid L1536. 
The observed Upper Scorpius fields are located north of all major concentrations of known sources. 
Most of the pointings were taken in sets of five 300~s exposures of adjacent fields, but a few observations were instead scheduled as full-orbit exposures of 1100-1500~s. The observations are summarized in Table~\ref{galexobs}.

The observations were reduced automatically by the GALEX data release 2 and 3 pipeline \citep{galexcalib} except for two orbits, TauAur\_CH45 and TauAur\_CH56, which were processed by an earlier pipeline version. We visually inspected the images and discarded those that were made unusable by clear artifacts or corrupted data.

Two fields, TauAur\_CH11\_0002\_sv03 and TauAur\_MOS23, had offsets in their astrometry solutions but no artifacts or signs of data corruption. We found mean astrometry corrections by comparing the positions of 8 bright NUV and 2MASS sources in each field, assuming a fixed offset independent of position on the detector. We found we needed to correct the source positions in TauAur\_CH11\_0002\_sv03 by $\Delta(\alpha, \delta) = (-11.54\pm0.26\arcsec, +14.97\pm0.21\arcsec)$ and those in TauAur\_MOS23 by $(-12.22\pm0.23\arcsec, -10.73\pm0.32\arcsec)$ for the GALEX and 2MASS positions to agree. Residuals from the fixed offset ($\sim 0.8$\arcsec) were larger than the scatter in GALEX-2MASS offsets in comparison fields ($\sim 0.5$\arcsec) but showed no pattern. The latter is similar to the absolute astrometric precision of 0.49\arcsec\ found by \citet{galexcalib}. We applied the corrections to the positions of all GALEX sources in these two fields before considering cross-matches to other catalogs.

We adopted photometry from the band-merged SExtractor \citep{SExtractor} catalog produced by the GALEX pipeline. Because GALEX observations have low backgrounds, the pipeline used a customized background algorithm \citep{galexcalib} rather than the one provided with SExtractor. The background was calculated as means over 128-pixel bins, iteratively clipping pixels having a probability of being Poisson fluctuations below $1.35 \times 10^{-3}$. Bright sources were masked out using an initial SExtractor run. The background map was then linearly interpolated back up to the resolution of the image. The detection threshold map was calculated from the background map as the count rate above which a pixel had less than a 31.7\% probability of being a Poisson fluctuation. Except where deblending was needed, SExtractor defined a source as a group of at least 10~adjacent pixels above the local detection threshold.

SExtractor found source fluxes by summing over an elliptical aperture with radius $R = 2.5 (\sum{rI(r)})/(\sum{I(r)})$, which should enclose  $\sim 94\%$ of a source's light for both stars and galaxies \citep{SExtractor}. Errors were calculated as $\sqrt{A \sigma^2 + F/g}$ where $A$ is the source area, $\sigma$ is the background rms, $F$ is the source flux, and $g$ is the gain\footnote{\url{http://www.ifa.hawaii.edu/\~rgal/science/sextractor\_notes.html}}.
We found that the catalog fluxes are consistent with wide (tens of arcseconds) aperture photometry performed on a trial field.
We investigated but did not apply additional nonlinearity or color corrections to the catalog output, as such corrections are still highly uncertain (Wyder 2008, private communication).

For the remainder of our work we selected only GALEX sources that:
\begin{itemize}
\item have $3\sigma$ flux measurements in at least one GALEX band
\item SExtractor determined to be a point source with 0.9 or higher probability in each detected band
\item are within 0.55\degr\ of the field center, where PSF variations are negligible, astrometric precision is uniformly high, detector efficiency is high, and artifacts are rare
\item do not have an artifact flag in the GALEX catalog
\item do not suffer any ambiguity in matching FUV and NUV sources
\end{itemize}
25389 ultraviolet sources in Taurus and 5863 in Upper Scorpius met these criteria. Over 99\% of these are unique detections; the rest are multiple detections of sources located in areas of overlap between two GALEX fields.\label{galex_matchable}

\section{UV Excesses}\label{excessproc}

\subsection{Reference photospheres}\label{intro2mass}

To characterize our sources, we matched them to data from the Two-Micron All-Sky Survey \citep[2MASS,][]{2MASS_cat}. We chose 2MASS photometry because it is a high quality, well-characterized data set, is available at low Galactic latitudes, and is less sensitive to extinction than visible photometry. We selected 2MASS sources that:
\begin{itemize}
\item have C quality or better photometry in both $J$ and $K$ band, i.e. repeatable detections, $5 \sigma$ flux measurements, and a source profile well fit by a single PSF
\item are not blended or contaminated in $J$ or $K$ band
\item are not in the 2MASS extended source catalog
\item are not associated with a known solar system object
\item are within 2\arcsec\ of a GALEX source, the matching radius recommended by \citet{galexcalib}
\end{itemize}

20045 of the 25389 Taurus GALEX sources and 4210 of the 5863 Upper Scorpius sources had 2MASS counterparts, after removing 184 duplicate GALEX observations from the Taurus sample. There were no duplicates in the Upper Scorpius sample because there were very few overlapping GALEX fields. Figure~\ref{unmatched} shows the GALEX-2MASS match fraction as a function of $FUV-NUV$ color, which is relatively insensitive to extinction: $E(FUV-NUV)/A_V = 0.05$ for $R_V = 3.1$ (Gil de Paz 2007, private communication). Many of the GALEX sources with no 2MASS counterpart have $FUV-NUV \lesssim 1$, the colors expected of star-forming galaxies or B or early A stars. Both types of objects are often too distant to be detected by 2MASS. 
Even among the redder GALEX sources $\sim 5\%$ lack 2MASS counterparts. These may simply be sources that were rejected by our quality control criteria.\label{tmass_unmatched}

After the merged sources were placed on $UV-J$ vs. $J-K$ color-color diagrams (section~\ref{whyccplots}), we found that the source distribution changes rapidly near $J \sim 14$. Brighter sources tend to form a clear main sequence, as shown in Figure~\ref{ccplots}, while fainter sources form a different distribution at constant $NUV-J \sim 5$. 
Since we already expected most sources with $J > 14$ to be unresolved galaxies \citep{galaxy14}, we imposed an additional constraint $J \le 14$ on all our sources. 

This left 14130 matched sources in Taurus and 2828 in Upper Scorpius.\label{tmass_matchnum}

\subsection{The UV-IR color locus}\label{whyccplots}

Colors calculated by Laget\footnote{\url{http://www.oamp.fr/people/laget/intg/intg0.html}} from Kurucz models suggest that stellar photospheres form a well-defined locus in the GALEX $FUV-J$ vs. $J-K$ and $NUV-J$ vs. $J-K$ color spaces, largely independent of surface gravity (figure~\ref{ccplots}). Ultraviolet excess sources should appear below this locus, with bluer UV colors relative to non-excess stars of the same photospheric infrared color. Since atmosphere models are not well tested in the ultraviolet, we did not rely on them to determine the precise position of the stellar locus. Instead we used the population of field stars, a mixture of dwarfs and giants, to define an empirical $UV-J$ vs. $J-K$ locus.

Since they can be detected at distances much farther than 140~pc, most of the early-type, intrinsically blue, stars in the field population are background objects, and are therefore extinguished by Taurus or Upper Scorpius. On the other hand, most of the detectable late-type, intrinsically red, stars are in the foreground and suffer little extinction (cf. Table~\ref{maxav}). 
In practice, since we cannot detect redder stars unless they have an ultraviolet excess, our measurement of the field star locus is determined by early-type stars, which have a similar extinction to those in Taurus or Upper Scorpius ($A_V \sim 2$).
In addition, the reddening vector is almost parallel to the main sequence, and as a result the locus of reddened stars inferred e.g. from the Kurucz model colors is much narrower than the stellar locus we observe. Since we are interested in the distance of a star from the locus, we infer that any systematic errors associated with reddening are much smaller than the effect of other errors or astrophysical variations.

Although we tried to characterize the main sequence by taking the median $NUV-J$ color of narrow $J-K$ bins, following \citet{mscolor}, this approach failed for bins with $J-K \gtrsim 0.5$ in the FUV and $J-K \gtrsim 0.9$ in the NUV, as there were too few field stars to give an unbiased median. Instead, we iteratively removed points that differed from the median locus by three standard deviations, then made linear fits to the remaining points with $J-K \leq 0.4$ in the FUV and $J-K \leq 0.8$ in the NUV. The resulting solutions were 
\begin{equation}\label{msfit}
\begin{array}{r@{}c@{}r@{}c@{}c@{}lll}
(FUV-J) & = & (20.5  \pm 0.5)  & (J-K) & + & (2.96 \pm 0.14) & \textrm{cov}(m,b) = -0.07 & \textrm{(Taurus)} \\
(NUV-J) & = & (10.36 \pm 0.07) & (J-K) & + & (2.76 \pm 0.04) & \textrm{cov}(m,b) = -2\times10^{-3} & \textrm{(Taurus)} \\
(NUV-J) & = & (10.05 \pm 0.16) & (J-K) & + & (2.64 \pm 0.08) & \textrm{cov}(m,b) = -0.013 & \textrm{(Upper Scorpius)}
\end{array}
\end{equation}
We defined a source's FUV and NUV excess as the difference between its observed FUV-J and NUV-J color, respectively, and the value predicted by equation~\ref{msfit}. The formal error in the excess, propagated from the photometric uncertainties and the uncertainties in the fit, is dominated by the error in the $J-K$ color. Because the stellar locus has a steep slope in ($UV-J$)-($J-K$) space, the mean uncertainty of 0.05 in $J-K$ propagates to an uncertainty of $\sim 1.1$ magnitudes in the FUV excess and $\sim 0.5$ magnitudes in the NUV excess. Since in section~\ref{cutofftests} we identify our sample of UV excess sources as those differing by a certain number of standard deviations from the main stellar locus, this translates to the smallest excess we can reliably detect: 3.2 FUV magnitudes and 1.6 NUV magnitudes if we select sources with a $3\sigma$ or larger excess, or 5.3 FUV magnitudes and 2.6 NUV magnitudes if we require a $5\sigma$ or larger excess. For stars with more precise $J-K$ colors, we can detect excesses below these limits.\label{minex}
% 3-sigma detection limit: 1.6/3.2 mags
% 4-sigma limit: 2.1/4.2 mags
% 5-sigma limit: 2.6/5.3 mags

\subsection{UV excess object selection}

We show in Figure~\ref{normex} the distribution of excesses divided by their errors. The observed distributions are roughly 1.9 times broader in the FUV than the errors imply, and 1.5 times broader in the NUV. This might suggest a source of error we have overlooked. An alternative is that the locus of field stars is broadened by variations in properties other than temperature and gravity, and we are seeing a convolution of such astrophysical broadening with our errors. 
We will continue to quote excesses in multiples of the formal errors as a measure of whether a source lies outside the locus, but the reader should bear in mind that this ratio is not a rigorous measure of statistical significance. However, in sparsely populated areas of the color-color diagram, 
especially at high $J-K$, 
there are too few sources for us to use alternative measures such as the locus width.

We now define our sample of UV excess sources as those having ultraviolet excess normalized by the formal error in that excess above a given cutoff value. For several choices of cutoff ($3\sigma$, $4\sigma$, ..., $7\sigma$) we estimate the number of sources erroneously included in the UV excess sample by assuming the distribution of field star colors is symmetric with respect to positive and negative excesses. For example, to estimate the number of erroneously included stars whose measured UV excess is $3\sigma$ or greater, we find the number of sources with a $3\sigma$ or greater UV deficit and subtract it from the number of UV excess sources, assuming the former are all outliers from the field locus.
We list in Table~\ref{exstats} the total number of UV excess sources, the expected number of erroneous sources, and the fraction of UV excess sources that appear legitimate for each cutoff value.\label{cutofftests}

Table~\ref{exstats} shows that the choice of cutoff carries a steep trade-off between completeness of the UV excess sample and reliability of an individual source's identification as UV excess. For example, only a third of the sources with a measured excess of over $3\sigma$ should be true UV excess sources, while around 86\% of the $5\sigma$ sources are reliable identifications. On the other hand, only 42\% of the genuine $3\sigma$ UV excess sources are expected to appear in a $5\sigma$ sample -- the rest are buried in the more heavily contaminated population with excess between $3\sigma$ and $5\sigma$. To support projects where completeness is a higher priority than reliability, we present in Table~\ref{catalogtable} all sources with a $3\sigma$ or better excess in at least one band. However, for many of our discussions in this paper we will concentrate on the more reliable $5\sigma$ subsample.

\subsection{Results}\label{thecatalog}

The spatial distribution of the $5\sigma$ excess sources is shown in figures~\ref{newmembers_tau} and~\ref{newmembers_sco}. Most of our sample of $5\sigma$ sources in both Taurus and Upper Scorpius is spread uniformly over the observed fields.

In Taurus we also find concentrations of sources near $\alpha = 80\degr, \delta = 24\degr$ and near $\alpha = 62\degr, \delta = 25\degr$, as well as some outlying members of the previously known group of sources associated with the cloud L1529 \citep{hsfr_tau}. The first group might not be a true physical association, appearing dense only because it is less than ten degrees from the Galactic plane. The group near (62\degr, 25\degr), consisting of 6 sources, is likely a Poisson fluctuation. Given the mean density of 1.5 $5\sigma$ sources per field, we find a 21\% probability that at least one of our 61 Taurus fields contains 6 sources or more. We conclude there is no statistical evidence for clustering of the UV excess sources.

There are very few sources in our Upper Scorpius sample, consistent with the observation by \citet{slesnick2} that only a small fraction of Upper Scorpius members of spectral types M3-M8 are north of $\delta = -17\degr$. The sample is too sparse for us to look for evidence of structure, nor has any previously been observed in the association \citep{hsfr_sco}.

We examine in section~\ref{membership} whether any of these UV excess sources are plausible members of Taurus or Upper Scorpius.

\section{Validation}

\subsection{Comparisons with previous surveys}\label{oldcheck}

To test our survey's recovery rate, we identified samples of Taurus and Upper Scorpius members compiled without reference to our observations. \citet{memberlist} and \citet{TAPcheck} together list 261 known Taurus members and \citet{memberlist} list in addition 401 Upper Scorpius members. Unfortunately, because of concerns that a bright UV source or concentration of sources might damage the GALEX detectors, our GALEX pointings avoid areas with large numbers of known young stars (figures~\ref{context_tau} and \ref{context_sco}). Only 5 of the previously known Taurus members were in the GALEX survey region, and none of the known Upper Scorpius members; the former are listed in Table~\ref{knownmembers}.

We recover AA~Tau and DN~Tau, both well known classical T Tauri stars, as UV excess objects to high significance. TAP~4, a naked T Tauri star, has only a $3\sigma$ UV excess. The remaining two sources, which we do not detect, are M~stars, and therefore strongly selected against by GALEX without strong UV excesses (section~\ref{maglimits}). While the detections of AA~Tau and DN~Tau are reassuring, the marginal detection of TAP~4 suggests that many of the more mature Taurus members -- our main goal -- may be buried in the UV-bright tail of the field star population, possibly below our $3\sigma$ limit.

\subsection{Control sample}

To investigate the amount of field contamination in our $3\sigma$ UV excess sample, we repeated our procedure on a set of 7 Medium Imaging Survey ($\sim 1400$~s exposure) fields, covering 6.5 square degrees, taken near ($l = 90$, $b = 30$). This latitude is slightly higher than that of Taurus or Upper Scorpius, but there were no GALEX observations at lower latitudes. The longitude was chosen as a point intermediate between the bulge (the direction of Upper Scorpius) and the anticenter (the direction of Taurus). IRAS images of this region show little emission aside from Galactic cirrus, and there are no known open clusters within ten degrees of the fields\footnote{From the WEBDA cluster database, \url{http://www.univie.ac.at/webda/}}. Based on these observations, we do not expect a population analogous to the Taurus or Upper Scorpius association in our comparison fields.

\begin{figure*}
\includegraphics[width=0.48\textwidth]{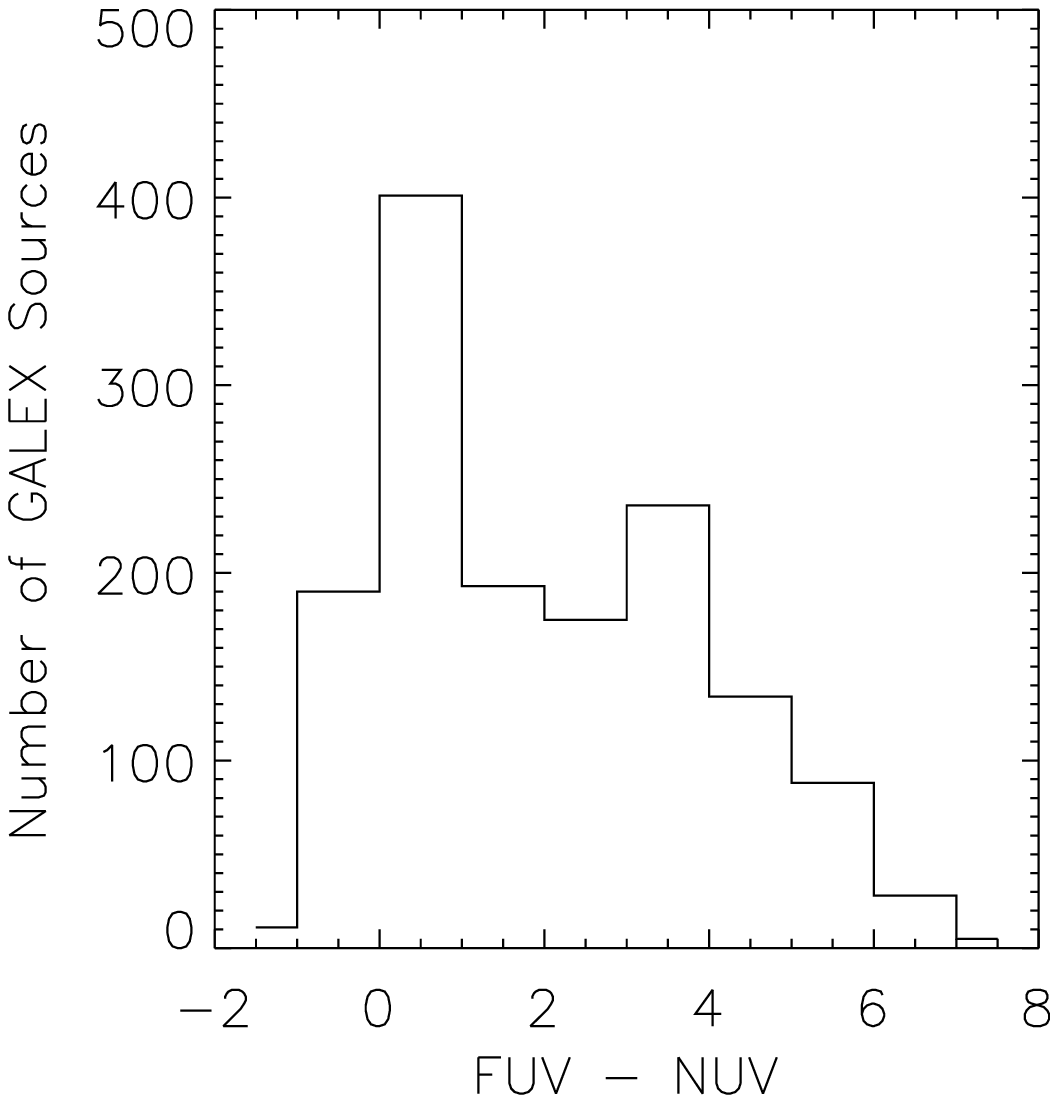}
\includegraphics[width=0.48\textwidth]{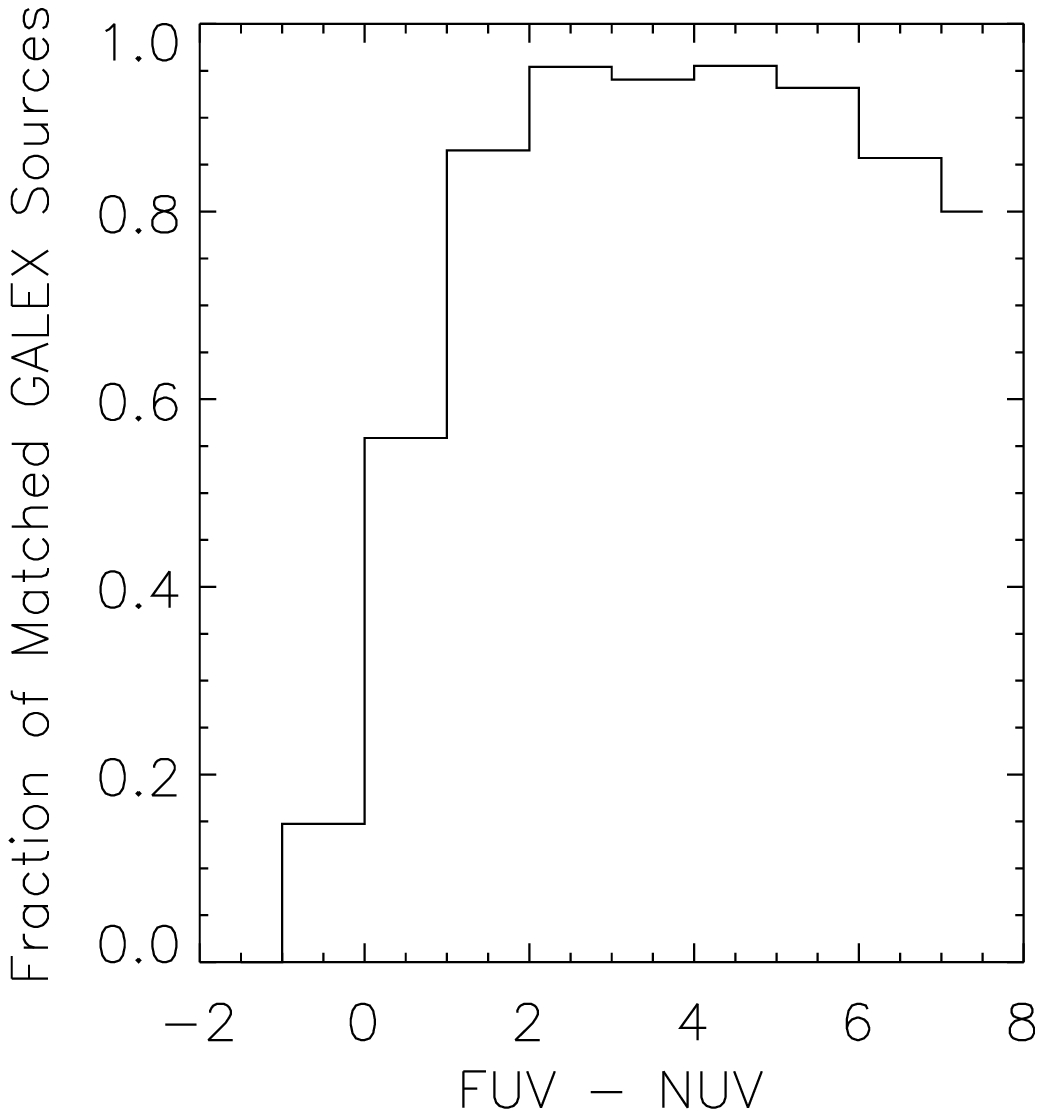}
\caption{Number of sources in our target fields as a function of $FUV-NUV$ color (left), along with the fraction of these sources that have 2MASS counterparts (right). Most of the unmatched sources are very blue objects, presumably too distant to be detected by 2MASS.}\label{unmatched}
\end{figure*}

Of the 10496 GALEX sources that met our quality cuts, 3148 had 2MASS matches. From examining the source distribution in color-color space we confirmed that $J \le 14$ discriminated stars from unresolved galaxies in these fields, but only 1461 sources met this last criterion.
The small fraction of matching sources, and stellar sources in particular, is expected from the higher galactic latitude and lower extinction of our control fields compared to our target fields. For reference, 14130 of our 25389 Taurus GALEX sources had a 2MASS counterpart with $J \le 14$, as did 2828 of our 5863 Upper Scorpius GALEX sources. Because of the low extinction, our sample of GALEX sources in the control fields contains on average more distant sources than the samples in the target fields do. We show in Figure~\ref{fuvnuvmatch} the distribution of sources as a function of $FUV-NUV$ color, analogous to Figure~\ref{unmatched} and the discussion in section~\ref{tmass_unmatched}. Very few of the sources with  $FUV-NUV \lesssim 2$ (either stars earlier than mid-A or unresolved star-forming galaxies) have 2MASS counterparts; the effect is much stronger here than in Figure~\ref{unmatched}. In addition, the distribution of $J$ magnitudes for matched sources (Figure~\ref{controlJ}) peaks over a magnitude fainter than in our target fields, suggesting again that we are probing greater distances in the control field.\label{uvcolortest}

We show in Figure~\ref{controlplot} color-color diagrams for the control fields analogous to Figure~\ref{ccplots}. These fields contain nine $5\sigma$ UV excess sources. This is the same fraction of GALEX-2MASS sources ($0.6\pm0.2\%$) as in Taurus ($0.63\pm0.07\%$) and Upper Scorpius ($0.42\pm0.12\%$). 
We cannot tell what fraction of the excess sources in the control fields are outliers from the main stellar locus, as there are no UV-deficit sources in Table~\ref{exstats} with which to repeat the analysis of section~\ref{cutofftests}.
Since these sources appear far from known clusters or star-forming regions, we assume they represent a population spread evenly throughout the Galactic disk. If this is the case, then the population should appear along any line of sight at similar Galactic latitude, including towards Taurus and Upper Scorpius, and in fact should account for a large fraction of the sources we see there. 

This inferred population of UV excess sources is qualitatively similar to the population of $\sim 100$~Myr old G and K stars invoked by \citet{ptts_xray} to explain ROSAT detection rates of X-ray luminous stars. However, they predicted a surface density of 0.2-0.3~deg$^{-2}$ for these stars, implying an average of 1.3-1.9 such stars in an area the size of our control fields. 
Because \citet{ptts_xray}'s work was based on ROSAT's sensitivity limits, most of the detectable stars in their model are within 100-200~pc. Three of the nine UV excess stars in the control sample have $J$ magnitudes too faint for a distance of 200~pc, so we drop these three to allow a fair comparison of our sample, now reduced to six, to their results. The probability that we would observe six or more sources when we expect 1.9, however, is only 1.3\%. 
Since to our knowledge there is no study on low-latitude ultraviolet populations analogous to \citet{ptts_xray}, we cannot explain why our background of UV excess sources is much larger than their background of X-ray sources.
\label{highbg}

\section{Spectroscopic Follow-Up}

\subsection{Target selection}

Like all youth indicators, ultraviolet excess is not perfectly correlated with age (Altenbach~2010, in prep.). As a result, like most young star surveys, we need spectroscopic follow-up to confirm the youth of our photometrically selected stars.
We carried out a spectroscopic observing program of some of our $3\sigma$ ultraviolet excess sources using the Double Spectrograph (DBSP) on the 200-inch Hale Telescope at Palomar. Our program had several goals. One was to obtain representative spectra of excess sources across a broad area of color-color and color-magnitude space. Another was to understand the diversity of sources that produce a UV excess. A third was to create as large a sample of spectroscopic candidate young stars as possible.

To balance these goals, we created a spectroscopic target list containing the following groups of sources, in decreasing order of population:
\begin{enumerate}
\item UV excess sources with $J-K > 0.7$ and $10 < J < 14$, as we found from our first run that this region of parameter space contained many emission line stars
\item UV excess sources with $0.3 < J - K < 0.5$ and $10 < J < 12$, as this region of color-magnitude space was dominated by sources with an excess only in the FUV, suggesting they represented a different population from sources in other regions
\item UV excess sources with $J - K > 0.5$ and an FUV excess, as these sources tended to have very large excesses with no clear pattern
\item obvious outliers in Figure~\ref{ccplots}, either sources with $J-K$ much lower or much higher than the main body of UV excess sources or sources with much larger excesses than other UV excess sources with similar $J-K$
\end{enumerate}
At the telescope, we would usually choose the brightest sources consistent with the above criteria to keep the exposure times reasonable. In addition, we tended to pick sources with large excesses ($\gtrsim 7$~mag in FUV or $\gtrsim 3$~mag in NUV) as a previous program of AIS-selected observations had shown that these sources are much more likely to have chromospheric activity indicators in their spectra (Altenbach~2010, in prep.). As a result, the spectroscopic sample has strong biases not present in the photometric sample of section \ref{thecatalog}.

\subsection{Observations}

We observed 20 of the 95 $3\sigma$ candidates in Upper Scorpius, including 7 of the 11 $5\sigma$ candidates, on 2008 June 5-6. We also observed Sco~X-1, but do not count it as a candidate as we were aware of its identity before the run (see section~\ref{scox1}). We likewise observed 43 of the 471 $3\sigma$ candidates in Taurus, including 20 of the 89 $5\sigma$ candidates, on 2008 November 28.
The spectra covered the ranges 3890-5430\AA\ at SNR $\sim 50-110$, $R \sim 6400$ and 6190-6750\AA\ at SNR $\sim 85-160$, $R \sim 8600$, allowing us to observe the key youth indicators Ca~II emission (3934, 3968\AA), H$\beta$ emission (4861\AA), H$\alpha$ emission (6563\AA), and Li absorption (6707\AA).
The images were reduced using the JHU astronomy library in IDL, and the spectra were extracted with the NOAO two-dimensional spectral package in IRAF.

% spectral ranges are overlap between June and Nov observations

\subsection{Results}

In Table~\ref{specresults} we list all the Palomar targets, together with equivalent widths of the youth indicators where we could measure them. In addition, where we did not detect lithium we give an upper limit on the equivalent width of $\textrm{10~\AA}/SNR$, where $SNR$ is the formal signal-to-noise ratio of the surrounding continuum, as propagated by the IRAF spectral extraction routines. The constant of proportionality was set by noting that, had the lithium line in 2MASS~J04505356+2139233 been weaker than its 0.15~\AA value, we would have taken it for a $3\sigma$ statistical fluctuation.

The spectra were classified visually using spectra of standards taken on a previous DBSP run, with the statistics shown in Table~\ref{spectypes}. 
As expected, our Upper Scorpius sample is dominated by G and K stars, the lowest-mass (and therefore most numerous) stars to which we are sensitive. On the other hand, our Taurus sample shows a large number of M stars but very few K stars, particularly with a $5\sigma$ or greater excess. This is surprising because while K~stars with weak excesses are not detectable by GALEX, those with a 3.5~mag ($\gtrsim 7\sigma$) excess or low extinction should be (section~\ref{maglimits}). Table~\ref{specresults} has many stars of other spectral types, including later types, with excesses of this level. We suspect the deficit of K~stars in our sample may represent an unexpected selection effect in our choice of spectroscopic targets.

We found that roughly a third of our sources had chromospheric activity indicators (Table~\ref{spectypes}). Three of those, all in Taurus, also show lithium absorption. We present spectra of these three in Figure~\ref{specli}. Among the three lithium stars, our survey recovers one previously known Taurus member, 1RXS~J044712.8+203809, and confirms a previous candidate, 1RXS~J045053.5+213927. The third star, 2MASS~J05122759+2253492, is a previously unknown lithium-rich M~dwarf toward the Galactic plane. The self-absorbed H$\alpha$ profile suggests this star is still accreting, and the high Li equivalent width of 530~m\AA\ implies an age under 10~Myr.

While many of our emission line stars in Taurus show both Balmer and calcium emission lines, all but one in Upper Scorpius show only calcium emission. Since, unlike Balmer emission, calcium emission is a chromospheric activity indicator often seen in field stars, this may mean that our Upper Scorpius fields are too far from the main body of the association to contain significant numbers of young stars and that nearly all of our Upper Scorpius candidates are $\gtrsim 100$~Myr old field stars.

Two thirds of the spectra, even of $5\sigma$ targets, show no unusual features, only photospheric absorption lines. This is surprising because we anticipated a good correlation between UV photometric excess and the presence of optical emission lines, aside from the population included from statistical fluctuations that we estimated in section~\ref{cutofftests}. From Table~\ref{exstats} we expect only one sixth of the $5\sigma$ targets to be erroneously identified as UV excess, and therefore for five sixths, not the observed one third, of the targets to show emission lines.
We conclude there is a significant population of UV excess sources with normal optical spectra; we explore this population more in section~\ref{phantomexcess}.\label{foundlines}

\section{Discussion}

\subsection{Membership in Taurus and Upper Scorpius}\label{membership}

With our spectroscopic data we can investigate how much of our UV excess sample, most of which lacks spectra, can be associated with Taurus or Upper Scorpius. 
A significant number of our spectra are of sources found below the main sequence at 140~pc, in part because we wanted to characterize our contaminants, in part because we were biased toward optically bright (and therefore blue, at fixed $J$) sources, and in part because we were concerned that our brightest sources were \emph{too} bright even for pre-main sequence stars and must instead be field giants. 
Of the 19 $3\sigma$ sources above the main sequence in Upper Scorpius, we obtained spectra of 9. Of the 101 $3\sigma$ sources above the main sequence, we followed up 23. In addition, we found information in the literature confirming or ruling out Taurus membership for 14 sources, 2 of which were also spectroscopic targets. We searched for but found no such information for our Upper Scorpius candidates.
We summarize these results in Table~\ref{extrapol}, classifying UV excess stars above the 140~pc main sequence as follows:
\begin{itemize}
\item stars confirmed as members of Taurus or Upper Scorpius in previous literature, usually by lithium detection
\item stars rejected as members in previous literature, usually by lithium non-detection or by proper motion association with the nearby Pleiades or Hyades
\item Palomar targets with no emission lines
\item Palomar targets with emission lines, but with a spectral type or luminosity class indicating they are actually either reddened background main sequence stars or background giants
\item Palomar targets with spectral type consistent with Taurus or Upper Scorpius membership and emission lines but no lithium line
\item Palomar targets with spectral type consistent with Taurus or Upper Scorpius membership and emission lines and a lithium detection
\item stars with neither membership information in previous literature nor Palomar spectra
\end{itemize}
For the 5 stars in Taurus and 5 in Upper Scorpius that had both emission lines and a spectral type consistent with Taurus or Upper Scorpius membership, we acquired preliminary proper motions based on published USNO and 2MASS positions (Kraus 2009, private communication). The results, presented in Table~\ref{pmresults}, indicate that 3 of the spectroscopic candidates in Taurus are likely members of the association, while only one of the candidates in Upper Scorpius is. However, we emphasize that these are very rough proper motions and should be used with caution.

We show in Figure~\ref{cmplots} color-magnitude diagrams of the Taurus, the Upper Scorpius, and the control fields, distinguishing both $5\sigma$ UV excess sources and spectroscopic targets (details in caption). We show the main sequence at 140~pc in blue. 
We expect low mass Taurus and Upper Scorpius members to lie above the indicated main sequence, as would extinguished early-type main sequence stars at a comparable distance. 
Figure~\ref{cmplots} shows that, as expected, almost all of our Palomar targets with emission lines are above the 140~pc main sequence, to within uncertainties in the stars' $J - K$ color.
The rarity of emission lines among the background UV excess stars suggests that the emission-line stars above the main sequence may represent a population associated with Taurus or Upper Scorpius, although we are limited by small numbers: a Fisher's exact test gives a probability of 20\% that we would see at least this large a discrepancy from statistical fluctuations if the fraction of emission stars were spatially uniform. 

In section~\ref{thecatalog}, we found many UV excess sources in the easternmost fields of our Taurus survey, but noted that these are the fields closest to the Galactic plane. To determine the nature of these fields we plot their sources in Figure~\ref{cmplots_tauparts}, with the fields in central Taurus for comparison. Clearly, the vast majority of the $5\sigma$ sources in the eastern fields are background objects; the high density of UV excess sources merely reflects the high density of stars toward the Galactic plane.

To see if the high density of \emph{emission-line stars} in eastern Taurus is likewise due to the fields' low Galactic latitude, we repeat a test from \citet{slesnick1}, who carried out an optical survey of roughly the same area as our GALEX survey. By selecting candidate pre-main-sequence stars from optical color-magnitude diagrams and using follow-up spectroscopy to look for gravity indicators, they found a broadly distributed population of pre-main sequence stars at $\alpha \gtrsim 80\degr$. They found no such population in western Taurus. Using Fisher's exact test they showed that the probability of observing so great a difference between the fraction of spectroscopically confirmed pre-main-sequence stars east and west of the known Taurus population is less than 3\% if in fact the two regions represent identical populations.

Applying Fisher's exact test to our own results (4 out of 15 Palomar targets have emission lines east of Taurus, 2 out of 8 have them west of Taurus), we find a probability of 100\% of observing fractions of emission-line stars differing by at least as much as they do. Repeating the test on samples restricted to $5\sigma$ sources, or samples above the main sequence, likewise gives no significant result.
We can explain the insignificance of our result in part as an effect of selecting sources only by their colors, making us more sensitive to background objects than \citet{slesnick1}, and in part by the small number of spectra we have. If we had a larger sample, we might be able to get a more useful constraint on the existence of an eastern extension to the Taurus association.

Despite the strong backgrounds, we have identified possible new members of Taurus and Upper Scorpius, with more likely to be identified once we have additional spectra. Since our GALEX fields do not cover the known areas of newly formed stars in either Taurus or Upper Scorpius, any new members we find must represent either previously unnoticed subgroups of Taurus (though we find no significant clustering), a more distributed mode of star formation than previously inferred in either region (though this is hard to reconcile with the classical theory that stars form in dense clumps), or migration from the actual sites of star formation \citep{ptts_intro, ptts_xray}. 
In principle, we can constrain this last hypothesis with the toy model we presented in the introduction to this paper.
However, as we find only one new candidate of type K2 or earlier in our Taurus survey, in addition to the three already known from the literature, and three new candidates of type K2 or earlier in Upper Scorpius, we do not yet have a sample large enough for good constraints. Our single detection in Taurus implies that 13\% of our $3\sigma$ sources might be K2 or earlier members. Extrapolating to the 36 $3\sigma$ sources in our central Taurus fields, where the calculations of the introduction apply, we predict a member density of 0.3~deg$^{-2}$. Our three K2 or earlier candidates in Upper Scorpius imply that 33\% of our unobserved sources might also be members, implying a member density of 0.5~deg$^{-2}$. 
Assuming that half of the sources in either region turn out to be ZAMS stars or other field contaminants, we find densities similar to the ones we predicted for constant star formation histories of 30~Myr or longer. 
Trying to apply formal upper limits rather than these highly uncertain values gives no constraint at all. 

Since most of our emission line stars lack lithium detections, our arguments that they truly are Taurus or Upper Scorpius members are statistical in nature. Better constraints on the ages of these stars could tell us whether they are co-eval with the known Taurus and Upper Scorpius populations or whether they are a distinctly older group -- since stars of given age show a broad scatter in their lithium equivalent widths, our non-detections are suggestive but do not by themselves prove an old population. In addition, more rigorous proper motions could confirm individual stars' membership and provide cleaner statistics, especially once we have an expanded spectroscopic sample.

\subsection{Unusual UV excess sources}\label{oddballs}

While our survey targeted chromospherically active young stars in Taurus and Upper Scorpius, GALEX is an ideal facility for finding a variety of UV-bright sources, potentially including new classes of objects. However, this diversity means that any search for a specific type of object must consider a variety of interlopers. As a demonstration of both the power of UV surveys to find unusual objects and the potential contaminants of such surveys, we note that our photometrically selected sample includes an X-ray binary, a possible mass-transfer binary, and a white dwarf M dwarf pair.

\paragraph{2MASS J16195506-1538250 (Sco X-1)}\label{scox1}

\begin{figure*}
\includegraphics[width=0.48\textwidth]{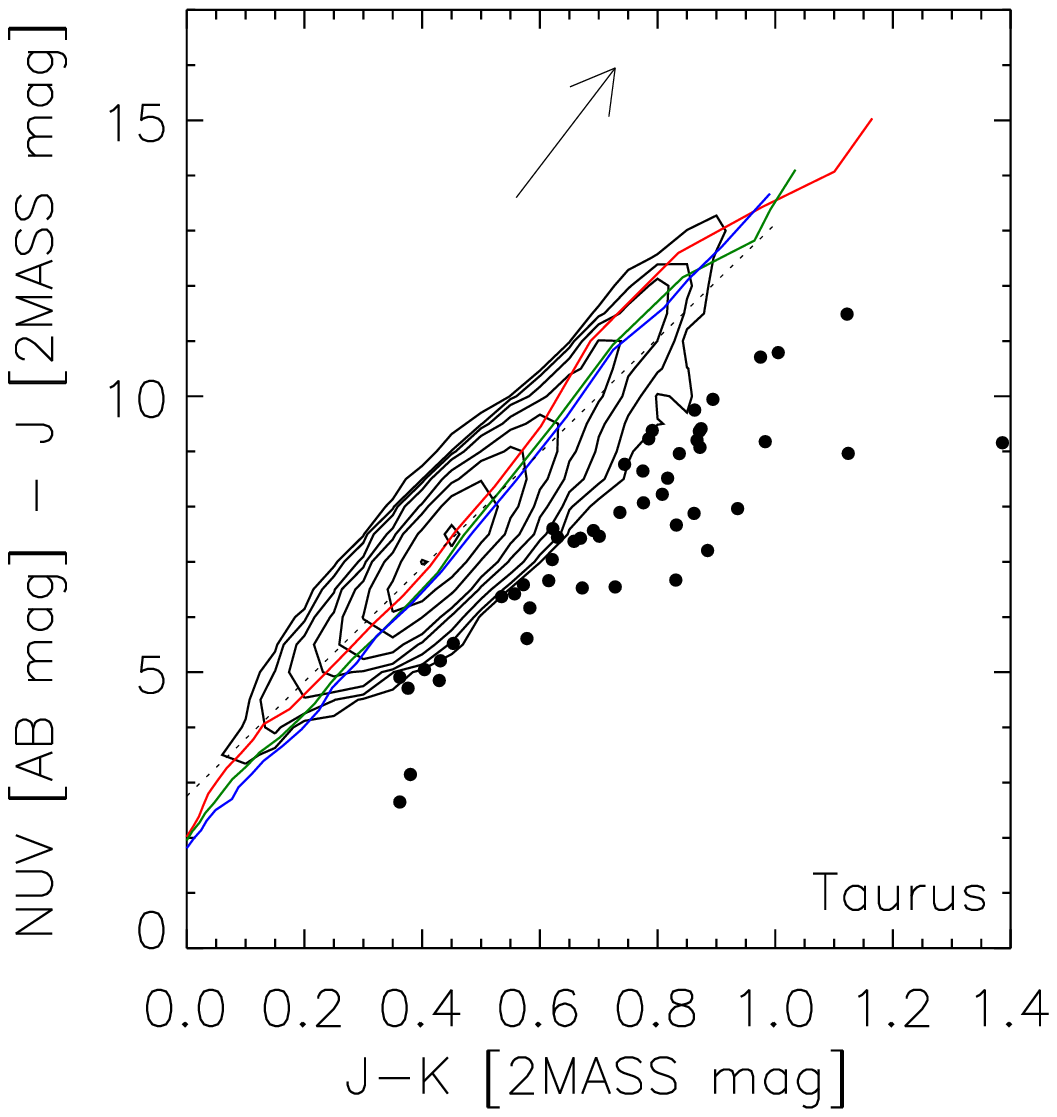}
\includegraphics[width=0.48\textwidth]{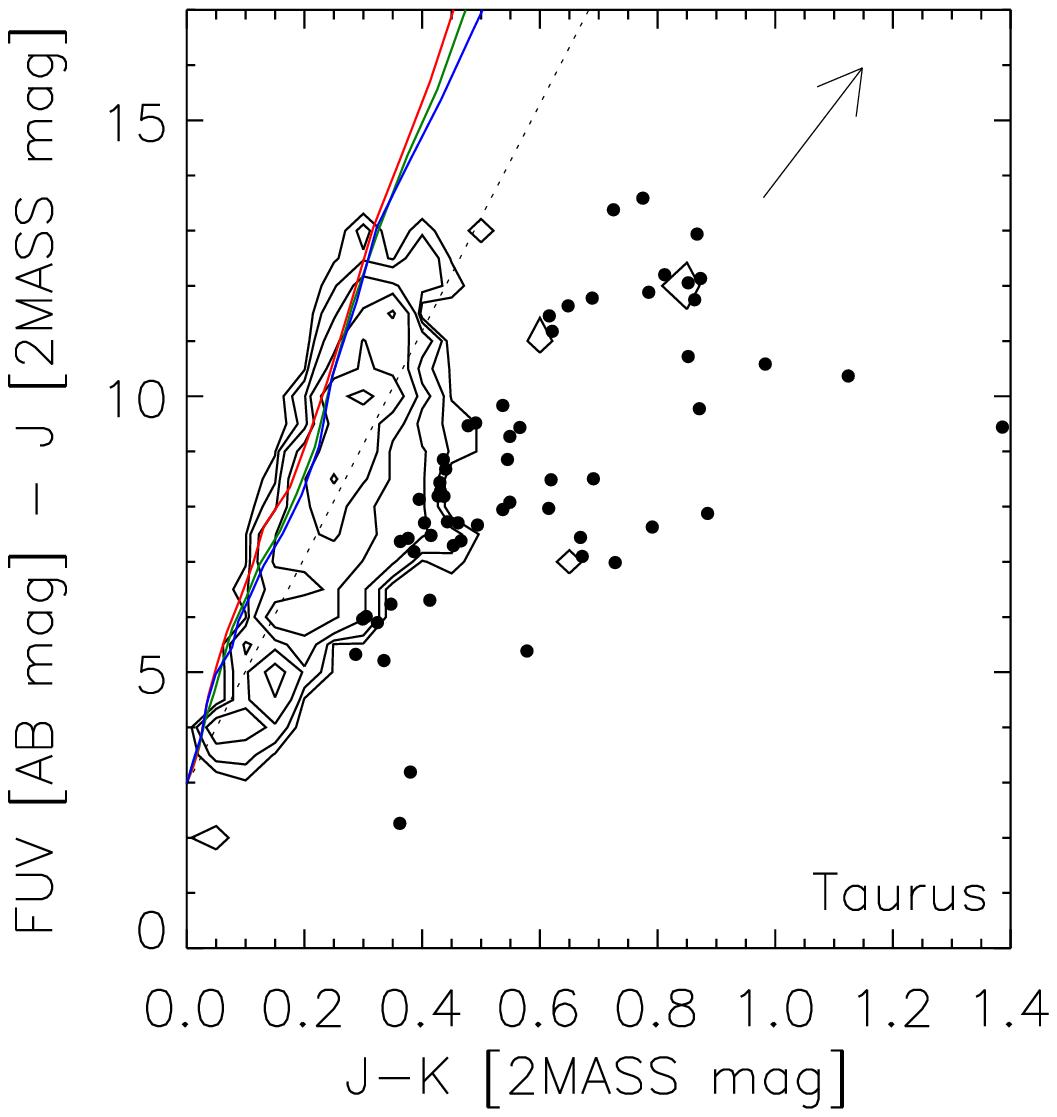}
\includegraphics[width=0.48\textwidth]{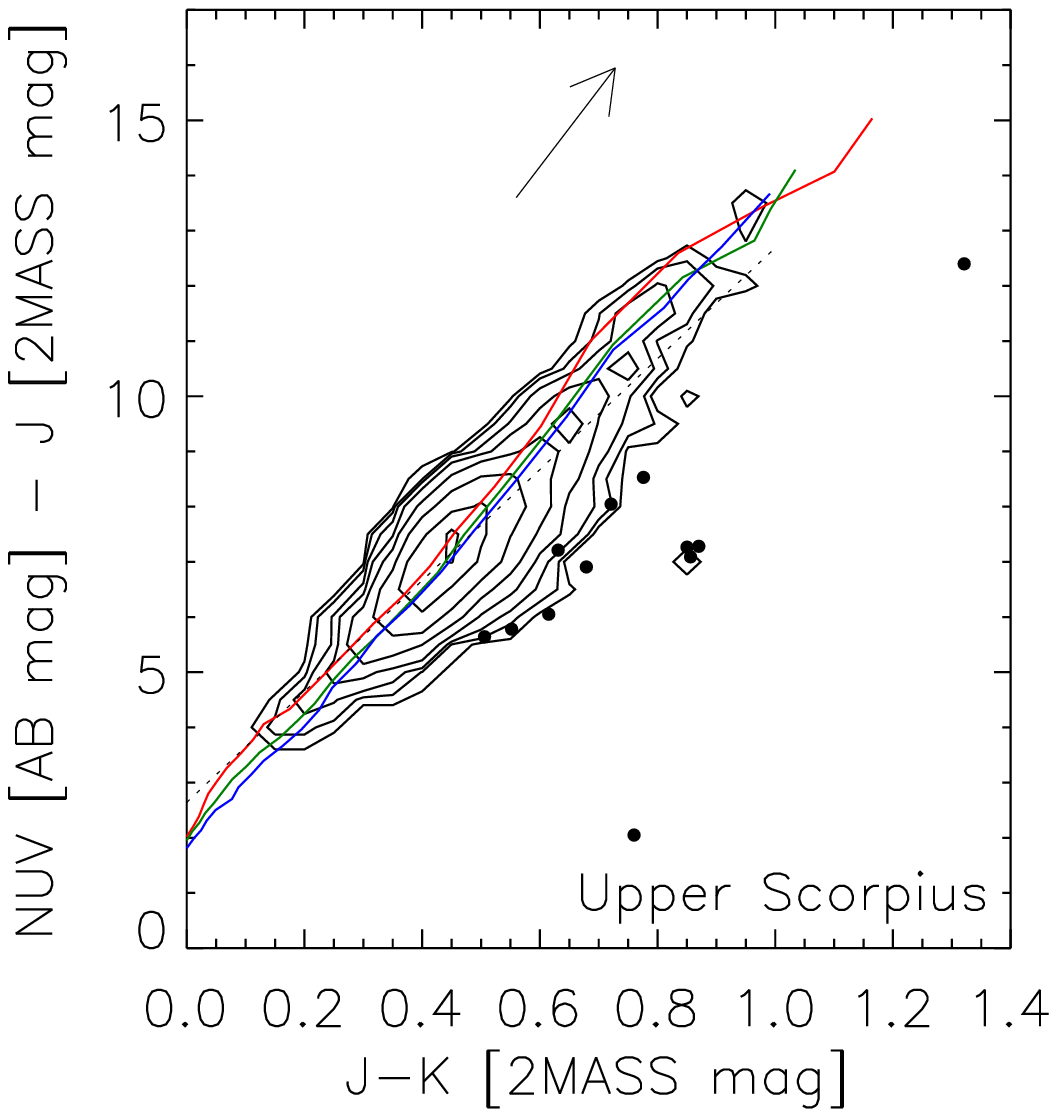}
\caption{$UV-J$ vs. $J-K$ diagrams of our two samples. The red, green and blue curves are colors integrated from Kurucz models with $\log g = 3.0$, 4.0, and 5.0, respectively. Contours represent the density of all matched sources with $J \le 14$ at 90\%, 50\%, 25\%, 12.5\%, ..., 0.78125\% of the peak source density in the NUV. In the FUV the lowest contour is at 3.125\% of the peak density because there are not enough sources to allow lower contours. The dashed line is our fit to the field stellar locus. Sources marked with a dot have a $5\sigma$ NUV excess (for $NUV-J$ plots) or $5\sigma$ FUV excess (for $FUV-J$ plots). The arrow represents $A_V = 1$.}\label{ccplots}
\end{figure*}

\begin{figure*}
\includegraphics[width=0.48\textwidth]{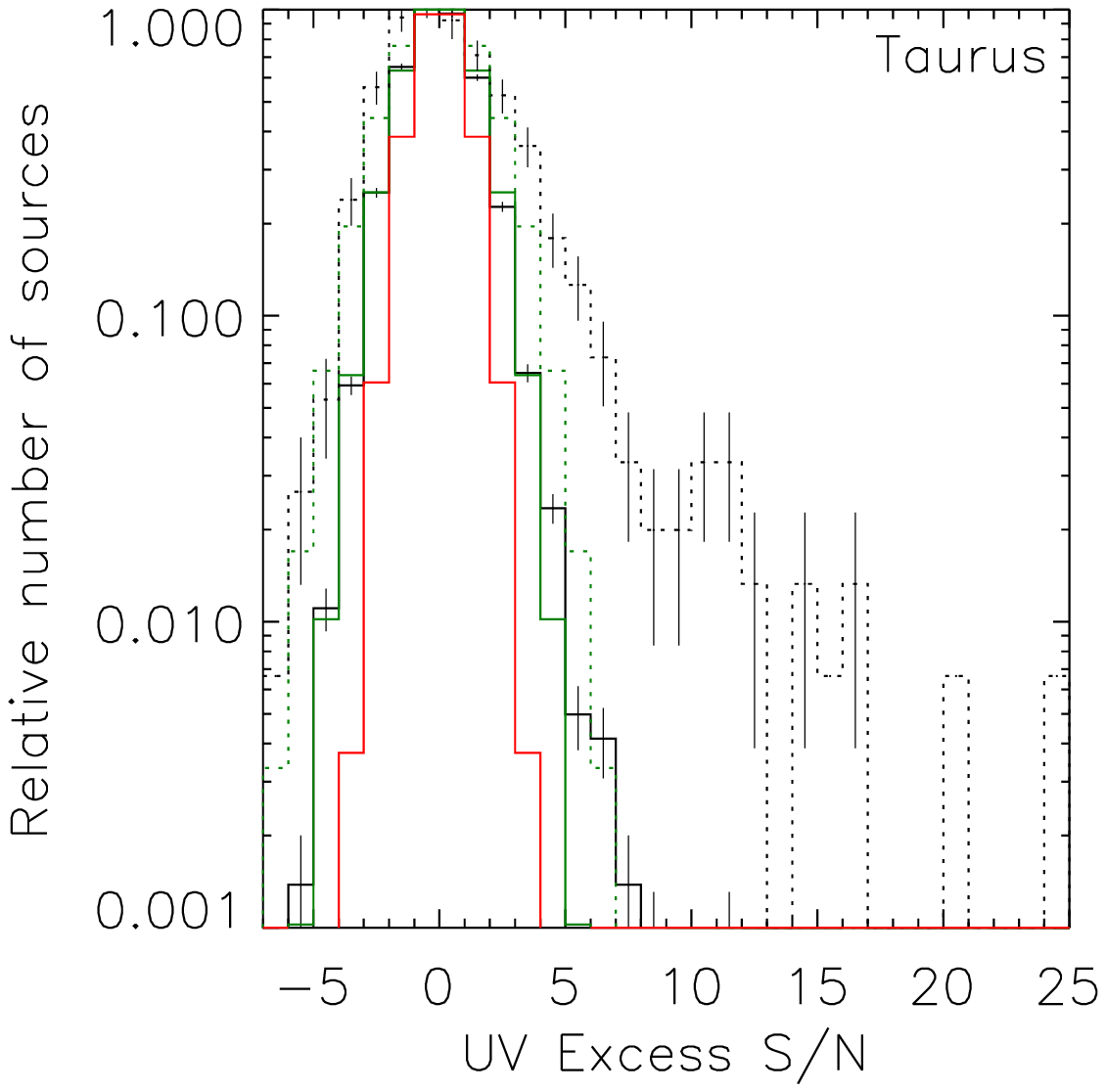}
\includegraphics[width=0.48\textwidth]{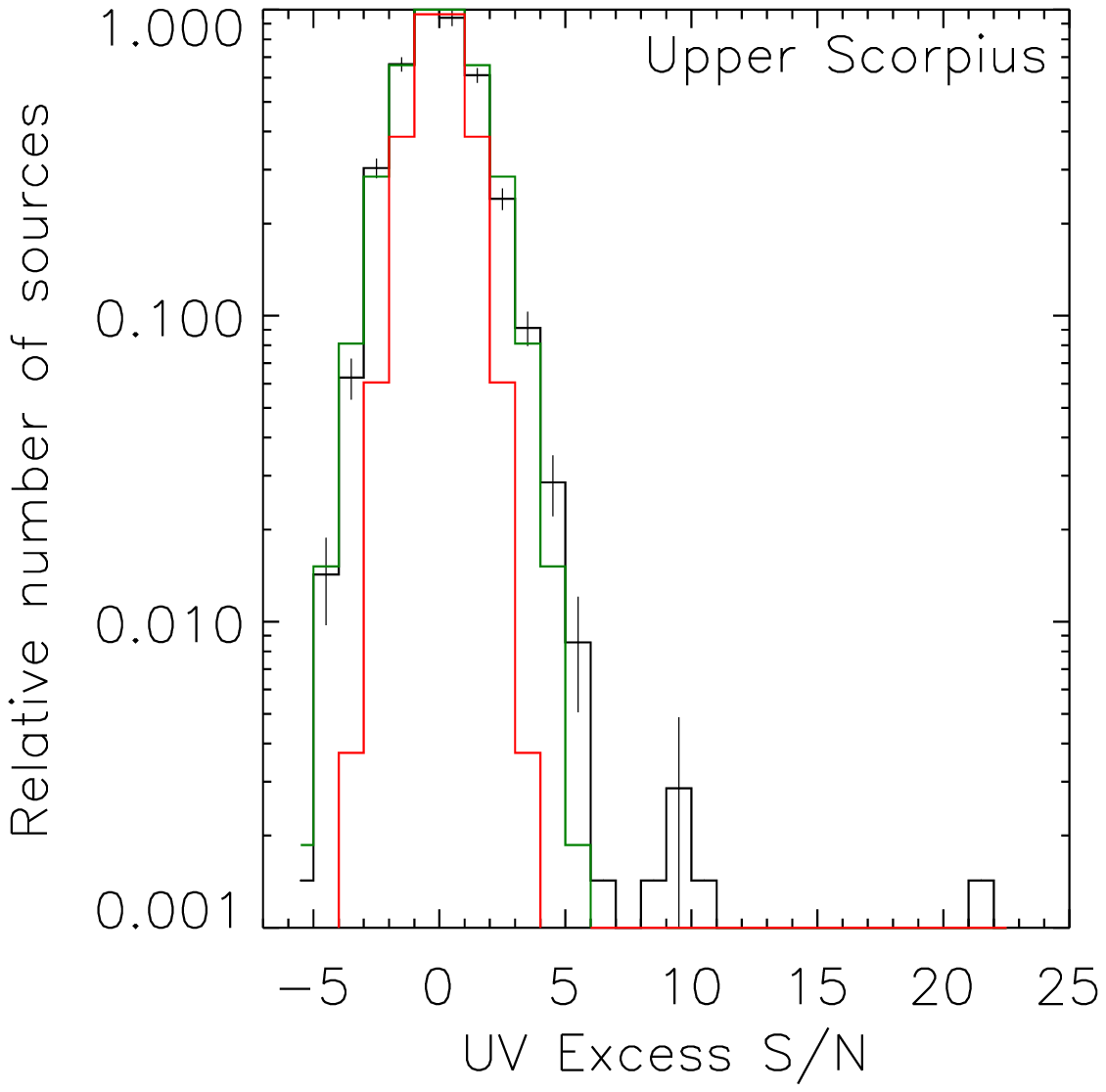}
\caption{Histograms of UV excesses divided by their errors for Taurus and Upper Scorpius data, plotted on a logarithmic scale. The solid black line shows the NUV excesses; the dotted line shows the FUV excesses for Taurus only.
The red curve is a normal distribution (i.e., $\sigma = 1$), the result expected if the spread in excesses is entirely due to Gaussian measurement errors.
To highlight the asymmetric wings, we plot in green Gaussian fits ($\sigma \sim 1.9$ in the FUV, $\sigma \sim 1.5$ in the NUV) to the histogram core, but we do not base our analysis on these fits.}\label{normex}
\end{figure*}

\begin{figure*}
\includegraphics[angle=90,width=0.9\textwidth]{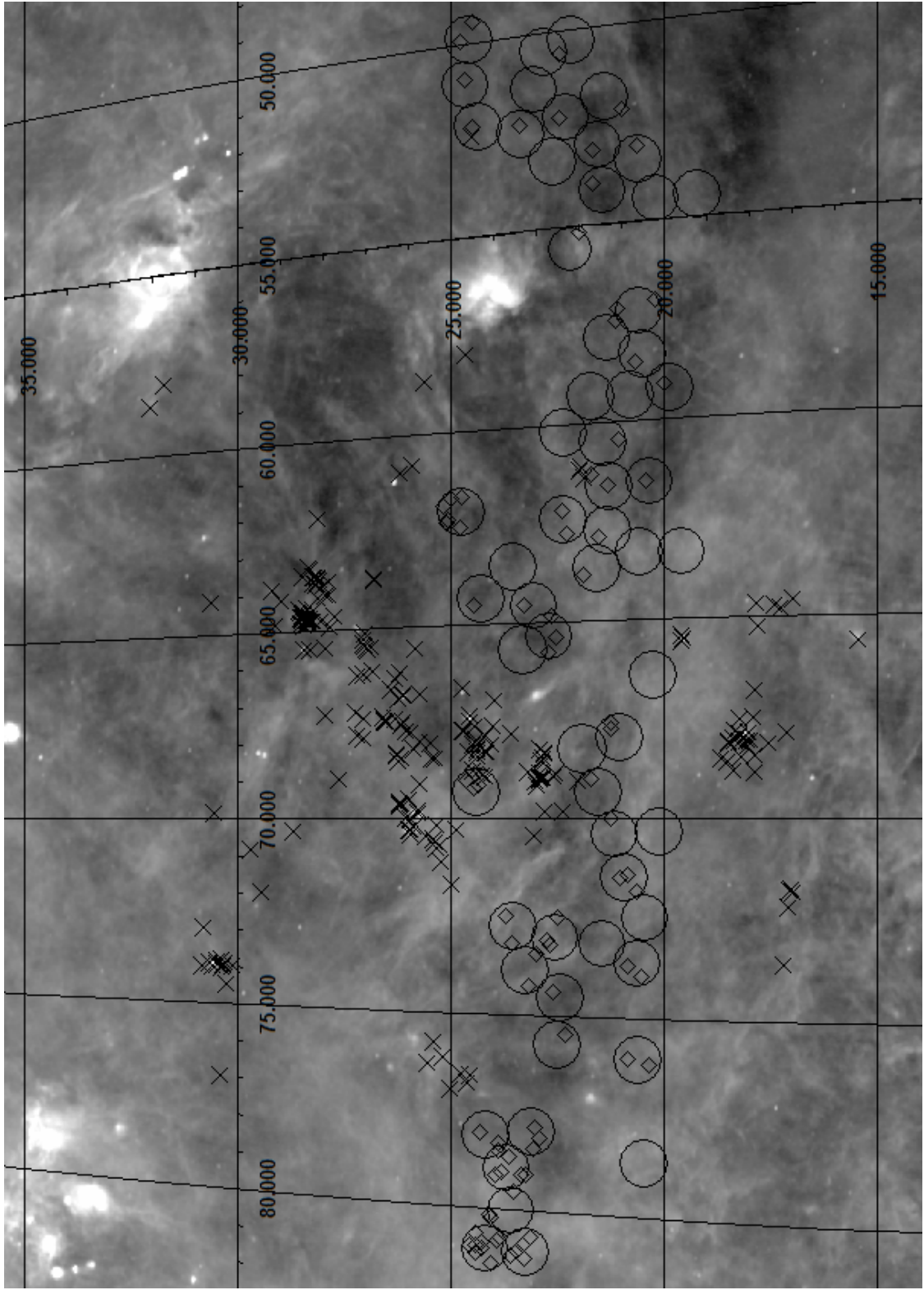}
\caption{Figure~\ref{context_tau}, with our $5\sigma$ UV excess sources in Taurus overplotted as diamonds. While to the eye there appears to be some clustering among the UV excess sources, we find no statistically significant concentrations.}\label{newmembers_tau}
\end{figure*}

\begin{figure*}
\includegraphics[width=\textwidth]{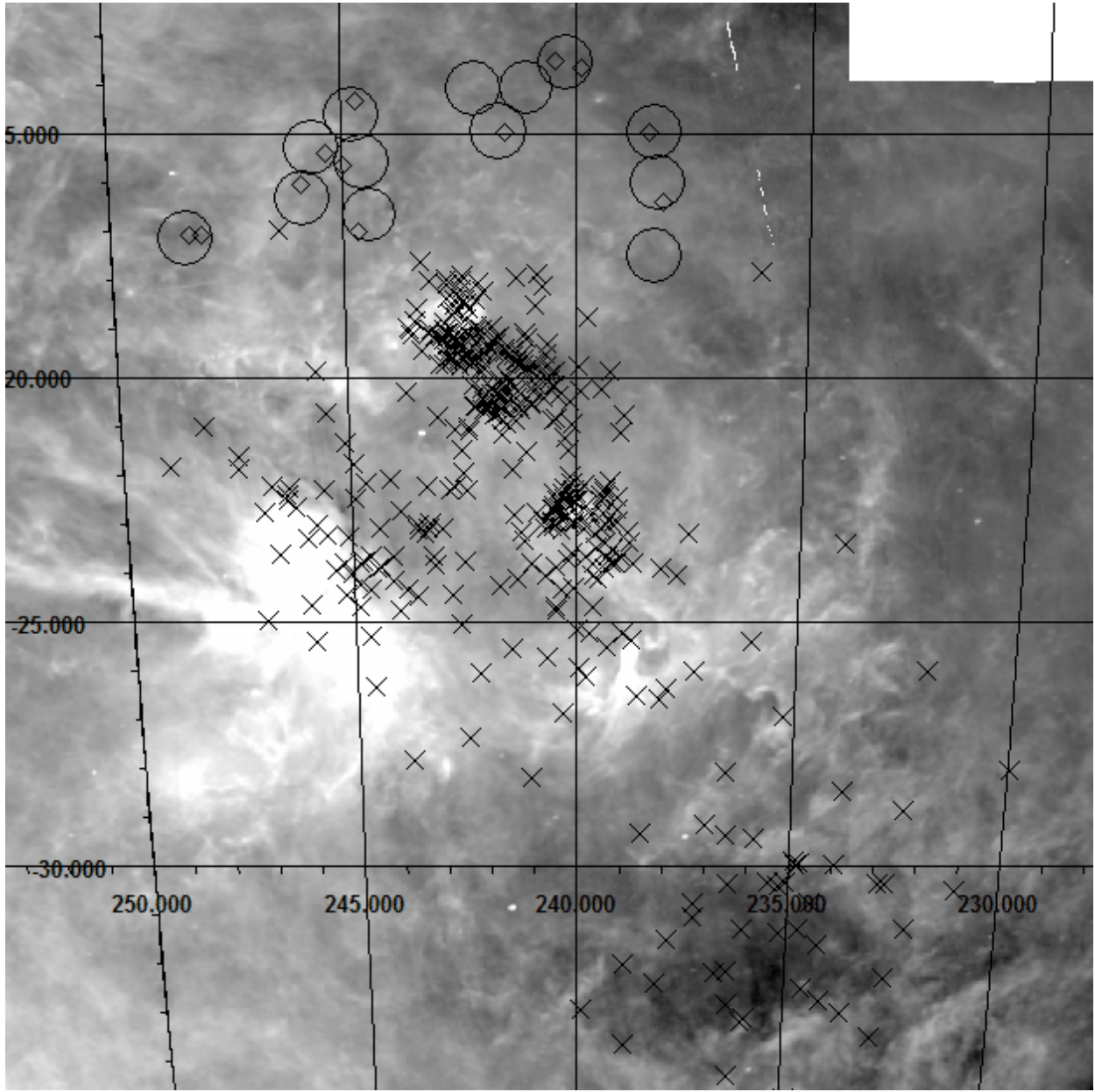}
\caption{Figure~\ref{context_sco}, with our $5\sigma$ UV excess sources in Upper Scorpius overplotted as diamonds. There are too few sources to test for structure.}\label{newmembers_sco}
\end{figure*}

\begin{figure*}
\includegraphics[width=0.48\textwidth]{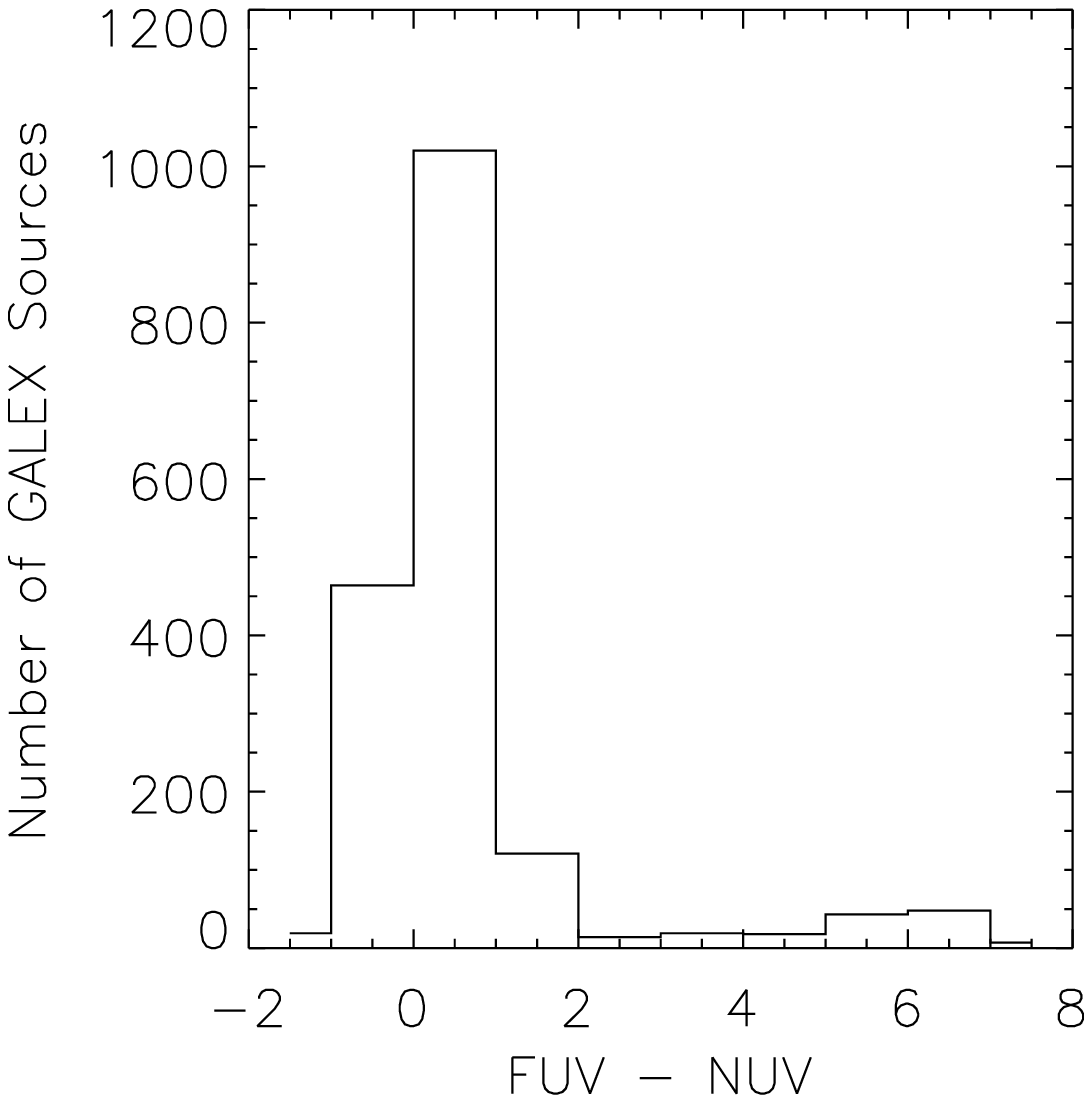}
\includegraphics[width=0.48\textwidth]{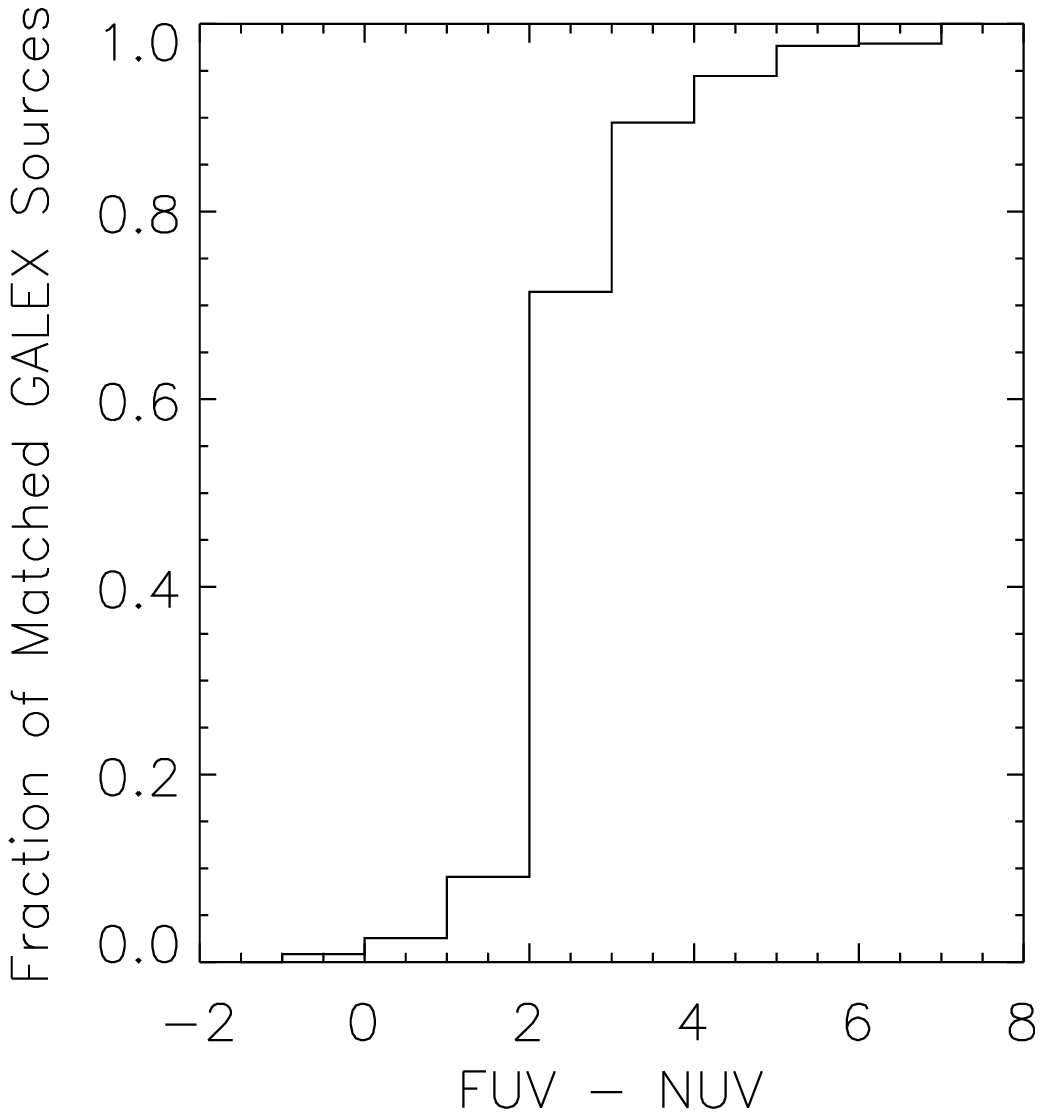}
\caption{Number of sources in our control field as a function of $FUV-NUV$ color (left), along with the fraction of these sources that have 2MASS counterparts (right). The population is dominated by $FUV-NUV \sim 0$ sources, but few of them have 2MASS counterparts, suggesting they are faint and either early-type stars or star-forming galaxies. The fraction of relatively blue, unmatched, and presumably distant GALEX sources is much higher than in Figure~\ref{unmatched}, which shows sources in Taurus and Upper Scorpius.}\label{fuvnuvmatch}
\end{figure*}

\begin{figure*}
\includegraphics[width=0.48\textwidth]{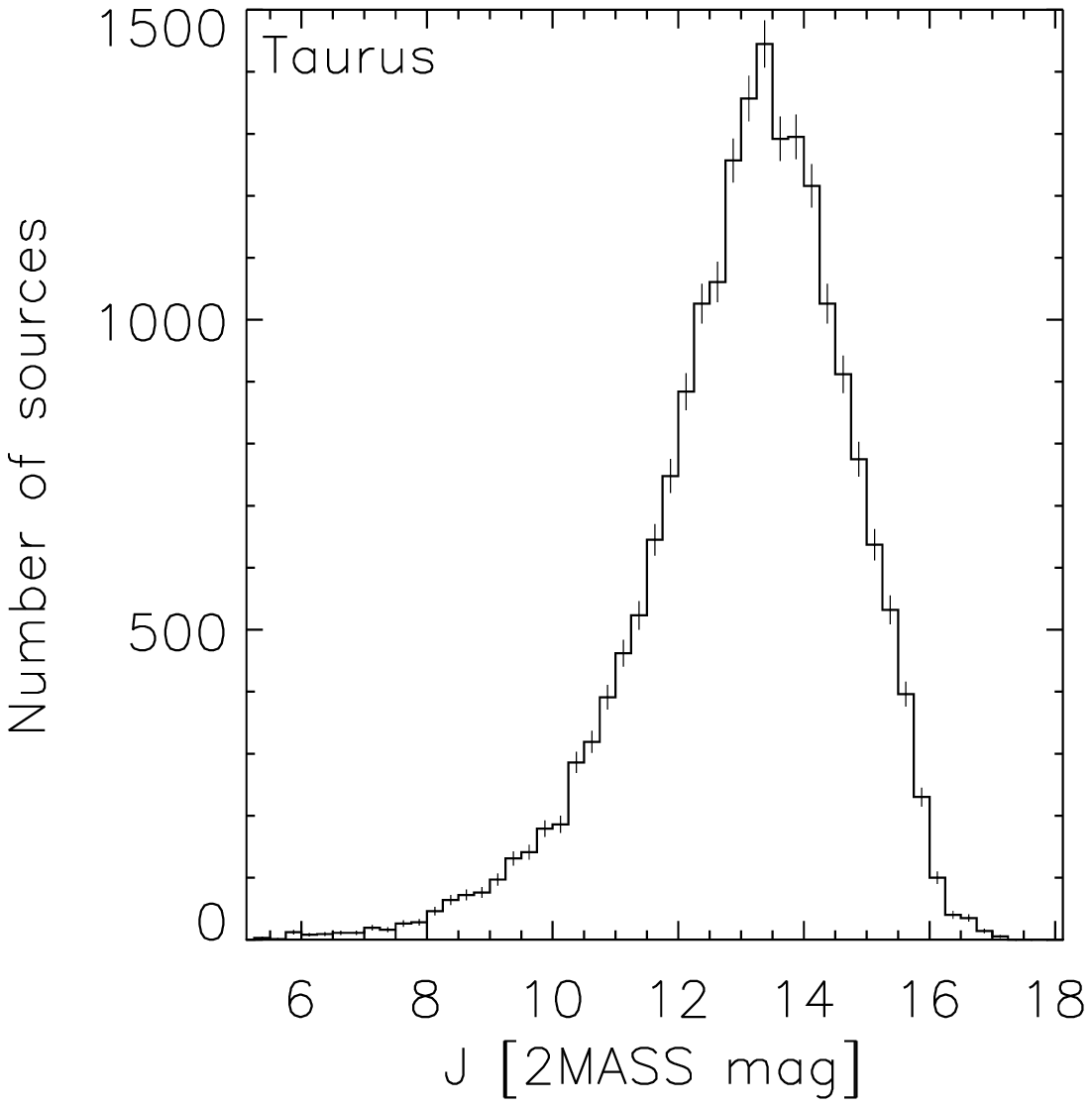}
\includegraphics[width=0.48\textwidth]{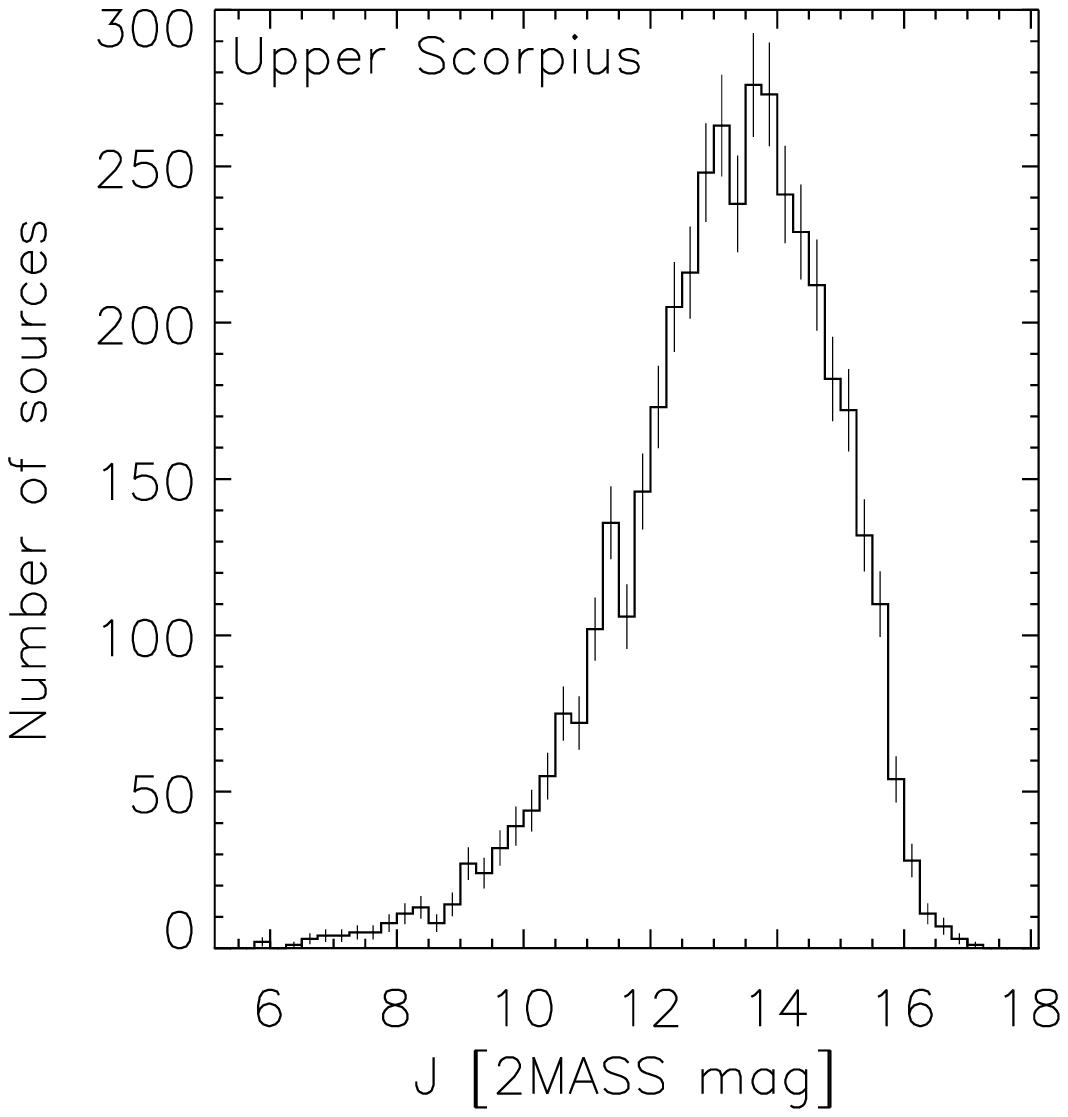}
\includegraphics[width=0.48\textwidth]{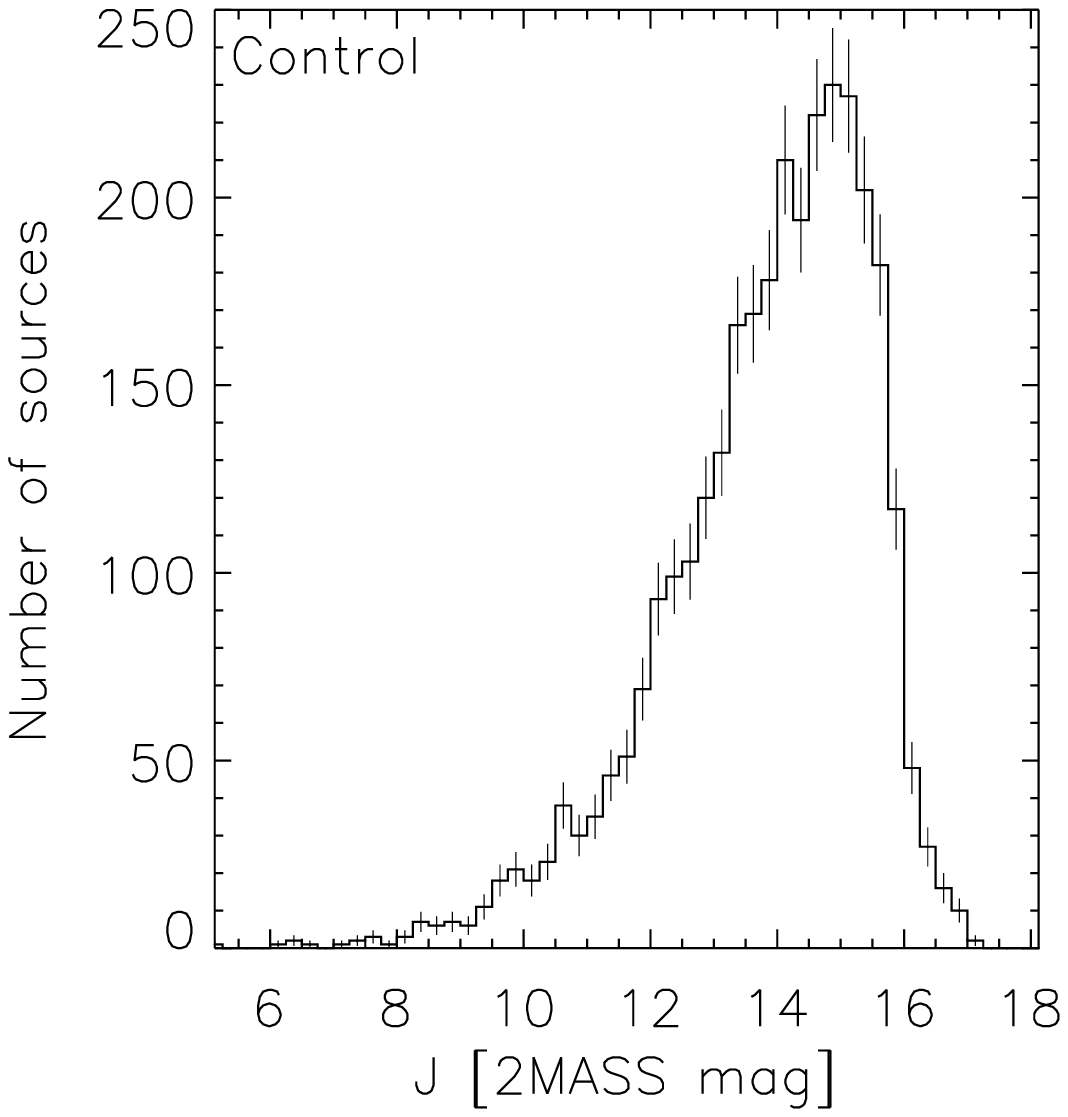}
\caption{Number of GALEX-2MASS matched sources in each of our three regions as a function of $J$ magnitude. The control field has many more faint sources, suggesting we are looking at more distant UV sources than in the other two fields.}\label{controlJ}
\end{figure*}

\begin{figure*}
\includegraphics[width=0.48\textwidth]{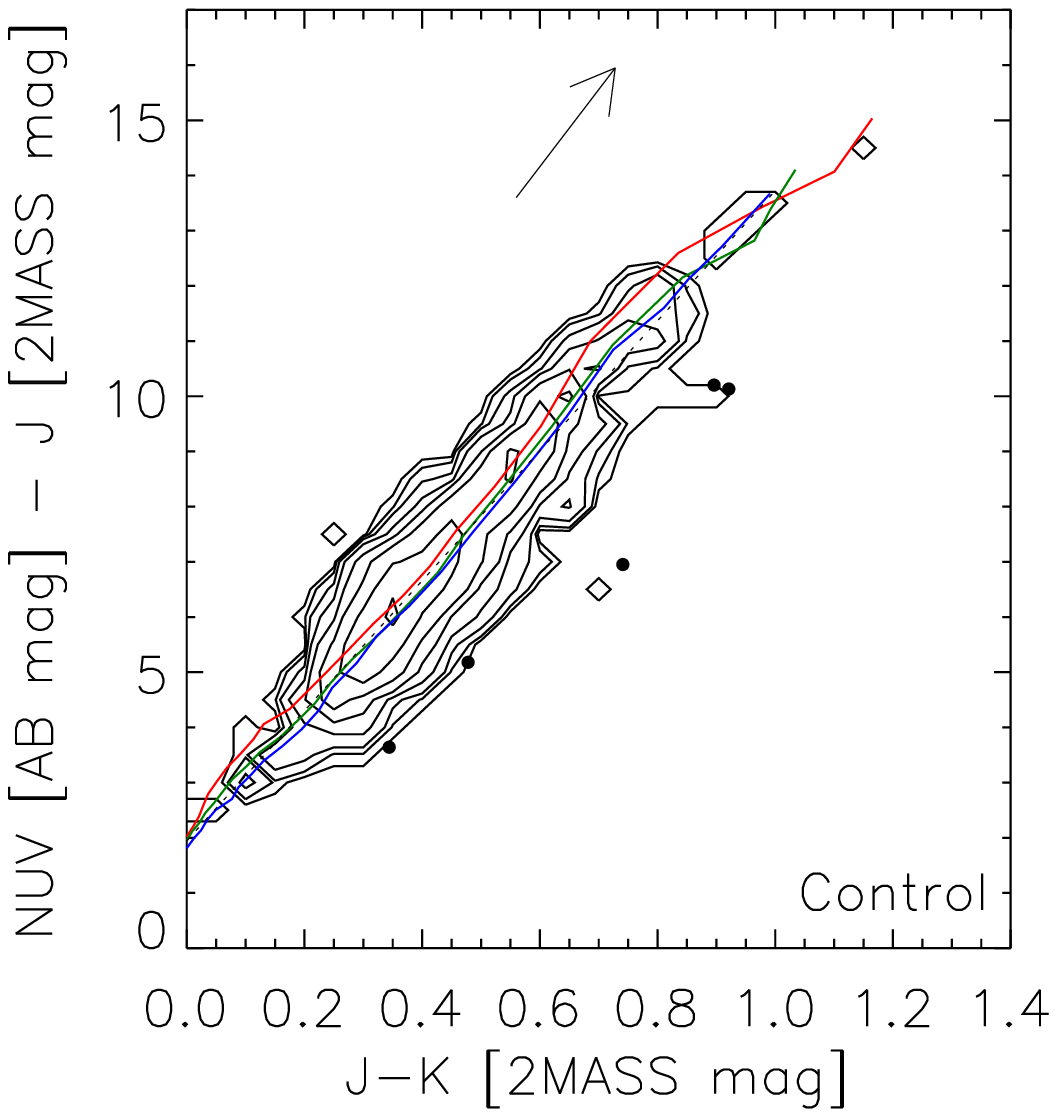}
\includegraphics[width=0.48\textwidth]{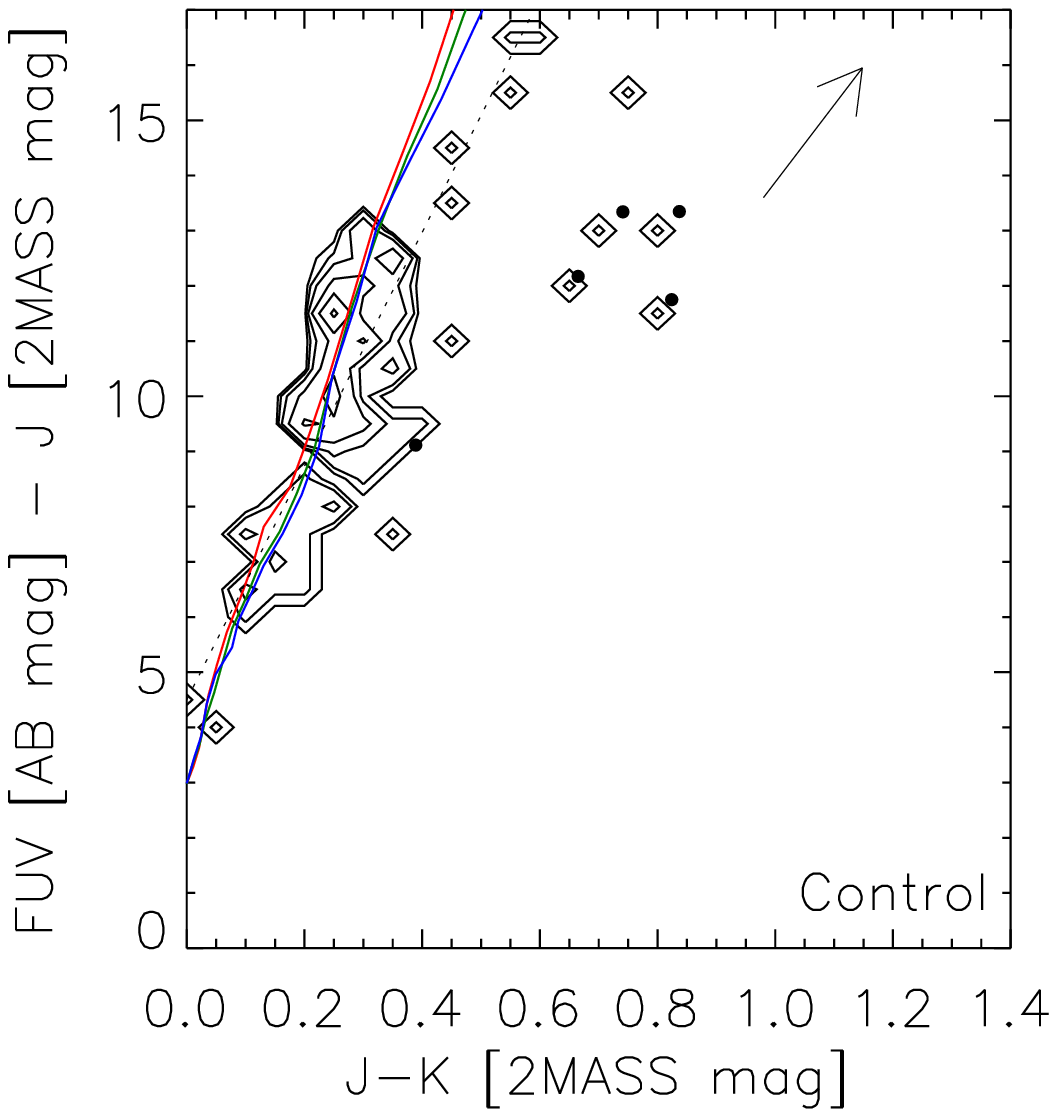}
\caption{$UV-J$ vs. $J-K$ diagrams for our control sample, analogous to Figure~\ref{ccplots}.}\label{controlplot}
\end{figure*}

\begin{figure*}
\includegraphics[width=0.48\textwidth]{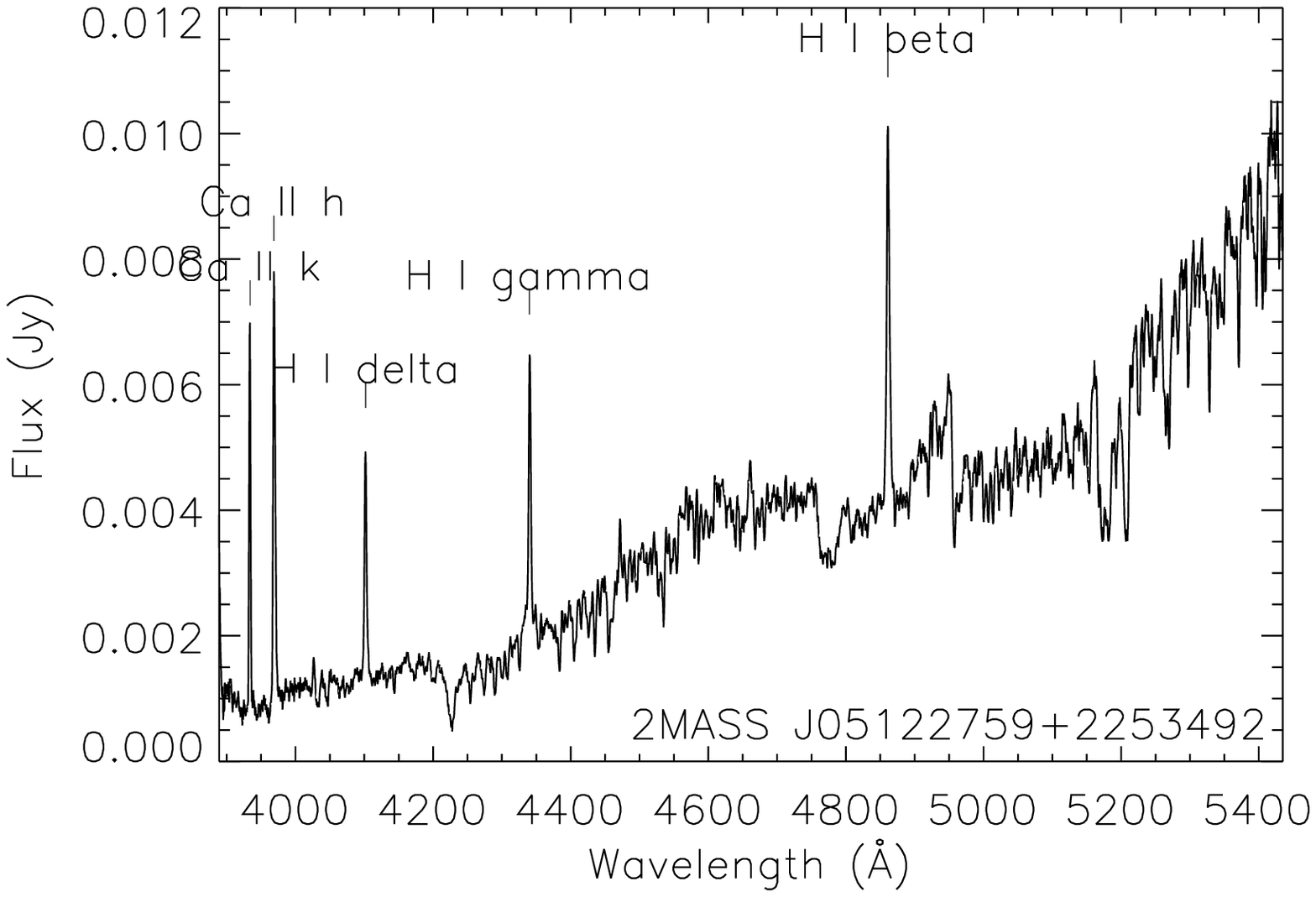}
\includegraphics[width=0.48\textwidth]{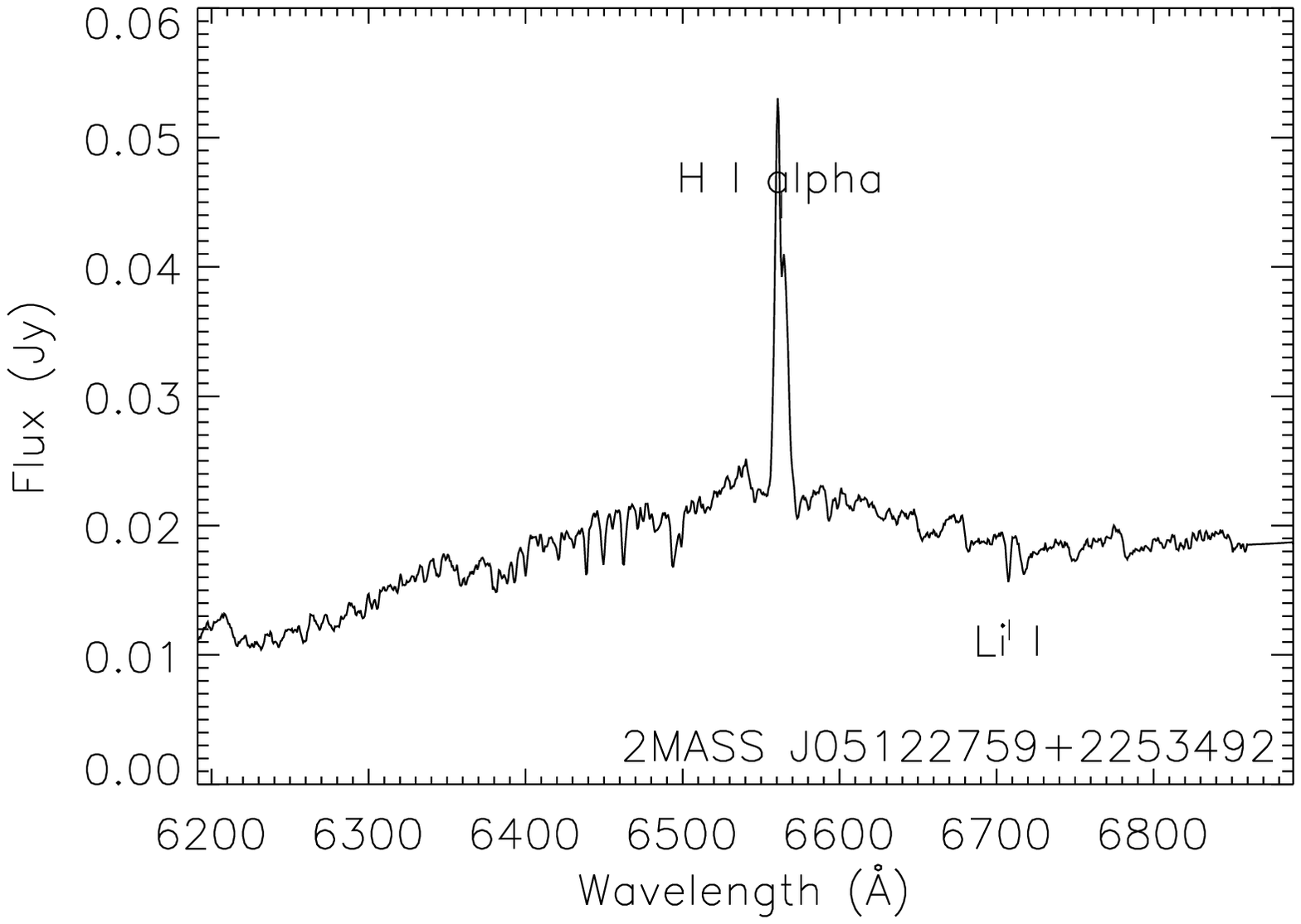}
\includegraphics[width=0.48\textwidth]{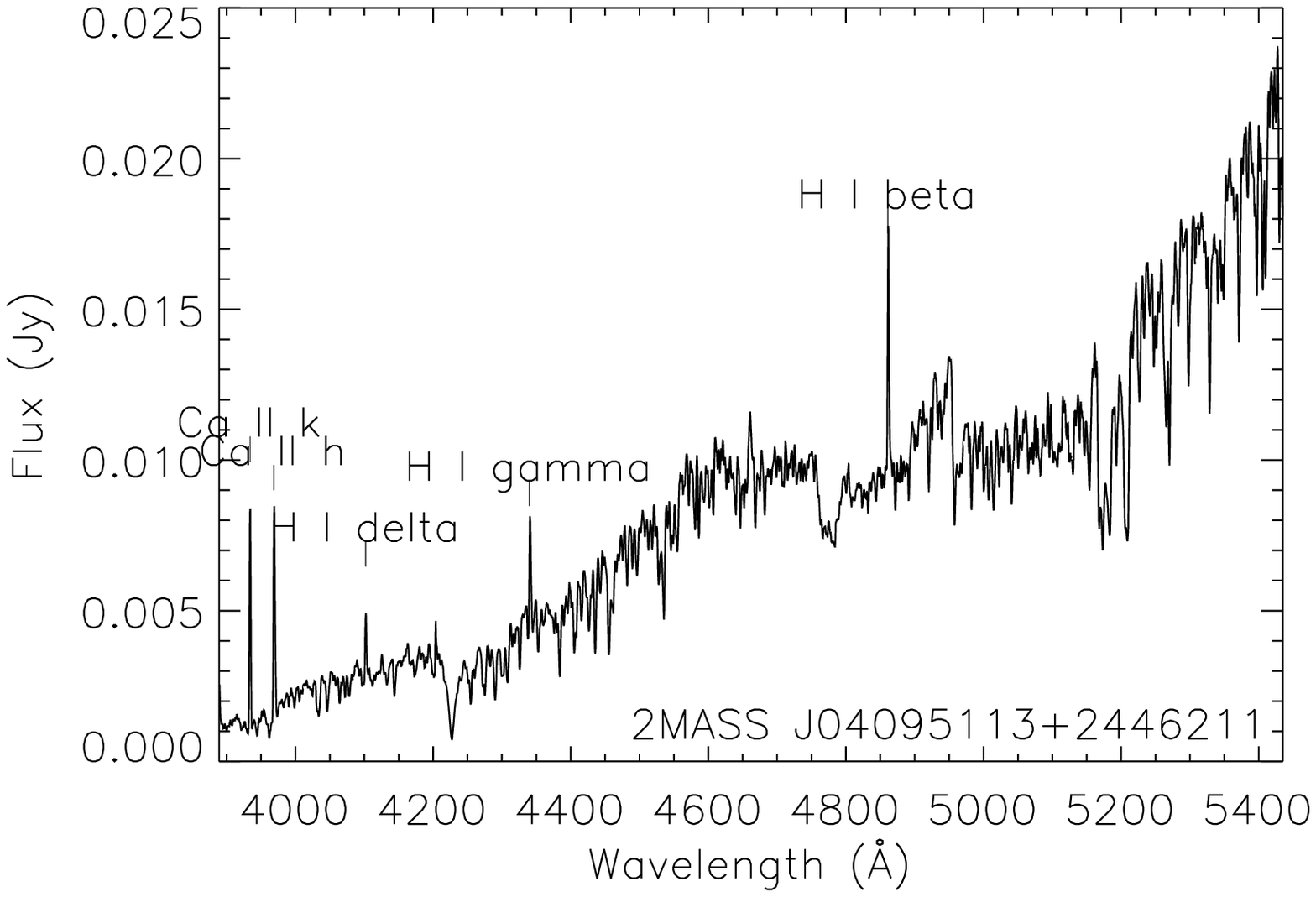}
\includegraphics[width=0.48\textwidth]{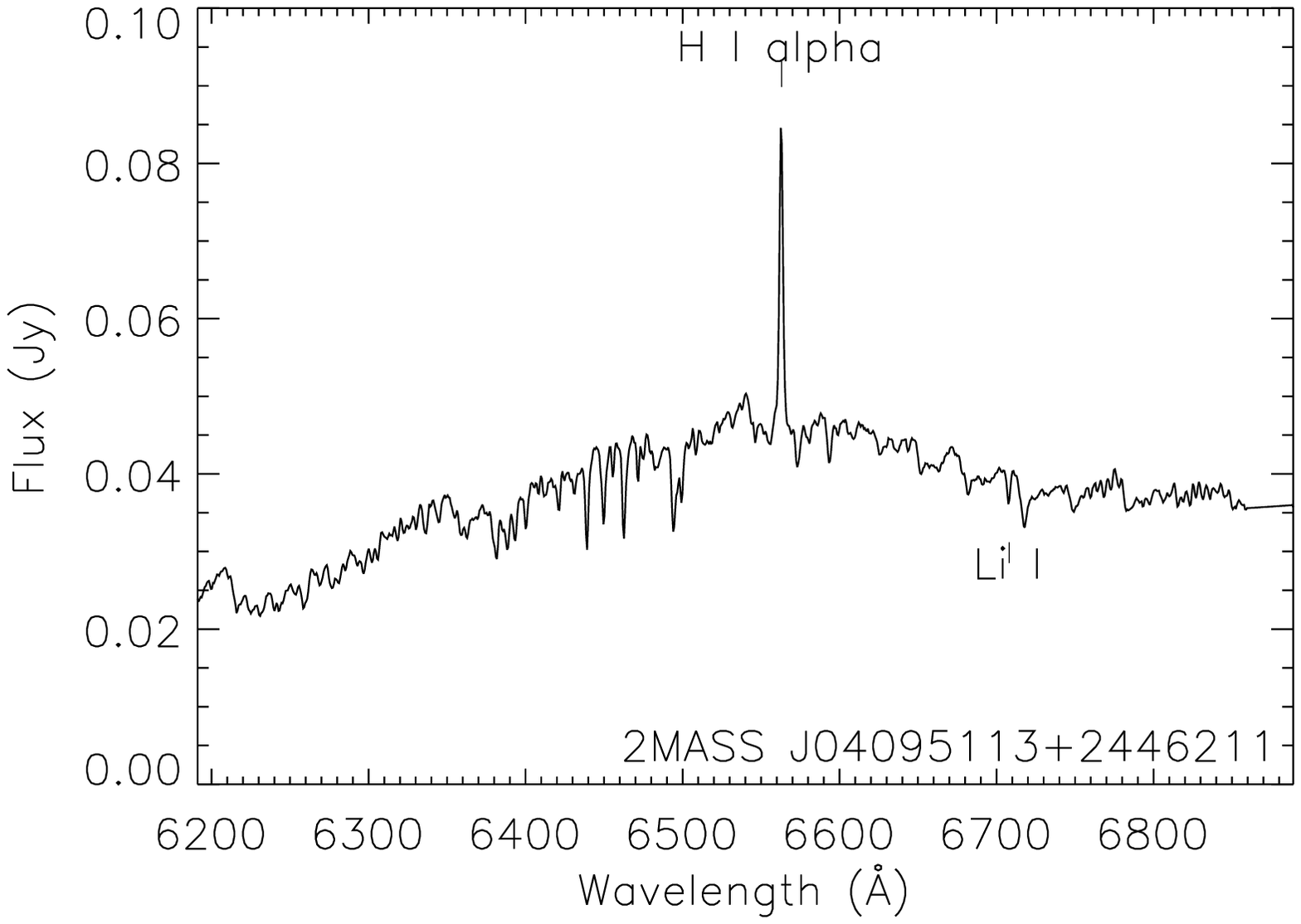}
\includegraphics[width=0.48\textwidth]{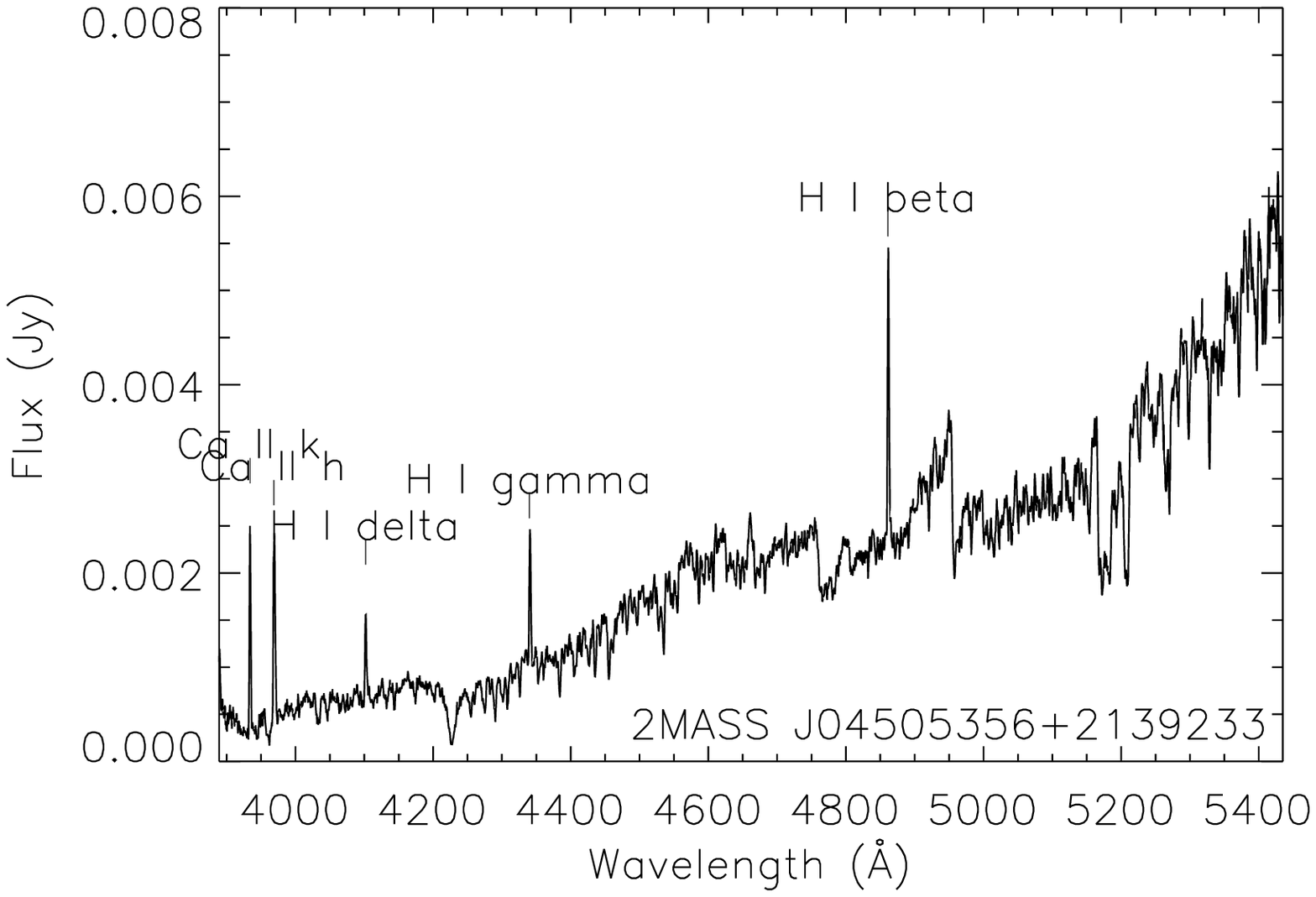}
\includegraphics[width=0.48\textwidth]{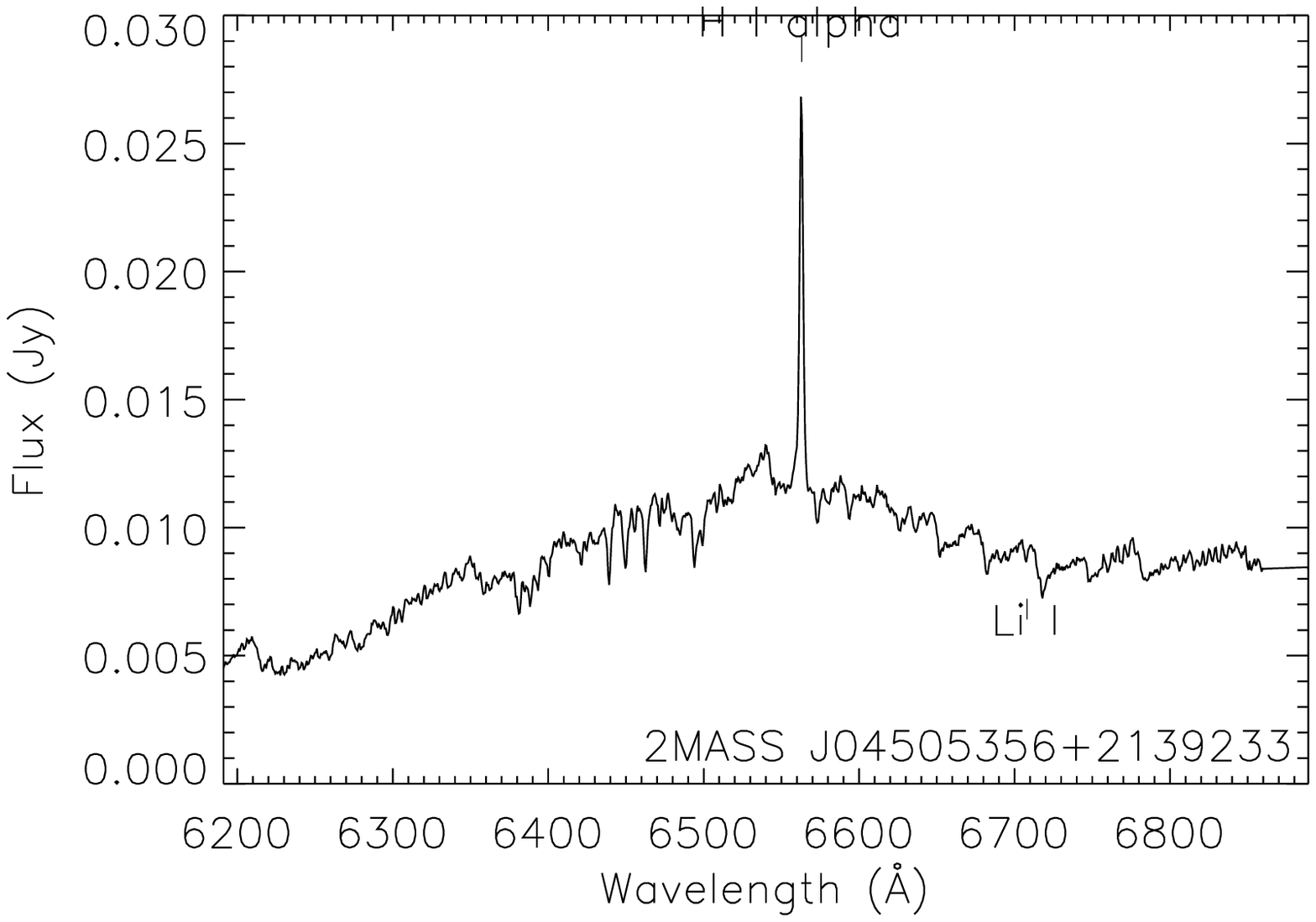}
\caption{The spectra of our three lithium-rich stars, with youth indicators marked. All three are active M dwarfs. 2MASS~J05122759+2253492 is a new lithium-rich dwarf near the Galactic plane. 2MASS~J04095113+2446211 is a rediscovery of the known Taurus member 1RXS~J044712.8+203809. Our spectrum of 2MASS~J04505356+2139233 is the first detection of lithium in the previously unstudied ROSAT source 1RXS~J045053.5+213927. }\label{specli}
\end{figure*}

\begin{figure*}
\includegraphics[width=0.48\textwidth]{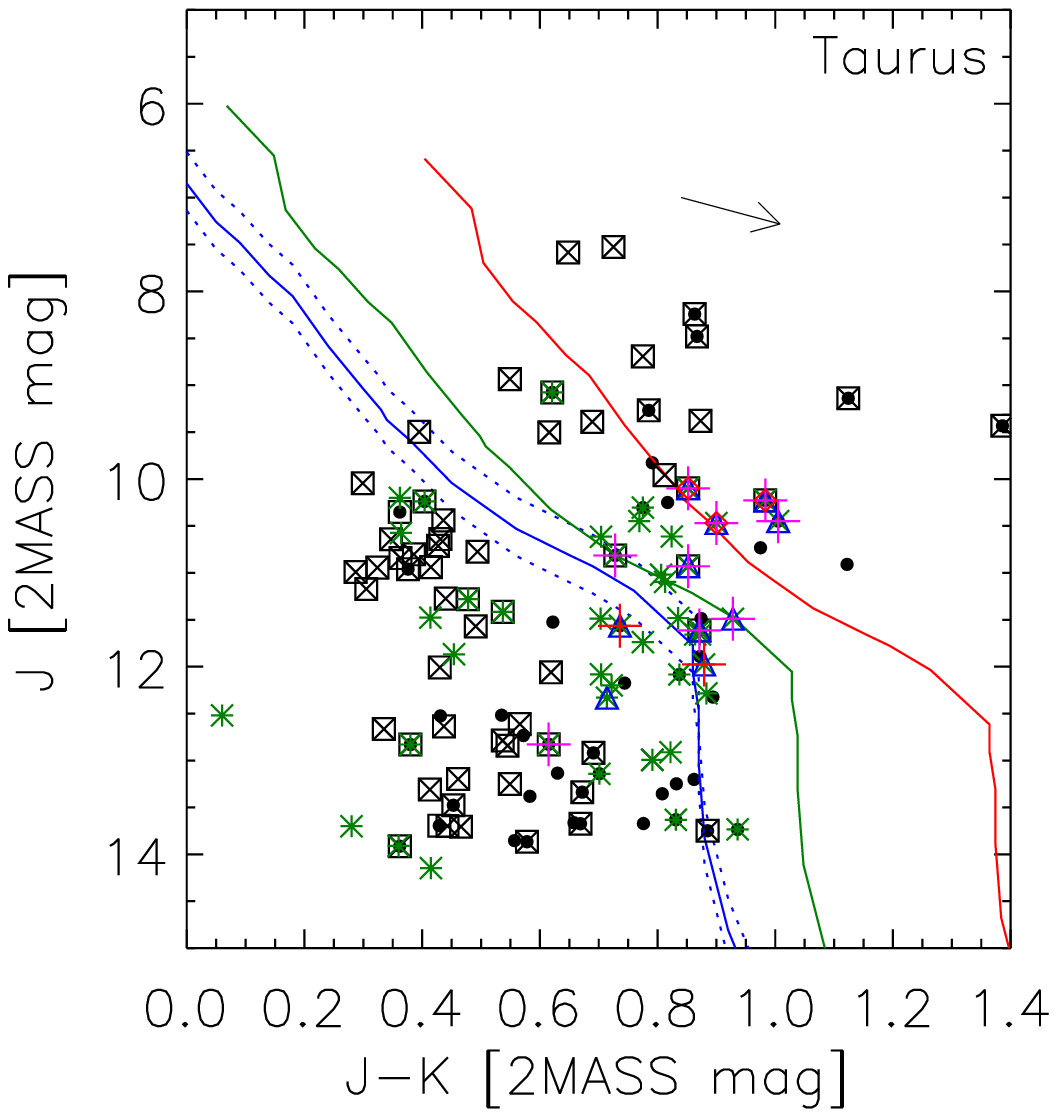}
\includegraphics[width=0.48\textwidth]{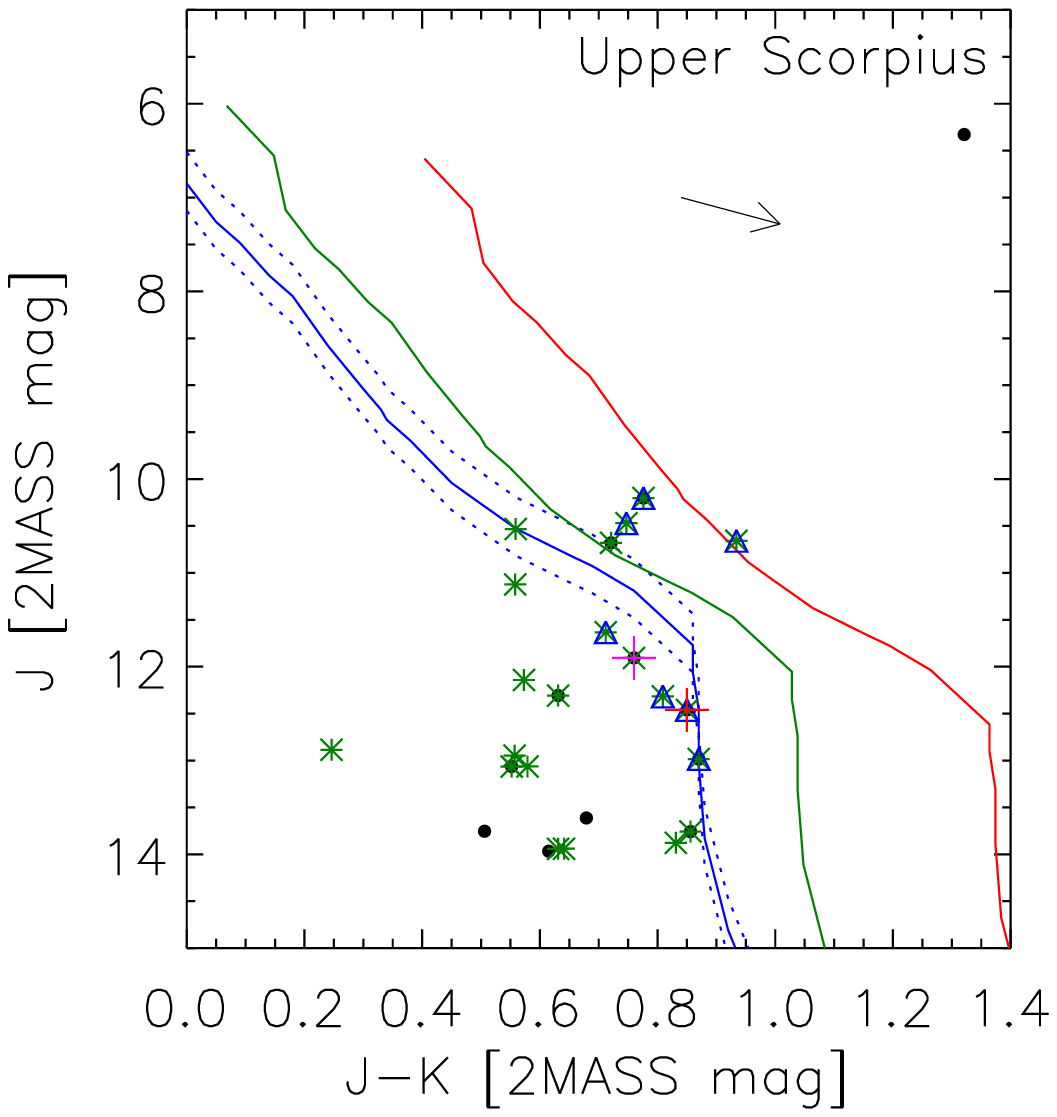}
\includegraphics[width=0.48\textwidth]{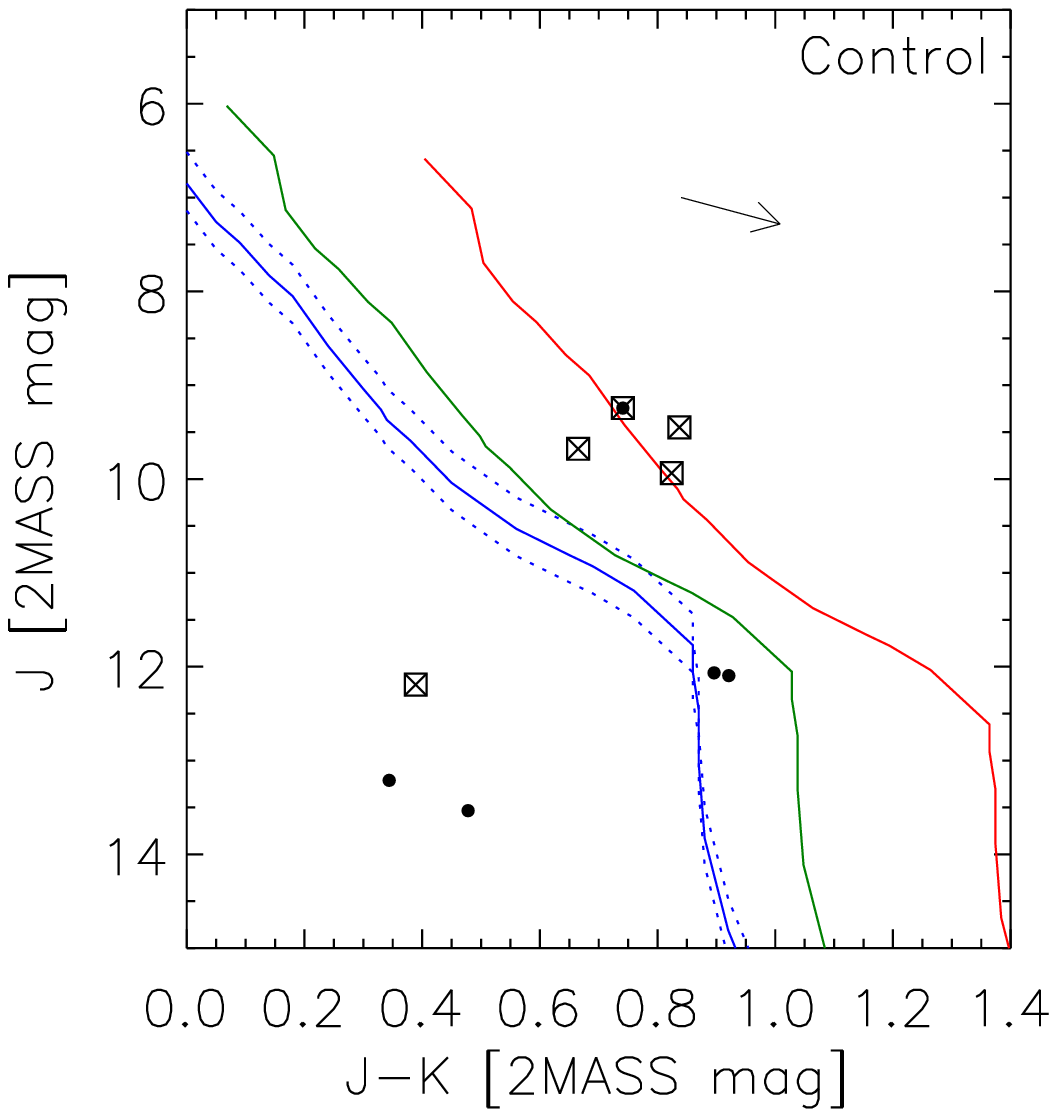}
\caption{Color-magnitude plots of our three survey regions. Black boxes represent $5\sigma$ excess sources detected in both NUV and FUV, regardless of in which band(s) they have an excess. Black dots represent $5\sigma$ sources with an NUV excess, regardless of whether they have an FUV detection. Black X represent sources with an FUV excess, regardless of whether they have an NUV detection. 
Green stars represent sources observed at Palomar; some are $3\sigma$ sources but not $5\sigma$ sources. Blue triangles, blue crosses, red crosses, and red diamonds represent Ca~II emission, H~I~$\beta$ emission, H~I~$\alpha$ emission, and Li~I absorption, respectively. 
The solid blue curve represents the unextinguished main sequence from \citet{mscolors} at 140~pc; the dotted curves are at 120~pc and 160~pc. The green curve is the main sequence at 140~pc with $A_V = 1$; the red curve is the same with $A_V = 3$.
The number of sources shown on each plot is roughly proportional to the area of each sample: 56~deg$^2$ in Taurus, 12~deg$^2$ in Upper Scorpius, and 6.5~deg$^2$ in our control fields.}\label{cmplots}
\end{figure*}

At $J - K = 0.76$, $NUV - J = 2.05$, Sco~X-1 is a clear outlier in Figure~\ref{ccplots}. We observed it at Palomar so that we would have a template spectrum for other X-ray binaries that might appear at smaller UV excess. Since there are no other sources in our Palomar sample with similar (emission-line dominated) spectra, and since we deliberately selected high-excess sources as Palomar targets, we are confident this is the only X-ray binary in our sample.

\paragraph{2MASS J05161463+2313174}\label{bgstar}

This A0e~V star has a self-absorbed H$\alpha$ emission line and an inverse P~Cygni H$\beta$ line (Figure~\ref{pcyg}), both signs of accretion. The difference between the peak and the trough of the H$\beta$ line is about 200~km~s$^{-1}$. The star's $J$ magnitude of 10.8 and a diffuse interstellar band at 4430\AA\ are difficult to reconcile with Taurus membership. It may instead be a background ($\gtrsim 600$~pc away) mass-transfer binary; however, we note that it has a much more modest UV excess than Sco X-1 (Table~\ref{specresults}).

\paragraph{2MASS J04015065+2103495}\label{wdmd}

This star's spectrum contains both broad Balmer absorption lines and strong TiO bands, identifying it as a white dwarf M~dwarf (WDMD) double. It is found at $J-K = 0.83$, $NUV-J = 6.67$ in Figure~\ref{ccplots}. While it is at the edge of the locus of UV excess sources, it is not an obvious outlier as was Sco~X-1. As a result, WDMD pairs are a potentially significant contaminant in our sample.

To identify the locus of WDMD doubles in $UV-J$ vs. $J-K$ space we find empirical absolute magnitudes of white dwarfs in appendix~\ref{mags_wd}. In particular, a white dwarf with $T_\textrm{eff} \sim 16000$~K has $M_\textrm{FUV} = 11.11, M_\textrm{NUV} = 11.41$. A double with an M0 primary and a 16000~K secondary, then, has $FUV - J \sim 5.1$, $NUV - J \sim 5.4$, $J-K \sim 0.86$ (or an apparent $FUV$ excess of 20 magnitudes and $NUV$ excess of 7.3 magnitudes over the M dwarf photosphere). As their white dwarfs cool, WDMD pairs form the WDMD locus by migrating along lines of constant $J-K$ toward the main sequence. 
Any extinction moves the WDMD locus redward almost parallel to the field locus.

\citet{wdmd} report that 1 out of every 2300 SDSS stars with $u < 20.5$ are WDMD pairs, while \citet{wdmd2} find that using $NUV$ data from the GALEX Early Release rather than the $u$ band doubles this number. Given 14130 and 2828 matched sources in Taurus and Upper Scorpius, respectively, we expect about 12 and 2 of them to be WDMD pairs. Since it is not clear to what excess level \citet{wdmd2} were able to probe, we cannot convert this figure to a number of UV excess sources in our sample; their study, benefiting from the high precision of SDSS photometry, may be sensitive to WDMDs to which we are not.

\subsection{UV excess sources without optical activity indicators}\label{phantomexcess}

We identified in section~\ref{oddballs} examples of sources other than chromospherically active stars that produce an ultraviolet excess, with spectra characteristic of X-ray binaries or WDMD pairs. However, many other UV excess sources remain unexplained, as their spectra are those of ordinary main sequence stars. In Taurus, 
%30 of our 43 $3\sigma$ Palomar targets and 
12 of our 20 $5\sigma$ targets show no signs of chromospheric activity in their spectra. Extrapolating to the rest of the $5\sigma$ sample, we infer a density of UV excess sources without optical activity indicators between 0.7~deg$^{-2}$ and 1.2~deg$^{-2}$ (90\% symmetric confidence interval).
In Upper Scorpius, the spectra of 
%13 of our 20 $3\sigma$ targets and 
4 of our 7 $5\sigma$ targets show no unusual features. We infer a density of UV excess sources without optical activity indicators between 0.4~deg$^{-2}$ and 0.7~deg$^{-2}$ with 90\% confidence.
We might have overlooked chromospheric activity indicators among A or early F stars, where the strong photospheric background makes it difficult to discern optical emission lines \citep[e.g.][]{bluerot}, but we should have seen any such indicators among the later spectral types.

We cannot attribute the entire population of stars with UV excesses but normal optical spectra to errors in our photometry or in our characterization of the field locus. Based on the statistics in Table~\ref{exstats}, we expect 3 of our $5\sigma$ Taurus spectroscopic targets to be included by such errors, and at most 1 in our Upper Sco sample. 
In addition, we see stars with normal spectra at up to $10\sigma$ excess, where such uncertainties should have no effect. 
We must seek other explanations for this result.

The first possibility is that we are looking at chance superpositions of UV and infrared sources. To find the expected number of such sources, we model the GALEX and 2MASS sources in our fields as three uniformly and independently distributed populations: one of UV-only sources with density $\Sigma_\textrm{UVonly}$, one of IR-only sources with density $\Sigma_\textrm{IRonly}$, and one of sources visible in both bands with density $\Sigma_\textrm{both}$. The observed densities of UV sources, IR sources, and matched sources should then be
\begin{equation}\label{matchmodel}
\begin{array}{rcl}
\Sigma_\textrm{UV} & = & \Sigma_\textrm{UVonly} + \Sigma_\textrm{both} \\
\Sigma_\textrm{IR} & = & \Sigma_\textrm{IRonly} + \Sigma_\textrm{both} \\
\Sigma_\textrm{match} & = & \Sigma_\textrm{both} + \Sigma_\textrm{UVonly}\Sigma_\textrm{IRonly}A_{m} \\
\end{array}
\end{equation}
where $A_m = \pi (\textrm{2\arcsec})^2$ is the area around a GALEX source in which a 2MASS source would be assumed a counterpart. Taking values for $\Sigma_\textrm{UV}$, $\Sigma_\textrm{IR}$, and $\Sigma_\textrm{match}$ from Table~\ref{galexobs} (note we only count 2MASS sources with $J \le 14$, as we have throughout the paper) and solving for the density of chance alignments $\Sigma_\textrm{UVonly}\Sigma_\textrm{IRonly}A_{m}$, we expect 6 such matches in our Taurus survey region and 2 in our Upper Scorpius fields. Since there are 89 $5\sigma$ UV-excess sources in Taurus and 11 in Upper Scorpius, and since some chance matches might produce a smaller excess than $5\sigma$, we can discount them as the source of most of these unexplained UV excesses.

Carpenter (2009, private communication) has suggested based on \citet{hoggbias} that some of our UV excess sources, those detected at low signal to noise ratio, have overestimated UV fluxes. If this flux bias is responsible for our unexplained UV excess sources, then we expect low SNR GALEX sources to be more likely to show a UV excess. However, when we measure the fraction of UV excess sources as a function of the GALEX SNR, we find no significant dependence on the SNR. SNR-related biases do not explain why so many UV excess sources appear inactive in spectroscopic follow-up.

\citet{galexmatching} have suggested that spurious UV excess sources may arise from confused matches in which a single GALEX source has multiple 2MASS counterparts. 
Our matching radius of 2\arcsec\ does not let us find more than one of a GALEX source's 2MASS 
counterparts, so we cannot easily identify sources where this occurs.
As a test of whether overlooked counterparts can account for our excess sources, we match our GALEX sample to the All Sky Combined Catalog \citep{ascc}, finding four close doubles detected by GALEX: HD~21392, HD~26514, HD~141959, and HD~242903. 
Of these, HD~26514 was excluded from our 2MASS sample because it had only E quality photometry in J~band [profile fit failed or intermittent detection], HD~242903 because it had F quality in K~band [error could not be determined], and HD~141959 because it had a confused K detection. 
Only HD~21392 was included in our 2MASS sample, and it had an $NUV-J$ color consistent with a single photosphere ($0.3\sigma$ excess). This test suggests that our quality cuts reliably exclude marginally resolved doubles and our procedure is not susceptible to this source of error.

As a second test of the confusion hypothesis, we examine the fraction of UV excess sources without spectroscopic activity indicators as a function of the offset between GALEX and 2MASS positions. If either the GALEX or 2MASS source is a marginally resolved double, we expect its measured position to have a greater offset from its true position than if it were a single source, and therefore a greater offset from an independent measurement in another band. If this is the cause of our spectroscopically unconfirmed UV excess sources, we expect the apparently inactive sources to be biased towards larger GALEX-2MASS separations. In Figure~\ref{phantom_r}, we show a histogram of the Palomar targets as a function of the separation. 32 of the 44 sources (73\%) with separations less than 1\arcsec\ lack optical activity indicators, whereas 11 of the 19 sources (58\%) with separations between 1\arcsec\ and 2\arcsec\ do. Assuming, as a prior, that the parent fraction of sources with no activity indicators is uniformly and independently distributed over the interval $[0,1]$ in either group, we find only a 12\% probability that the high-separation targets were drawn from a population with a larger such fraction than the low-separation targets. 

Having considered several forms of experimental error, we now consider what contaminating sources could produce an ultraviolet excess but a normal optical spectrum. One possibility is a binary consisting of a G- or K-type primary and a flare star secondary. In appendix~\ref{mags_flare} we show that the flare star CR~Dra has $M_\textrm{NUV} = 14.09$ in quiescence and $M_\textrm{NUV} = 13.27$ if observed during a flare. A K0 star with a companion similar to CR~Dra would show an apparent UV excess of 0.15 magnitudes in quiescence and 0.29 magnitudes if we caught it during a flare. Only for a K4 or later primary would a companion flare star produce apparent excesses of the right order (1.7 magnitudes in quiescence, 2.4 in flare) to be detectable by our procedure. Allowing for a flare $\sim 2$ magnitudes brighter makes this a potential source of contamination for K2 and later stars. However, since most of our UV excess sources with normal spectra are around G or earlier stars (cf. Table~\ref{spectypes}), we cannot explain the NUV excesses as flux from a companion flare star. We do not have any data with which to estimate the effect of a flare star companion on a system's FUV magnitude.

A similar alternative is a binary consisting of a G- or K-type primary and a white dwarf secondary. A G or K star has an NUV flux comparable to that of a white dwarf, so such binaries are most prominent in the FUV. Taking values from the appendices, a K0 star with a 16000~K white dwarf companion would have an apparent NUV excess of only 1~magnitude, but a clear apparent FUV excess of 10~magnitudes. A G0 star and a 16000~K white dwarf would have a somewhat more marginal apparent FUV excess of 5~magnitudes. 
Since a $0.6~M_\sun$ white dwarf cools to 16000~K within 150~Myr, and even a $1.0~M_\sun$ dwarf cools within 450~Myr \citep{bergeron_tracks}, binaries with white dwarfs hotter than 16000~K should be too rare to contribute a significant population of NUV excess sources. 
G and K stars with white dwarfs might explain some of our FUV-only excess sources, but nothing with an NUV excess.
\label{wdgd}

% 7 and 3 if we don't drop the early-type stars
5 of our 20 $5\sigma$ spectroscopic targets in Taurus and 1 of our 7 $5\sigma$ targets in Upper Scorpius are late F- or G-type stars with an NUV excess and normal optical spectra. None of the explanations discussed here are plausible for these stars, and they will form the subject of a future paper once additional follow-up is obtained.

\section{Summary and Future Work}

By combining a photometric, wide-field survey of ultraviolet excess objects with an optical follow-up search for emission lines and lithium absorption, we have confirmed two new members of Taurus through lithium detections, and identified 3 additional probable members of Taurus and 5 of Upper Scorpius on the basis of emission lines. We expect to identify 8-20 more such sources in our Taurus survey fields and 3-6 in our Upper Scorpius fields once we complete our follow-up program. Because our survey probes areas away from the molecular clouds in Taurus and away from the association body in Upper Scorpius, our detections represent tentative evidence for a distributed population \citep[cf.][]{slesnick1}. However, without lithium detections we will need additional evidence to confirm our proposed members.

We have found, in addition to possible new Taurus and Upper Scorpius members, over 400 background UV-excess stars in our fields. We can detect such stars to distances of several hundred parsecs (see section~\ref{bgstar}) along sight lines away from the Taurus molecular clouds. 
While some of these stars show optical emission lines and are presumably older chromospherically active stars, two thirds show no evidence of stellar activity. We have not yet identified the source of the ultraviolet excess in the latter stars.

Most Galactic work using GALEX has been based on the rough classification of large GALEX-SDSS matched data sets \citep[e.g.][]{uvcolors, galexmatching, galexsdsscat} or on the detailed analysis of a small number of previously known sources \citep[e.g.][]{mflares, agbs}. Very few studies have used GALEX colors to identify specific populations of stars (notable exceptions are \citet{galexsurvey1} and \citet{galexsurvey2}) and as far as we know ours is the first to include fields near the Galactic plane. As such, our work explores a previously overlooked part of the ultraviolet sky in more detail than is typical for large-area GALEX surveys. A new domain means new problems: the broader than expected field locus in $UV-J$ vs. $J-K$ space, the high density of UV-excess sources even in fields distant from known star-forming regions, the small fraction of excess sources with emission lines. In the absence of constraints from population models or field observations, we do not yet have a robust way of dealing with any of these complications.

At present, the Galactic population of ultraviolet sources is constrained only in very rough terms. Main sequence stars are among the most common UV sources in the Galaxy, but the variation of UV properties among main sequence stars as a function of age or rotation speed, or the diversity of UV properties at a fixed mass, metallicity, age, or rotation speed, has been constrained only through indirect tracers. 
The contributions of white dwarfs and of white-dwarf-main-sequence binaries are well known from both models and observations \citep[and references therein]{galexmatching}, but discussions of ultraviolet populations make no mention of accreting systems, young active stars, RS~CVn binaries, flaring M~dwarfs, or other objects expected to be prominent ultraviolet sources. Understanding these details of the Galactic ultraviolet background would make UV-selected samples of rare objects much more complete and reliable.

\acknowledgments

We would like to thank John Carpenter for his discussions of the initial GALEX proposal and for his helpful comments on the manuscript. We also thank Karl Forster for his help planning the observations, Ted Wyder for his advice on GALEX data reduction, and Adam Kraus for his proper motion analysis.


%% in parentheses, after the keyword, but they will not be verified.

{\it Facilities:} \facility{GALEX}, \facility{Hale (DBSP)}

\appendix

\section{Main Sequence Absolute Magnitudes}\label{mags_ms}

In their table~5, \citet{mscolors} compiled empirical magnitudes in the 2MASS bands for main sequence stars covering spectral types B8-L0. We reproduce these magnitudes in Table~\ref{ourms}. To our knowledge, no one has made a similar compilation for the GALEX bands. Instead, we used $UV - J$ colors calculated from Kurucz models by Laget\footnote{\url{http://www.oamp.fr/people/laget/intg/intg0.html}}, linearly interpolated to the appropriate temperature and surface gravity for each spectral type as given by \citet{mscolors}. We list the resulting $FUV$ and $NUV$ magnitudes in Table~\ref{ourms}.

\section{White Dwarf Absolute Magnitudes}\label{mags_wd}

We found ultraviolet absolute magnitudes for white dwarfs by cross-matching a sample of white dwarfs with parallaxes \citep{wd_mag} to the GALEX GR5 public data. We found approximate temperatures by comparing the $FUV-NUV$ colors of these white dwarfs to \citet{wdwarf}. A selection of white dwarfs is given in Table~\ref{ourwds}; a more thorough study will be presented in a separate paper.

\section{Flare Star Absolute Magnitudes}\label{mags_flare}

Unfortunately, there is no systematic study of the absolute magnitudes of flare stars in the GALEX bands. For example, \citet{mflares} observed only one star, CR~Dra, that had a known distance (20.7~pc). 
Many other papers observed none at all.

From GALEX time-series data, CR~Dra had a quiescent NUV flux of 58 counts per second, corresponding to an absolute magnitude of 14.09. 800~s into the 1500~s exposure GALEX observed a flare; the NUV flux had not yet returned to the quiescent level by the end of the orbit. The SExtractor source catalog for the orbit gave an average magnitude $NUV = 14.85$, corresponding to an absolute magnitude of 13.27. Had CR~Dra been in our survey area, we would have used this catalog magnitude in our analysis.

\bibliographystyle{apj}
\bibliography{tausco}

\clearpage

\begin{figure*}
\includegraphics[angle=-90,width=\textwidth]{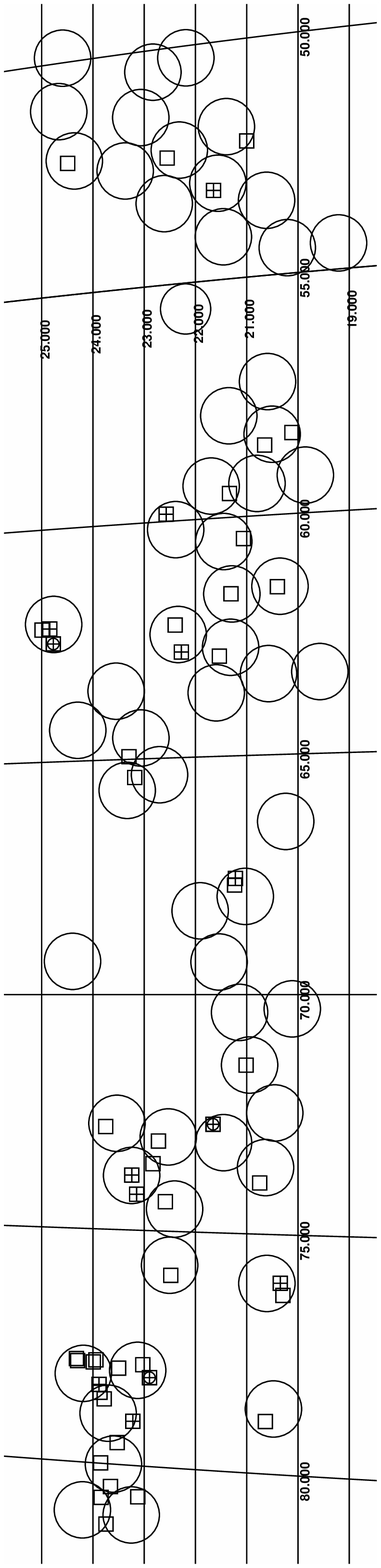}
\includegraphics[angle=-90,width=\textwidth]{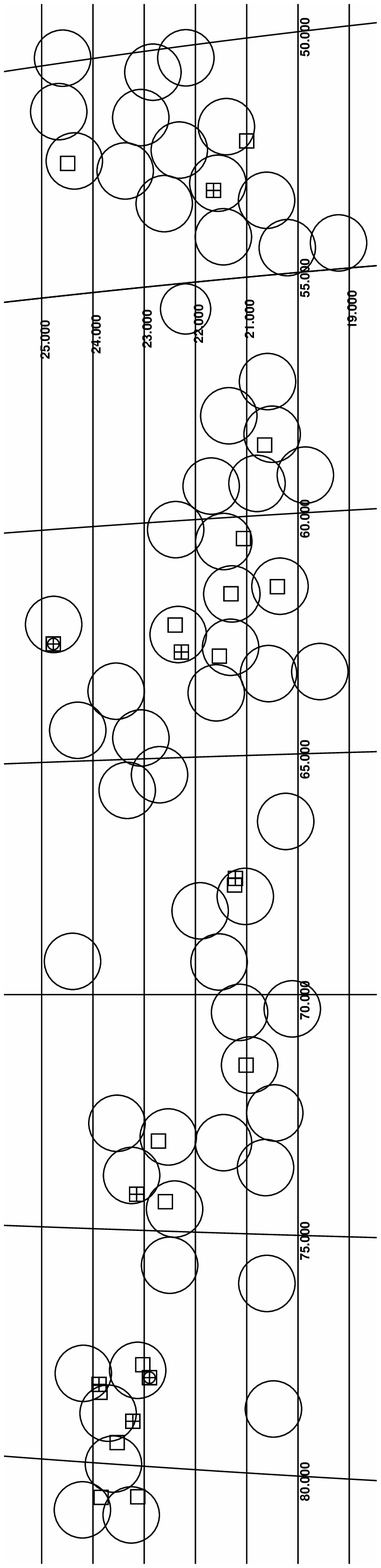}
\caption{Spatial distribution of the $3\sigma$ (top) and $5\sigma$ (bottom) UV excess sources of which we acquired spectra at Palomar. Squares mark Palomar targets. Crosses inscribed in the squares mark stars with emission lines. Circles inscribed in the squares mark stars with lithium absorption. There is no pattern apparent in the spatial distribution of emission-line stars.}\label{palmap}
\end{figure*}

\begin{figure*}
\includegraphics[width=0.48\textwidth]{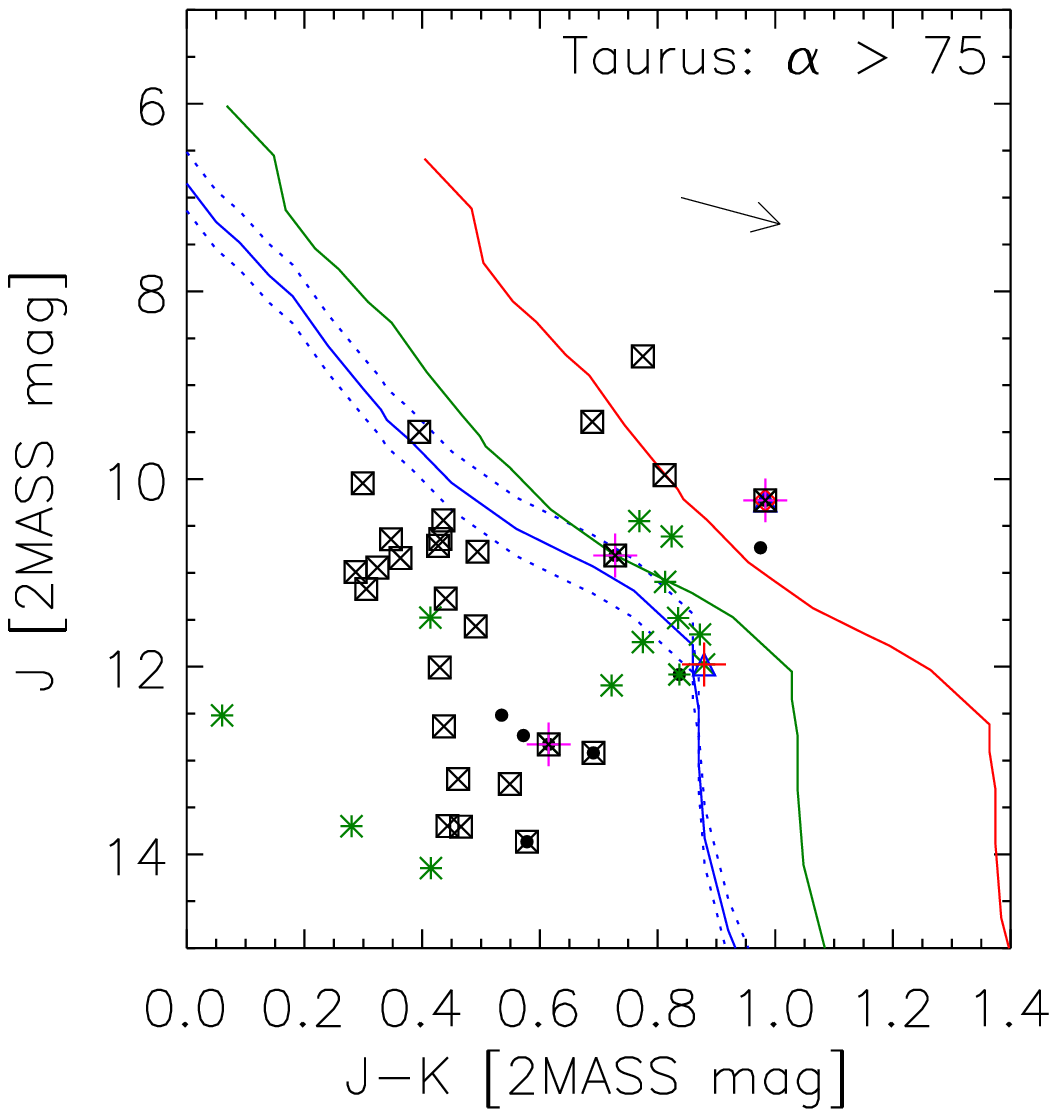}
\includegraphics[width=0.48\textwidth]{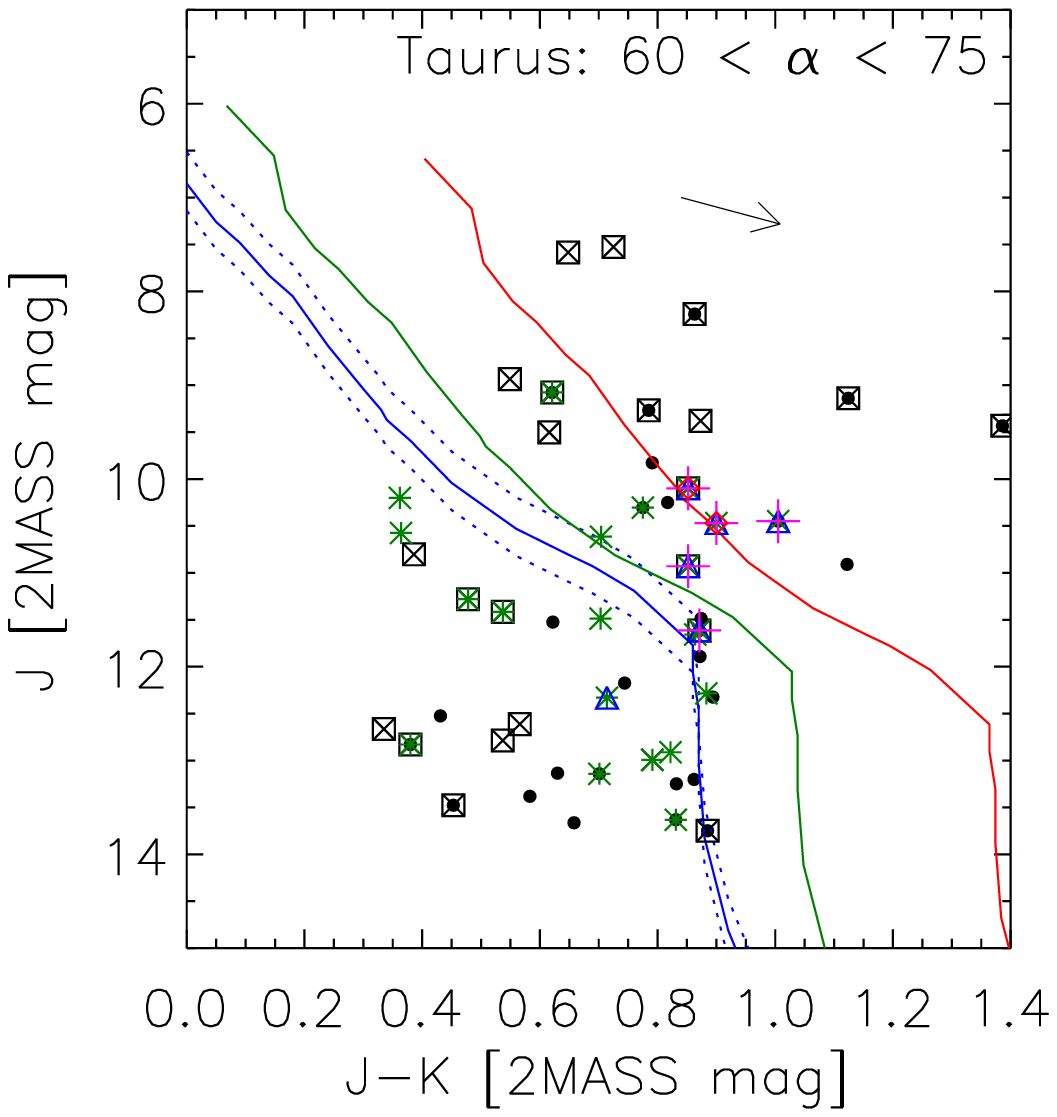}
\caption{Same as Figure~\ref{cmplots}, but for the easternmost (left) and central (right) fields of Taurus. Most of the UV excess sources in the eastern fields are background stars, while the central fields have a much larger population with color and magnitude consistent with Taurus membership.}\label{cmplots_tauparts}
\end{figure*}

\begin{figure*}
\includegraphics[width=0.48\textwidth]{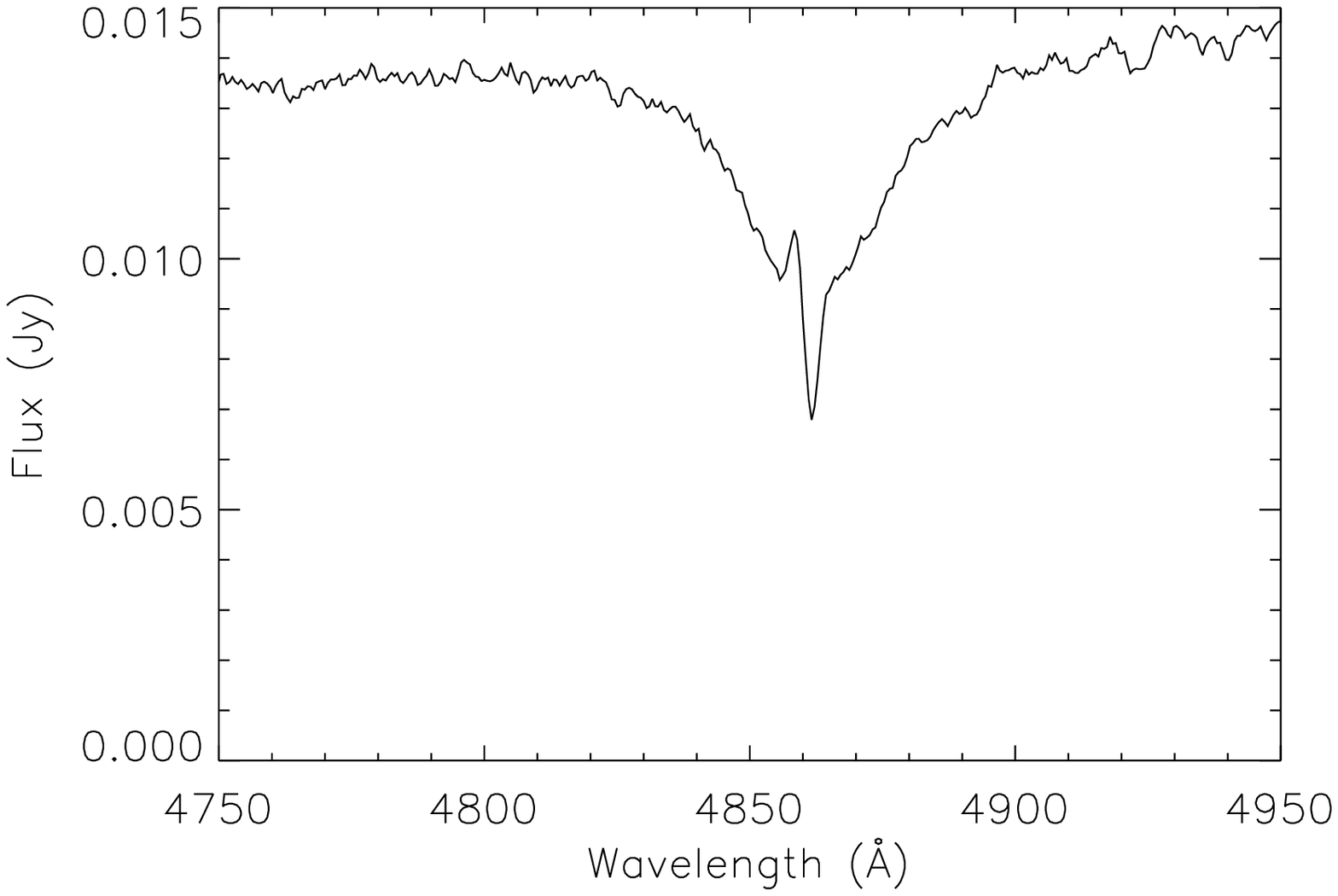}
\includegraphics[width=0.48\textwidth]{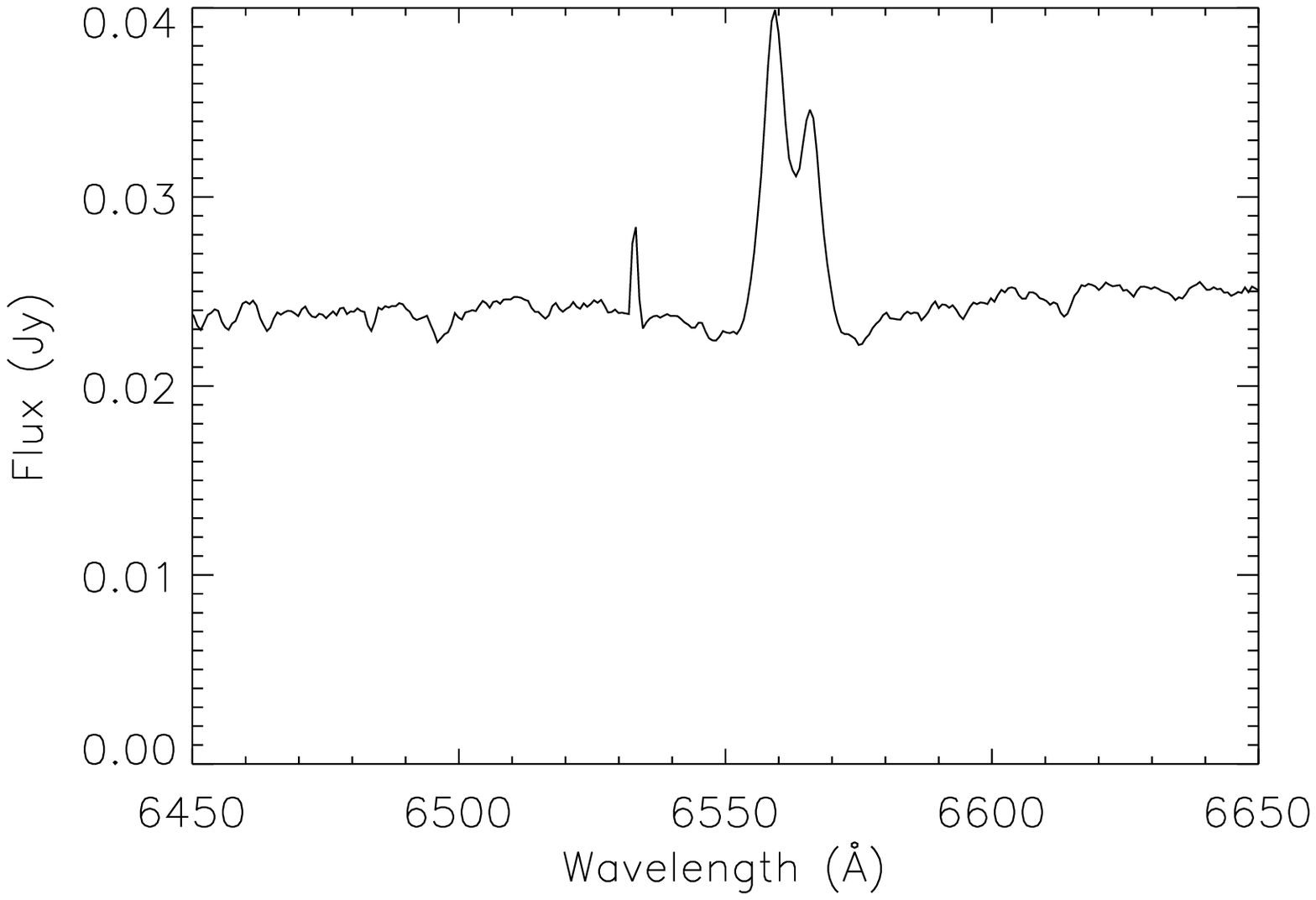}
\caption{Sections of our spectrum of 2MASS~J05161463+2313174 centered on the H$\beta$ (left) and H$\alpha$ (right) lines.}\label{pcyg}
\end{figure*}

%\begin{figure*}
%\includegraphics[width=0.48\textwidth]{f16a.eps}
%\includegraphics[width=0.48\textwidth]{f16b.eps}
%\includegraphics[width=0.48\textwidth]{f16c.eps}
%\caption{The fraction of sources that show a $3\sigma$ (solid line) or $5\sigma$ (dotted line) excess as a function of the GALEX source's signal to noise ratio. The distribution of UV excess sources does not appear to be biased towards low SNR sources. Because we selected sources that had a $3\sigma$ flux measurement in either the FUV or NUV, not neccessarily both, the Taurus plots include a $2\sigma$ bin that is too sparsely populated to provide any constraint.}\label{hoggbiastest}
%\end{figure*}

\begin{figure*}
\includegraphics[width=\textwidth]{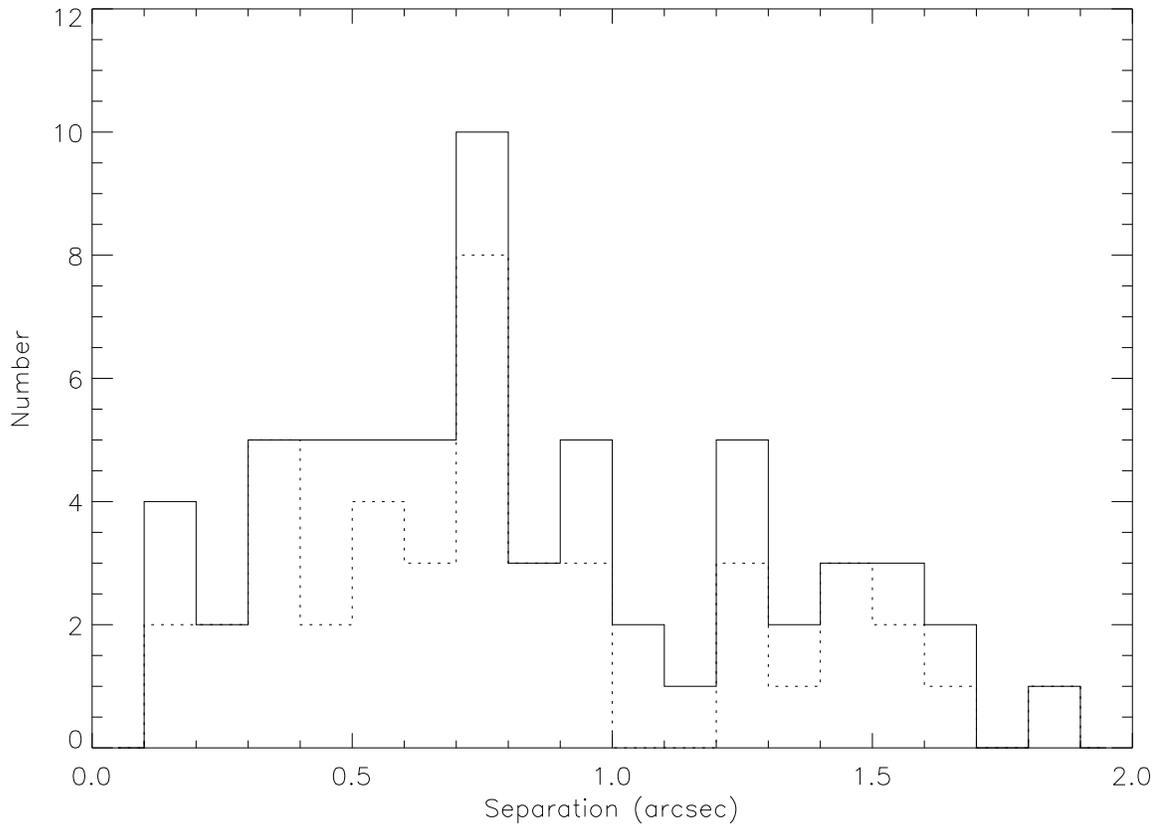}
\caption{Number of Palomar targets (solid line) and the number of targets with no youth indicators (dashed line) as a function of separation between the GALEX and 2MASS source positions. Were some of the UV excesses from marginally resolved binaries or chance matches, one would expect a larger fraction of the high-separation sources to lack youth indicators.}\label{phantom_r}
\end{figure*}

%% Tables should be submitted one per page, so put a \clearpage before

%% Tables may also be prepared as separate files. See the accompanying
%% sample file table.tex for an example of an external table file.
%% To include an external file in your main document, use the \input
%% command. Uncomment the line below to include table.tex in this
%% sample file. (Note that you will need to comment out the \documentclass,
%% \begin{document}, and \end{document} commands from table.tex if you want
%% to include it in this document.)

\clearpage

\begin{deluxetable}{lcccc}
\tablewidth{0pt}
\tablecaption{$3\sigma$ sensitivity limits for GALEX observations\label{gxdetect}}
\tablehead{\colhead{} & \colhead{100~s} & \colhead{300~s} & \colhead{1000~s} & \colhead{1500~s}\\
\colhead{} & \colhead{(mag)} & \colhead{(mag)} & \colhead{(mag)} & \colhead{(mag)}}
\startdata
FUV 	& 21.1  & 22.1	& 22.9   & 23.2		\\
NUV 	& 22.0  & 22.8	& 23.5   & 23.8
\enddata
\tablecomments{\ All magnitudes AB.}
\end{deluxetable}
\begin{deluxetable}{lcc}
\tablewidth{0pt}
\tablecaption{99\% completeness limits for 2MASS sources\label{2mdetect}}
\tablehead{\colhead{}	& \colhead{A quality} & \colhead{All sources}\\
\colhead{}	& \colhead{(mag)} & \colhead{(mag)}}
\startdata
J 	& 15.7	& 16.1	\\
H 	& 14.7	& 15.5	\\
K 	& 14.4	& 15.1
\enddata
\tablecomments{\ Limits are for the high photometric quality (``A'') 2MASS sources and for the entire point source catalog. Since we required C quality photometry, our 2MASS completeness limits should lie between these two values.}
\end{deluxetable}

\begin{deluxetable}{lcccc}
\tablewidth{0pt}
\tablecolumns{5}
\tablehead{\colhead{} & \multicolumn{2}{c}{Photosphere} & \multicolumn{2}{c}{Minimum Excess}	\\
\colhead{} & \colhead{FUV} & \colhead{NUV} & \colhead{FUV} & \colhead{NUV}\\
\colhead{Type} & \colhead{(mag)} & \colhead{(mag)} & \colhead{(mag)} & \colhead{(mag)}}
\tablecaption{Maximum extinction ($A_V$) at which GALEX can detect a main sequence star at 140~pc\label{maxav}}
\startdata
F0   & 2.5 & 4.2	& 4.4 & 5.2	\\
F2   & 2.0 & 4.0	& 4.0 & 5.0	\\
F5   & 1.1 & 3.6	& 3.1 & 4.5	\\
F8   & 0.4 & 3.2	& 2.4 & 4.2	\\
\hline
G0   &\nodata& 3.0	& 2.0 & 4.0	\\
G2   &\nodata& 2.8	& 1.6 & 3.8	\\
G5   &\nodata& 2.7	& 1.3 & 3.7	\\
G8   &\nodata& 2.3	& 0.8 & 3.3	\\
\hline
K0   &\nodata& 1.9	& 0.2 & 2.9	\\
K2   &\nodata& 1.3	&\nodata& 2.3	\\
K4   &\nodata& 0.6	&\nodata& 1.6	\\
K5   &\nodata& 0.1	&\nodata& 1.1	\\
K7   &\nodata& \nodata	&\nodata& 0.7	\\
\hline
M0   &\nodata& \nodata	&\nodata& 0.4	\\
M1   &\nodata& \nodata	&\nodata& 0.1	\\
M2   &\nodata& \nodata	&\nodata& \nodata	\\
\enddata
\tablecomments{The first two columns assume that a star produces only photospheric UV flux, the second two that a star has the smallest UV excess that can be reliably identified (see section~\ref{minex}). Values are for a 300~s GALEX exposure; the maximum $A_V$ for 1500~s exposures is about 0.5~mag higher in both bands. An entry marked by \nodata\ denotes a star that cannot be detected even if $A_V = 0$.}
\end{deluxetable}

\begin{deluxetable}{lcc}
\tablewidth{0pt}
\tablecaption{Basic properties of our GALEX observations\label{galexobs}}
\tablehead{\colhead{} & \colhead{Taurus} & \colhead{Scorpius}}
\startdata
%Number of fields 	& 61	& 13	\\
Number of FUV fields 	& 61	& 0	\\
Number of NUV fields 	& 61	& 13	\\
Number of 300s fields 	& 54	& 12	\\
Number of 1500s fields 	& 7	& 1	\\
Area (deg$^2$)		& 56.1	& 12.2	\\
\hline
GALEX Sources		& 25389	& 5863	\\
2MASS Sources with $J\le 14$ & $\sim 45000$	& $\sim 11400$	\\
Matched Sources 	& 14130	& 2828	\\
\enddata
\end{deluxetable}

\begin{deluxetable}{lrccc}
\tablewidth{0pt}
\tablecaption{UV excess sample properties as a function of the UV excess cutoff\label{exstats}}
\tablehead{\colhead{Region} & \colhead{Cutoff} 	& \colhead{Total Sources} & \colhead{Field Sources} & \colhead{Reliability}}
\startdata
Taurus & $3\sigma$	& 471	& 306	& 0.35	\\
Taurus & $4\sigma$	& 191	& 60	& 0.69	\\
Taurus & $5\sigma$	& 89	& 13	& 0.85	\\
Taurus & $6\sigma$	& 58	& 4	& 0.93	\\
Taurus & $7\sigma$	& 36	& 1	& 0.97	\\
\hline
Upper Sco & $3\sigma$	& 96	& 55	& 0.43	\\
Upper Sco & $4\sigma$	& 32	& 11	& 0.66	\\
Upper Sco & $5\sigma$	& 12	& 1	& 0.92	\\
Upper Sco & $6\sigma$	& 6	& 0	& \nodata	\\
Upper Sco & $7\sigma$	& 5	& 0	& \nodata	\\
\hline
Control & $3\sigma$	& 50	& 24	& 0.52	\\
Control & $4\sigma$	& 21	& 0	& \nodata	\\
Control & $5\sigma$	& 9	& 0	& \nodata	\\
Control & $6\sigma$	& 4	& 0	& \nodata	\\
Control & $7\sigma$	& 4	& 0	& \nodata	\\
\enddata
\tablecomments{The number of field sources is estimated from the number of apparent UV-deficit sources; see section~\ref{cutofftests}. The last column in the table gives the expected fraction of UV excess sources that are \emph{not} contaminants from the field.}
\end{deluxetable}

\begin{deluxetable}{l cc c rlrl rlrlrl c rlcrlc lc}
\rotate
\setlength{\tabcolsep}{0.015in}
\tabletypesize{\scriptsize}
\tablecaption{Sources with a $3\sigma$ or greater UV excess in at least one band\label{catalogtable}}
\tablehead{\colhead{} & \colhead{$\alpha$} & \colhead{$\delta$} & 
	\colhead{} & 
	\colhead{$FUV$} & \colhead{$\sigma_\textrm{FUV}$} & \colhead{$NUV$} & \colhead{$\sigma_\textrm{NUV}$} & 
	%\colhead{$J$} & \colhead{$\sigma_{J}$} & \colhead{$H$} & \colhead{$\sigma_{H}$} & \colhead{$K$} & \colhead{$\sigma_{K}$} & 
	%\colhead{} & 
	\colhead{$\Delta_\textrm{FUV}$} & \colhead{$\sigma_{\Delta_\textrm{FUV}}$} & \colhead{} & 
	\colhead{$\Delta_\textrm{NUV}$} & \colhead{$\sigma_{\Delta_\textrm{NUV}}$} & \colhead{} %& 
%	\colhead{} & \colhead{}
	\\
	\colhead{2MASS} & \colhead{(deg)} & \colhead{(deg)} & 
	\colhead{GALEX Tile} &
	\colhead{(mag)} & \colhead{(mag)} & \colhead{(mag)} & \colhead{(mag)} & 
	%\colhead{(mag)} & \colhead{(mag)} & \colhead{(mag)} & \colhead{(mag)} & \colhead{(mag)} & \colhead{(mag)} & 
	%\colhead{JHK Quality} & 
	\colhead{(mag)} & \colhead{(mag)} & \colhead{$\Delta_\textrm{FUV}/\sigma$} & 
	\colhead{(mag)} & \colhead{(mag)} & \colhead{$\Delta_\textrm{NUV}/\sigma$} %& 
	%\colhead{Other Identifiers} & \colhead{Previously Known Member}
	}
\startdata
03171986+2447143&49.332784&24.787331&TauAur\_MOS44\_v2&\nodata&\nodata&20.187&0.066
%&13.940&0.025&13.940&0.033&13.463&0.036&AAA
&\nodata&\nodata&\nodata&1.5&0.5&3.1%&\nodata&\nodata
\\
03174690+2429233&49.445453&24.489820&TauAur\_MOS44\_v2&\nodata&\nodata&20.273&0.068
%&13.854&0.026&13.854&0.027&13.297&0.028&AAA
&\nodata&\nodata&\nodata&2.1&0.4&5.0%&\nodata&\nodata
\\
03180156+2411297&49.506523&24.191601&TauAur\_MOS44\_v2&\nodata&\nodata&20.085&0.061
%&13.962&0.024&13.962&0.030&13.469&0.031&AAA
&\nodata&\nodata&\nodata&1.7&0.4&4.1%&\nodata&\nodata
\\
03181321+2447174&49.555079&24.788185&TauAur\_MOS44\_v2&\nodata&\nodata&18.324&0.021
%&12.186&0.021&12.186&0.020&11.731&0.020&AAA
&\nodata&\nodata&\nodata&1.3&0.3&4.2%&\nodata&\nodata
\\
03181487+2429114&49.561995&24.486513&TauAur\_MOS44\_v2&\nodata&\nodata&19.905&0.052
%&13.751&0.024&13.751&0.027&13.245&0.029&AAA
&\nodata&\nodata&\nodata&1.8&0.4&4.5%&\nodata&\nodata
\\
\enddata
\tablecomments{Table \ref{catalogtable} is published in its entirety in the electronic edition of \textit{The Astronomical Journal}. A portion is shown here for guidance regarding its form and content. The electronic table contains more rows and columns.
Positions are in J2000.0 and taken from the 2MASS Point Source Catalog. FUV and NUV photometry in AB magnitudes is taken from the GALEX SExtractor catalogs. J, H, and K photometry in Vega-based magnitudes is taken from 2MASS. The notation $\Delta_\textrm{FUV}$ denotes the FUV excess. If we know of an association between the GALEX source and a known object, and if the literature confirms or rules out membership in Taurus or Upper Scorpius, we list the information in the last two columns. Our list of associations is not complete.}
\end{deluxetable}

\begin{deluxetable}{lllcc}
\tablewidth{0pt}
\tablecaption{Previously known Taurus members in our GALEX fields\label{knownmembers}}
\tablehead{\colhead{} & \colhead{} & \colhead{Spectral} & \colhead{FUV Excess} & \colhead{NUV Excess}\\
\colhead{ID} & \colhead{2MASS} & \colhead{Type} & \colhead{(mag)} & \colhead{(mag)}}
\startdata
TAP~4		& 03293837+2430379	& K1	& Not Detected & 1.2 (3.3$\sigma$)	\\
AA~Tau		& 04345542+2428531	& K7	& 22.0 (24$\sigma$)	& 8.0 (21$\sigma$)	\\
DN~Tau		& 04352737+2414589	& M0	& 15.7 (21$\sigma$)	& 5.4 (17$\sigma$)	\\
Co~Ku~Tau~3	& 04354093+2411087	& M1	& Not Detected	& Not Detected	\\
2MASS~J04361030+2159364&04361030+2159364& M8.5	& Not Detected	& Not Detected	\\
\enddata
\tablecomments{Members are taken from \citet{memberlist} and \citet{TAPcheck}. Spectral type for TAP~4 from Simbad. Types for other sources taken from \citet{memberlist}.}
\end{deluxetable}

\begin{deluxetable}{l l cc cc rr c ccccc}
\tablecolumns{14}
\rotate
\setlength{\tabcolsep}{0.015in}
\tabletypesize{\scriptsize}
\tablecaption{UV excess stars observed at Palomar\label{specresults}}
\tablehead{\colhead{} & \colhead{} & \colhead{FUV} & \colhead{NUV} & \colhead{J} & \colhead{K} & 
	\colhead{FUV Excess} & \colhead{NUV Excess} & 
	\colhead{} & \colhead{$W_\textrm{Ca II k}$} & \colhead{$W_\textrm{Ca II h}$} & \colhead{$W_\textrm{H$\beta$}$} & \colhead{$W_\textrm{H$\alpha$}$} & \colhead{$W_\textrm{Li}$}\\
\colhead{2MASS} & \colhead{Previous Literature} & \colhead{(mag)} & \colhead{(mag)} & \colhead{(mag)} & \colhead{(mag)} & 
	\colhead{(mag)} & \colhead{(mag)} & 
	\colhead{Spectral Type} & \colhead{(\AA)} & \colhead{(\AA)} & \colhead{(\AA)} & \colhead{(\AA)} & \colhead{(\AA)}}
\startdata
\cutinhead{Taurus Sources}
03283651+2100015&&17.94&15.29&10.24&9.84&3.6 (6.2$\sigma$)&1.9 (6.4$\sigma$)&A1-3 V&&&&&$<0.04$ \\
03284114+2429333&&&21.70&13.74&12.80&&4.5 (8.4$\sigma$)&M2 IV-V&&&&&$<0.29$ \\
03291688+2233071&&&20.91&11.02&10.21&&1.2 (3.3$\sigma$)&K1 IV?&&&&&$<0.06$ \\
03322691+2138499&&&19.46&11.56&10.83&&2.5 (7.9$\sigma$)&G5e III?&\tablenotemark{a}&\tablenotemark{a}&&-0.6454&$<0.07$ \\
03531583+2007078&&&20.82&12.08&11.38&&1.3 (3.8$\sigma$)&G5 V&&&&&$<0.11$ \\
03540872+2039003&&16.17&16.56&13.91&13.55&8.1 (8.6$\sigma$)&3.9 (8.0$\sigma$)&F1-2 V&&&&&$<0.13$ \\
03575997+2120125&&&17.78&11.87&11.41&&1.6 (4.9$\sigma$)&F2-3 V&&&&&$<0.07$ \\
03592289+2234169&Mel 22 DH 875, Non-Member&&21.99&11.49&10.56&&1.9 (3.8$\sigma$)&K5e V&-8.364&-8.992&-0.5841&-1.087&$<0.08$ \\
04015065+2103495&&&20.30&13.63&12.80&&4.7 (11.3$\sigma$)&WDMD&&&&&$<0.33$ \\
04060135+2024074&&20.62&&12.99&12.20&11.6 (14.5$\sigma$)&&G0 IV-V&&&&&$<0.11$ \\
04062461+2118284&&&20.61&13.14&12.44&&2.6 (6.5$\sigma$)&G2 IV&&&&&$<0.13$ \\
04083270+2450332&&&22.21&12.91&12.09&&2.0 (4.9$\sigma$)&F6 V&&\tablenotemark{a}&&&$<0.19$ \\
04083606+2459336&&&22.11&11.66&10.79&&1.3 (3.3$\sigma$)&K0-1? III?&&&&&$<0.11$ \\
04084754+2223470&&16.02&15.98&12.83&12.45&7.6 (11.0$\sigma$)&3.5 (10.1$\sigma$)&F1-2 IV&&&&&$<0.07$ \\
04095113+2446211&1RXS~J040951.0+244639, Member&22.15&20.27&10.10&9.25&8.4 (12.1$\sigma$)&1.4 (4.7$\sigma$)&M1e IV-V&-16.97&-13.59&-1.986&-2.505&0.2583 \\
04110570+2216313&&21.65&20.92&10.93&10.08&9.8 (12.7$\sigma$)&1.6 (4.1$\sigma$)&M4e V&-20.97&-17.62&-6.211&-5.276&$<0.18$ \\
04113544+2132023&&21.25&19.95&11.42&10.88&4.2 (5.8$\sigma$)&-0.2 (-0.6$\sigma$)&G5-8 V&&&&&$<0.06$ \\
04213355+2311060&&21.79&15.72&10.20&9.84&-1.2 (-1.7$\sigma$)&1.0 (3.3$\sigma$)&F2 V&&&&&$<0.05$ \\
04301583+2113173&&21.39&20.98&11.61&10.74&11.1 (15.0$\sigma$)&2.4 (6.7$\sigma$)&M2e-3e III-V&-10.99&-9.063&-3.031&-2.839&$<0.12$ \\
04455392+2100427&&&18.95&10.31&9.53&&2.1 (6.9$\sigma$)&G3-5 II-III&&&&&$<0.08$ \\
04505356+2139233&1RXS~J045053.5+213927&&20.88&10.47&9.57&&1.7 (4.4$\sigma$)&M2e V&-14.94&-16.9&-3.224&-3.791&0.1542 \\
04511488+2345034&&&20.14&11.49&10.79&&1.4 (4.4$\sigma$)&F7 IV-V&&&&&$<0.09$ \\
04522346+2243195&&20.75&17.75&11.28&10.80&3.3 (5.3$\sigma$)&1.2 (4.1$\sigma$)&A3 V&&&&&$<0.05$ \\
04541932+2249418&&&19.47&10.62&9.91&&1.2 (4.0$\sigma$)&G0 IV&&&&&$<0.09$ \\
04551985+2314331&&&21.00&12.33&11.62&&1.5 (4.2$\sigma$)&G5 III&\tablenotemark{a}&&&&$<0.09$ \\
04554339+2044503&&20.61&16.04&10.58&10.21&0.4 (0.7$\sigma$)&1.1 (3.7$\sigma$)&F2 IV-V&&&&&$<0.06$ \\
04565683+2308514&&&21.24&10.45&9.44&&2.4 (6.1$\sigma$)&K2e-4e IV&\tablenotemark{a}&\tablenotemark{a}&-0.4059&-3.193&$<0.07$ \\
04573053+2235135&HD~284988&20.25&16.12&9.08&8.46&4.5 (5.6$\sigma$)&2.2 (5.3$\sigma$)&F2&&&&&$<0.07$ \\
05034320+2228535&&&19.92&10.45&9.68&&1.3 (3.9$\sigma$)&G2 III-IV&&&&&$<0.07$ \\
05040207+2020419&&&21.85&11.98&11.10&&2.0 (4.6$\sigma$)&G8e-K0e III&\tablenotemark{a}&\tablenotemark{a}&&-1.215&$<0.14$ \\
05050209+2017375&&&21.99&11.66&10.79&&1.5 (3.8$\sigma$)&G8-K1 V&&&&&$<0.12$ \\
05111074+2419209&&14.67&14.61&12.52&12.46&2.0 (2.9$\sigma$)&1.3 (3.7$\sigma$)&B1 V&&&&&$<0.05$ \\
05111165+2356327&&17.23&17.26&13.70&13.43&5.1 (4.9$\sigma$)&2.1 (3.9$\sigma$)&B3 V&&&&&$<0.12$ \\
05112064+2417542&&&21.04&11.48&10.65&&1.9 (3.8$\sigma$)&K0 V&&&&&$<0.06$ \\
05112299+2359353&&&21.19&11.74&10.96&&1.3 (3.2$\sigma$)&G8-K0? III-IV?&&&&&$<0.10$ \\
05114780+2329554&&&20.57&10.61&9.79&&1.3 (3.9$\sigma$)&G3-5 IV-V&&&&&$<0.09$ \\
05122759+2253492&&20.81&19.41&10.23&9.25&12.6 (16.5$\sigma$)&3.8 (11.3$\sigma$)&M2e IV-V&-19.36&-25.57&-7.008&\tablenotemark{b}&0.5303 \\
05131650+2352552&&20.80&19.49&12.83&12.21&7.6 (11.3$\sigma$)&2.5 (7.6$\sigma$)&A2e V&&&\tablenotemark{b}&\tablenotemark{a}&$<0.10$ \\
05135696+2351438&&&21.05&12.08&11.25&&2.5 (5.1$\sigma$)&F8 V&&&&&$<0.14$ \\
05153663+2038127&&&21.02&11.10&10.28&&1.3 (3.5$\sigma$)&F7 V&&&&&$<0.08$ \\
05161463+2313174&&17.80&17.36&10.82&10.09&10.9 (16.5$\sigma$)&3.8 (11.6$\sigma$)&A0e V&&&\tablenotemark{c}&\tablenotemark{ab}&$<0.06$ \\
05215550+2339297&&&21.13&12.20&11.48&&1.3 (3.3$\sigma$)&F0 IV-V&&&&&$<0.10$ \\
05250924+2344444&&22.14&21.83&14.15&13.73&3.5 (3.3$\sigma$)&-0.6 (-1.1$\sigma$)&A0-2&&&&&$<0.19$ \\
\cutinhead{Upper Scorpius Sources}
15522878-1551090&&&19.68&11.63&10.92&&1.8 (4.0$\sigma$)&K0-1 V&\tablenotemark{a}&\tablenotemark{a}&&&$<0.07$ \\
15523589-1555386&&&21.35&12.32&11.51&&1.7 (3.7$\sigma$)&K0 IV-V&\tablenotemark{a}&\tablenotemark{a}&&&$<0.09$ \\
15525340-1504400&&&17.26&10.53&9.97&&1.5 (4.3$\sigma$)&G1-3 V&&&&&$<0.05$ \\
15530107-1458325&&&19.74&13.06&12.48&&1.8 (4.0$\sigma$)&G8p&&&&&$<0.15$ \\
15534689-1457364&&&20.27&12.99&12.12&&4.1 (10.0$\sigma$)&K7 V?&\tablenotemark{a}&\tablenotemark{a}&&&$<0.14$ \\
15534807-1740333&&&17.80&11.12&10.57&&1.6 (4.1$\sigma$)&F7-G0 IV&&&&&$<0.06$ \\
15535071-1436057&&&20.59&13.94&13.31&&2.3 (3.5$\sigma$)&F2-5&&&&&$<0.30$ \\
16014778-1328293&&&19.52&12.31&11.68&&1.8 (5.0$\sigma$)&F8-G2 III&&&&&$<0.08$ \\
16060455-1456342&&&18.84&13.06&12.51&&2.4 (6.3$\sigma$)&F0-5&&&&&$<0.10$ \\
16155520-1634512&&&16.36&12.89&12.64&&1.6 (3.6$\sigma$)&A0 IV?&&&&&$<0.07$ \\
16163981-1527048&&&19.47&12.95&12.39&&1.7 (4.5$\sigma$)&F3-5 V?&&&&&$<0.14$ \\
16181899-1526538&&&18.67&12.14&11.57&&1.9 (4.8$\sigma$)&F3-5 IV&&&&&$<0.08$ \\
16183457-1654379&&&21.12&10.66&9.73&&1.6 (3.6$\sigma$)&K2-3 III&\tablenotemark{a}&\tablenotemark{a}&&&$<0.08$ \\
16184513-1419042&&&20.85&13.76&12.90&&4.2 (8.1$\sigma$)&M1e-2e II-III?&&&&&$<0.33$ \\
16184596-1659455&&&19.73&12.46&11.61&&3.9 (9.1$\sigma$)&K4e V&\tablenotemark{a}&\tablenotemark{a}&&-0.4271&$<0.09$ \\
16185702-1632303&&&20.80&13.94&13.30&&2.2 (3.7$\sigma$)&G1-3 V&&&&&$<0.13$ \\
16195506-1538250&Sco~X-1, Non-member&&13.96&11.91&11.15&&8.2 (21$\sigma$)&LMXB&&&-0.92&-5.14& \\
16212481-1523160&&&18.73&10.20&9.43&&1.9 (5.4$\sigma$)&K2 III&\tablenotemark{a}&\tablenotemark{a}&&&$<0.06$ \\
16220590-1505118&&&18.93&10.47&9.72&&1.7 (4.8$\sigma$)&G3 V&\tablenotemark{a}&&&&$<0.06$ \\
16232902-1601156&&&18.73&10.68&9.96&&1.8 (6.0$\sigma$)&F2-3 IV&&&&&$<0.05$ \\
16333769-1711347&&&20.00&13.88&13.05&&4.9 (4.9$\sigma$)&F5-7&&&&&$<0.13$ \\
\enddata
\tablecomments{GALEX magnitudes are in AB, infrared magnitudes are in the 2MASS system. If we know of an association between the GALEX source and a known object, and if the literature confirms or rules out membership in Taurus or Upper Scorpius, we list the information in the ``Previous Literature'' column.}
\tablenotetext{a}{Core emission line in an absorption line. Could not measure emission equivalent width.}
\tablenotetext{b}{Self-absorbed profile. Could not measure emission equivalent width.}
\tablenotetext{c}{Inverse P Cygni profile. Could not measure emission equivalent width.}
\end{deluxetable}

\begin{deluxetable}{l ccccc ccccc}
\tablewidth{0pt}
\tablecaption{Spectral types and youth indicators among our Palomar targets\label{spectypes}}
\tablecolumns{11}
\tablehead{	& \multicolumn{5}{c}{Taurus} & \multicolumn{5}{c}{Upper Scorpius} \\
\colhead{Type} 	& \colhead{Sources} & \colhead{Ca II} & \colhead{H I} & \colhead{Li I} & \colhead{Young} 
		& \colhead{Sources} & \colhead{Ca II} & \colhead{H I} & \colhead{Li I} & \colhead{Young}}
\startdata
\cutinhead{$3\sigma$ Excess Sources}
B	& 2 & 0 & 0 & 0 & 0	& 0 & 0 & 0 & 0 & 0 	\\
A	& 5 & 0 & 2 & 0 & 2	& 1 & 0 & 0 & 0 & 0 	\\
F	& 11& 0 & 0 & 0 & 0	& 7 & 0 & 0 & 0 & 0 	\\
G	& 13& 3 & 2 & 0 & 3	& 5 & 1 & 0 & 0 & 1 	\\
K	& 5 & 2 & 2 & 0 & 2	& 6 & 6 & 1 & 0 & 6 	\\
M	& 7 & 5 & 5 & 3 & 5	& 1 & 0 & 0 & 0 & 0 	\\
\hline
Total	& 43& 11& 11& 3 & 13	& 20 & 7 & 1 & 0 & 7 	\\
\cutinhead{$5\sigma$ Excess Sources}
A 	& 4 & 0 & 2 & 0 & 2 	& 0 & 0 & 0 & 0 & 0 	\\
F 	& 4 & 0 & 0 & 0 & 0 	& 2 & 0 & 0 & 0 & 0 	\\
G 	& 5 & 1 & 1 & 0 & 1 	& 1 & 0 & 0 & 0 & 0 	\\
K 	& 1 & 1 & 1 & 0 & 1 	& 3 & 2 & 1 & 0 & 2 	\\
M 	& 6 & 4 & 4 & 2 & 4 	& 1 & 1 & 0 & 0 & 1 	\\
\hline
Total	& 20& 6 & 8 & 2 & 8 	& 7 & 3 & 1 & 0 & 3 	\\
\enddata
\tablecomments{Number of spectroscopically observed stars by spectral type, followed by total number of Ca-emission, Balmer-emission, and Li-absorption stars. The last column in each section lists the number of stars that had at least one of these three indicators.}
\end{deluxetable}

\begin{deluxetable}{lcccc}
\tablewidth{0pt}
\tabletypesize{\scriptsize}
\tablecaption{Classification of photometric candidate members of Taurus and Upper Scorpius\label{extrapol}}
\tablehead{\colhead{} 	& \colhead{East Taurus} & \colhead{Central Taurus} & \colhead{West Taurus} & \colhead{Upper Scorpius}}
\startdata
\cutinhead{$3\sigma$ Sources}
total sources				& 20	& 54	& 27	& 19	\\
\hline
literature members			& 0	& 4	& 1	& 0	\\
literature non-members			& 0	& 5	& 4	& 0	\\
\hline
Palomar target with no emission lines	& 6	& 5	& 3	& 4	\\
Palomar background target with emission	& 2	& 0	& 0	& 0	\\
Palomar non-background target with emission, no lithium	& 0	& 3	& 0	& 5	\\
Palomar non-background target with emission, lithium	& 1	& 1	& 0	& 0	\\
\hline
unidentified stars			& 11	& 36	& 19	& 10	\\
expected emission-line stars in unidentified group& 2-3	& 6-13	& 0-4	& 3-6	\\
\cutinhead{$5\sigma$ Sources}
total sources				& 8	& 25	& 3	& 6	\\
\hline
literature members			& 0	& 3	& 0	& 0	\\
literature non-members			& 0	& 4	& 0	& 0	\\
\hline
Palomar target with no emission lines	& 1	& 3	& 2	& 2	\\
Palomar background target with emission	& 1	& 0	& 0	& 0	\\
Palomar non-background target with emission, no lithium	& 0	& 3	& 0	& 3	\\
Palomar non-background target with emission, lithium	& 1	& 0	& 0	& 0	\\
\hline
unidentified stars			& 5	& 12	& 1	& 1	\\
expected emission-line stars in unidentified group& 3	& 5-6	& 0	& 0	\\
\enddata
\tablecomments{Total number of UV excess sources above the 140~pc main sequence, broken down by whether we determined membership from previous literature or from our Palomar observations. We do not count Palomar targets that are also identified in the literature. Because our Taurus fields cover a large region we divide them into western ($\alpha \leq 60\degr$), central ($60\degr \leq \alpha \leq 75\degr$), and eastern ($75\degr \leq \alpha$) groups, following \citet{slesnick1}.}
\end{deluxetable}

\begin{deluxetable}{l cl rrr c}
\tablewidth{0pt}
\tablecolumns{7}
\setlength{\tabcolsep}{0.015in}
\tabletypesize{\scriptsize}
\tablecaption{Spectroscopic candidate members of Taurus and Upper Scorpius\label{pmresults}}
\tablehead{\colhead{} & \colhead{} & \colhead{Youth} & \colhead{PM RA} & \colhead{PM Dec.} & \colhead{PM Error} & \colhead{}\\
\colhead{2MASS} & \colhead{Spectral Type} & \colhead{Indicators} & \colhead{(mas/yr)} & \colhead{(mas/yr)} & \colhead{(mas/yr)} & \colhead{PM Member?}}
\startdata
\cutinhead{Taurus Sources}
04110570+2216313 & M4e V   & Ca, H	& +2.8	& -19.3	& 3.4 & Yes	\\
04301583+2113173 & M2e-3e III-V& Ca, H	& -33.6	& +9.2	& 3.3 & No	\\
04505356+2139233 & M2e V   & Ca, H, Li	& +0.7	& -20.3	& 3.6 & Yes	\\
04565683+2308514 & K2e-4e IV& Ca, H	& +8.9	& -6.6	& 3.7 & No	\\
05122759+2253492 & M2e IV-V& Ca, H, Li	& +0.8	& -15.4	& 3.1 & Yes	\\
\cutinhead{Upper Scorpius Sources}
15534689-1457364 & K7 V?   & Ca		& -19.3	& -10.6	& 3.5 & Yes	\\
16183457-1654379 & K2-3 III& Ca 	& +1.8	& -0.4	& 3.1 & No	\\
16184596-1659455 & K4e V   & Ca, H	& +1.9	& -38.9	& 3.1 & No	\\
16220590-1505118 & G3 V    & Ca		& +2.6	& +12.9	& 3.4 & No	\\
\enddata
\tablecomments{These are the emission-line stars with spectral type and CMD location consistent with Taurus or Upper Scorpius membership. We also present proper motions and errors calculated by A. Kraus (2009, private communication) from published positions along with a qualitative indication of whether the proper motion is consistent with membership in the Taurus or Upper Scorpius association, respectively.}
\end{deluxetable}

% Appendix tables

\begin{deluxetable}{lcccc}
\tablewidth{0pt}
\tablecaption{Absolute magnitudes for main sequence stars\label{ourms}}
\tablehead{\colhead{} & \colhead{$M_\textrm{FUV}$} & \colhead{$M_\textrm{NUV}$} & \colhead{$M_{J}$} & \colhead{$M_{K}$}\\
\colhead{Type} & \colhead{(mag)} & \colhead{(mag)} & \colhead{(mag)} & \colhead{(mag)}}
\startdata
B8&1.47&0.88&0.01&0.11	\\
\hline
A0&3.52&2.49&0.54&0.56	\\
A2&4.88&3.40&1.12&1.12	\\
A5&6.88&4.39&1.53&1.48	\\
A7&8.52&5.20&1.75&1.66	\\
\hline
F0&9.77&5.94&2.10&1.96	\\
F2&10.91&6.54&2.32&2.14	\\
F5&13.46&7.73&2.85&2.61	\\
F8&15.25&8.59&3.31&3.01	\\
\hline
G0&16.43&9.13&3.53&3.20	\\
G2&17.40&9.60&3.64&3.30	\\
G5&18.07&10.01&3.86&3.48	\\
G8&19.45&10.94&4.31&3.86	\\
\hline
K0&21.23&11.99&4.49&4.00	\\
K2&24.11&13.65&4.80&4.24	\\
K4&27.15&15.50&5.08&4.43	\\
K5&28.75&16.69&5.20&4.51	\\
K7&29.98&17.80&5.46&4.70	\\
\hline
M0&30.84&18.73&6.04&5.18	\\
M1&31.74&19.47&6.33&5.47	\\
M2&33.02&20.46&6.73&5.86	\\
M3& \nodata & \nodata &7.31&6.44	\\
M4& \nodata & \nodata &8.10&7.22	\\
M5& \nodata & \nodata &9.08&8.16	\\
M6& \nodata & \nodata &10.15&9.16	\\
M7& \nodata & \nodata &10.76&9.69	\\
M8& \nodata & \nodata &11.19&10.03	\\
M9& \nodata & \nodata &11.49&10.26	\\
\hline
L0& \nodata & \nodata &11.76&10.44	\\
\enddata
\tablecomments{GALEX magnitudes are given on the AB system; 2MASS magnitudes are Vega-based.}
\end{deluxetable}

\begin{deluxetable}{lccccc}
\tablewidth{0pt}
\tablecaption{Absolute AB magnitudes of some local white dwarfs\label{ourwds}}
\tablehead{\colhead{} & \colhead{} & \colhead{Distance} & \colhead{$T_\textrm{eff}$} & \colhead{$M_\textrm{FUV}$} & \colhead{$M_\textrm{NUV}$}\\
\colhead{ID} & \colhead{Type} & \colhead{(pc)} & \colhead{(K)} & \colhead{(mag)} & \colhead{(mag)}}
\startdata
WD~1134+300 & DA2 & 15.32	& 16000	& 11.11 & 11.41	\\
WD~2326+049 & DA4 & 13.62	& 13000	& 14.41 & 13.13	\\
WD~2105-820 & DA6 & 17.06	& 11000	& 15.80 & 13.21	\\
WD~1609+135 & DA6 & 18.35	&  9000	& 18.78 & 14.86	\\
WD~1953-011 & DA6 & 11.39	&\nodata& 20.48 & 15.10	\\
\enddata
\tablecomments{The temperatures are estimated from comparing the $FUV-NUV$ color to \citet{wdwarf}.}
\end{deluxetable}

\end{document}